\documentclass{aa}  

\usepackage{natbib}
\usepackage{amsmath}
\usepackage{graphicx}
\usepackage{svg}
\usepackage{txfonts}
\usepackage{ctable}
\usepackage{bbold}
\usepackage{placeins}
\usepackage[colorlinks]{hyperref}
\usepackage{xcolor}
\definecolor{dark-red}{rgb}{0.9,0.0,0.0}
\definecolor{dark-blue}{rgb}{0.15,0.15,0.9}
\definecolor{dark-green}{rgb}{0.15,0.8,0.15}
\definecolor{medium-blue}{rgb}{0,0,0.9}
\hypersetup{linkcolor={dark-blue},citecolor={dark-blue}, urlcolor={medium-blue}
}

\makeatletter
\renewcommand*\aa@pageof{, page \thepage{} of \pageref*{LastPage}} 
\makeatother

\newcommand\gaia{\textit{Gaia}}
\newcommand\gdrtwo{\gaia~DR2}
\newcommand\gedrthree{\gaia~EDR3}
\newcommand\gdrthree{\gaia~DR3}
\newcommand\hip{\textsc{Hipparcos}}
\newcommand\twomass{\textsc{2MASS}}
\newcommand\wise{\textsc{WISE}}

\newcommand\secref[1]{Sect.~\ref{#1}}
\newcommand\figref[1]{Fig.~\ref{#1}}

\newcommand\equref[1]{Eq.~\eqref{#1}}
\newcommand\tabref[1]{Table~\ref{#1}}
\newcommand\appref[1]{Appendix~\ref{#1}}

\def\gmag{$G$}

\def\parallax{$\varpi$}
\def\parallaxerror{$\sigma_{\varpi}$}

\def\magrm{~{\rm mag}}

\bibpunct{(}{)}{;}{a}{}{,} 

\usepackage{mathtools}
\newcounter{mysubequations}

\renewcommand{\themysubequations}{(\roman{mysubequations})}

\newcommand{\mysubnumber}{\refstepcounter{mysubequations}\themysubequations}

\begin{document} 

   \title{The Fifth Catalogue of Nearby Stars (CNS5)}

   \author{Alex Golovin\inst{1}\thanks{Fellow of the International Max Planck Research School for Astronomy and Cosmic Physics at the University of Heidelberg (IMPRS-HD)}
          \and
          Sabine Reffert\inst{1}
          \and
          Andreas Just\inst{2}
          \and
          Stefan Jordan\inst{2}
          \and
          Akash Vani\inst{2}
          \and 
          Hartmut Jahrei{\ss}\inst{2}}

   \institute{
   Landessternwarte,
   Zentrum f\"{u}r Astronomie der
       Universit\"{a}t Heidelberg, 
       K\"{o}nigstuhl 12, 69117 Heidelberg, Germany
   \and
   Astronomisches Rechen--Institut, Zentrum f\"{u}r Astronomie der
       Universit\"{a}t Heidelberg, 
       M\"{o}nchhofstr.~12--14, 69120 Heidelberg, Germany\\
     \email{agolovin@lsw.uni-heidelberg.de}}

  \date{Version: \today}

  \abstract{
We present the compilation of the Fifth
Catalogue of Nearby Stars (CNS5), based on astrometric and photometric data from \textit{Gaia}~EDR3 and \textsc{Hipparcos}, and supplemented with parallaxes from ground-based astrometric surveys carried out in the infrared. 
}{
The aim of the CNS5 is 
to provide the most complete sample of objects in the solar neighbourhood. For all known stars and brown dwarfs in the 25\,pc sphere around the Sun,
 basic astrometric and photometric parameters are given. Furthermore, we provide the colour-magnitude  diagram and various luminosity functions of the stellar content in the solar neighbourhood, and characterise the completeness of the CNS5 catalogue.
}{
We compile a sample of stars and brown dwarfs which most likely are located within 25~pc of the Sun, taking space-based parallaxes from \gedrthree{} and \hip{} as well as ground-based parallaxes from Best et al.\ (2021), Kirkpatrick et al.\ (2021), and from the CNS4 into account. We develop a set of selection criteria to clean the sample from spurious sources. Furthermore, we show that effects of blending in the \textit{Gaia} photometry, which affect mainly the faint and red sources in \textit{Gaia}, can be mitigated, to reliably place those objects in a colour-magnitude diagram.
We also assess the  completeness of the CNS5 using a Kolmogorov-Smirnov test and derive observational optical and mid-infrared luminosity functions for the main-sequence stars and white dwarfs in the solar neighbourhood.
}{The CNS5 contains 5931 objects,
including 5230 stars (4946 main-sequence stars, 20 red giants and 264 white dwarfs) and 701 brown dwarfs. 
We find that the CNS5 catalogue is statistically complete down to 19.7~mag in G-band and
 11.8~mag in W1-band absolute magnitudes, corresponding to a spectral type of L8. The stellar number density in the solar neighbourhood is $(7.99\pm0.11)\times10^{-2}\,\text{stars pc}^{-3}$, and 
about 72\% of stars in the solar neighbourhood are M dwarfs. Furthermore, we show that the white dwarf sample in CNS5 is statistically complete within 25~pc.
The derived number density of white dwarfs is $(4.03\pm0.25)\times10^{-3}\,\text{stars pc}^{-3}$. The ratio between stars and brown dwarfs within 15\,pc is $4.6\pm0.4$, whereas within 25\,pc it is $7.5\pm0.3$. 
Thus, we estimate that about one third of brown dwarfs is still missing within 25~pc, preferentially those with spectral types later than L8 and distances close to the 25~pc limit.
}{}


   \maketitle

   \authorrunning{A.\,Golovin\,et\,al.}
 \titlerunning{The Fifth Catalogue of Nearby Stars (CNS5)}

\section{Introduction}
\label{sec:intro}

\citet{hertzsprung22} was probably the first to notice that the systematic study of the stars in the solar neighbourhood is scientifically very valuable, for a multitude of reasons. First and foremost, the solar neighbourhood stars are the brightest (in terms of apparent magnitude) and largest (in terms of angular size) representatives of each spectral type, which allows us to observe them more easily while they can serve as proxies for more distant stars of the same type. Furthermore, the 
stellar sample in the solar neighbourhood provides the most complete census of the 
Galactic disc population, covering more than 20\,mag in absolute visual brightness.

Often, the solar neighbourhood stars are used to develop concepts that are later applied elsewhere in the universe.
Applications span a diverse range of astronomical topics, such as the number of stars in the Galaxy and Galactic disc modelling \citep{pritchet83, bahcall86,Bin10,Just10}, the present-day and the initial mass functions 
\citep{miller79,rana92,sollima19}, the stellar luminosity function and its fine structure \citep{reid02,chabrier03,bahcall86,upgren81,wielen83,kroupa90,jao18}, constraints on star formation \citep{kirkpatrick19}, the local stellar surface density \citep{mckee15}, stellar multiplicity \citep{duquennoy91,dieterich12}, brown dwarfs \citep{meisner20}, and target lists for exoplanet searches \citep{reiners18,turnbull03}. 
This compilation is certainly not exhaustive; the cited works merely
serve as an illustration of studies that have been conducted in each field. The solar neighbourhood sample hereby often complements other large photometric and spectroscopic datasets which have become available, such as RAVE \citep{steinmetz20}, SDSS \citep{ahumada20}, and PanSTARRS \citep{kaiser02}.

The number of surveys which are based on the nearby stars sample is expected to grow considerably in the future,
especially due to the many endeavours to discover
and characterise exoplanets. 
NASA has acknowledged the fact that many of the
upcoming exoplanet search programs overlap in their target lists, especially for the nearby stars, and has established a working group (SAG~22) on the topic of a target star archive \citep{hinkel21}.
Nearby stars possess
several advantages over more distant stars as exoplanet hosts, such as better resolution for direct imaging 
(HabEx, \citealt{gaudi20}; LUVOIR, \citealt{luvoir19}), or better stellar characterisation.
Continued interest in the nearby stars sample is thus expected for the foreseeable future. 

For these reasons we present here an update
of the {\it Catalogue of Nearby Stars}, denoted
CNS5. An update is well in order; since the publication of the {\it Preliminary Version of the Third Catalogue of Nearby Stars} \citep{cns3}, a fourth version of the catalogue, CNS4, had been prepared by Hartmut Jahrei{\ss}, but never published.

While CNS4 was for the most part based on
space-based parallaxes from \hip{} \citep{hipcat, vanleeuwen07, vanL07pap}, CNS5 incorporates
\textit{Gaia} data (from the release \gedrthree{}, \citealt{gaiaedr3_summary, gaiaedr3_astromertry}). Due to \textit{Gaia}'s superb
astrometric precision and its large number
of catalogue sources down to $G\sim20$\,mag and fainter, 
the completeness of the nearby star
sample and the accuracy of astrometric and photometric
parameters is tremendously improved. At the same
time, the number of falsely included stars is
also dramatically reduced. 

As opposed to the recently published {\it Gaia Catalogue of Nearby Stars} (GCNS, \citealt{gaiaedr3_gcns}), which lists all stars
within 100~pc, the CNS5  sticks to its traditional volume of 25~pc around the Sun. CNS5 is not
purely based on \textit{Gaia} data, but also incorporates data from \hip{} where favourable, as well as from CNS4 and from ground-based astrometric surveys carried out in the near-infrared. 
It aims for completeness and cleanliness of the final nearby stars sample to the extent possible, in order to facilitate statistical studies based on the volume-limited sample as a whole. In contrast, the GCNS includes all stars with a non-zero probability to lie within 100~pc, so it aims more at completeness rather than cleanliness near the distance limit. 
Naturally, completeness is higher in CNS5 compared to the GCNS, but this comes at the expense of a smaller volume. 
The applications of the two catalogues are thus to some extent complementary. 

The compilation of the CNS5 is much more complicated than just selecting all sources with parallaxes larger than 40~mas. 
The reasons include a number of issues: i) spurious catalogue entries with apparently large parallaxes \citep{gaiaedr3_astromertry,gaiaedr3_gcns} which have to be filtered out, ii) blended sources whose photometry has to be deblended so that they can be reliably placed into a Hertzsprung-Russell diagram, and iii) the various components of multiple systems which have to be treated in a consistent way, even if reliable parallaxes do not exist for all of the components. 
Furthermore, we do not only use \gedrthree{}, but also the \hip{} Catalogue \citep{hipcat} for the brighter stars missing in \textit{Gaia} as well as ground-based infrared parallax surveys \citep{Best_2021AJ....161...42B, Kirkpatrick_2021ApJS..253....7K} for the red and optically faint objects such as brown dwarfs. 
We also derive completeness limits and white dwarf and main-sequence star luminosity functions. 

The paper is organised as follows. 
In \secref{sec:history} we describe various other catalogues of nearby stars, both previous and current. In \secref{sec:construction}
we present the philosophy behind the compilation of the CNS5 as well as our methods.
Section~\ref{sec:description} provides a description of the catalogue content, while \secref{sec:content} addresses completeness limits as well as the luminosity function; it also presents the colour-magnitude diagram (CMD) of all catalogue stars. 
Section~\ref{sec:discussion} provides our conclusions and a summary. Detailed descriptions of algorithms are provided in the appendices.

\section{Overview of the catalogues of nearby stars}
\label{sec:history}

\subsection{Catalogue of Nearby Stars Series (CNS)}
\label{sec:hist_CNS}

\subsubsection{Previous editions}

The first version of the Catalogue of Nearby Stars (CNS1) was published by \citet{CNS1_1957}; it contained 1094 stars (915 systems) within 20 pc.
The subsequent update of the catalogue (CNS2) expanded the coverage to 22.2 pc and contained 1627 stars in 1313 systems \citep{CNS2_1969}. 
It lists photometric, spectroscopic and trigonometric parallaxes as well as the `resulting parallax', an estimate of the best value considering all measurements.  

CNS3, entitled \textit{Preliminary Version of the Third Catalogue of Nearby Stars}, lists 3403 stars within the 25~pc distance limit \citep{cns3}.
In contrast to CNS2,
the resulting parallax in CNS3 is the trigonometric parallax, unless it is either not available or has unusually large errors. 
Although this version of the catalogue was denoted as `preliminary', it is still the most recent publicly available edition of the CNS (a `final' version was never published).
Therefore, many recent publications are still based on the CNS3 (e.g.\ \citet{Tamazian_2014_AcA....64..359T, Bar_2017_ApJ...850...34B, Price_2020_AJ....159...86P} to name a few).

The 4th version
(CNS4) was created by \citet{Jahr97}. It contained ground-based parallaxes as well as
trigonometric parallaxes from \hip{}. However, the data from previous versions of the CNS series was not
fully homogenised with the new data, which lead to inconsistencies.

Over decades, one of us (Hartmut Jahrei{\ss}) has continuously scanned the literature to update astrometry, photometry and other supplementary data (radial velocities, binary information) of potential nearby stars. This information has also been used in the construction of the CNS5.

\subsubsection{Numbering scheme in the CNS}
\label{sec:hist_numbering}

Stars in the CNS1 were designated as ~\texttt{Gl~NNN}, where \texttt{NNN} represents the consecutive integer number ordered by right ascension. In the CNS2 designations were formatted as ~\texttt{Gl~NNN.N}, adding a decimal place for the new stars. This allowed the original Gliese star numbers to be preserved and the ordering system by right ascension to be kept. 

Since the publication of an extension to CNS2 \citep{CNS2_Suppl_1979}, the stars were referred to as ~\texttt{GJ~NNNN}. 
In addition, the CNS3 had listed
Woolley numbers \citep{Woolley_1970ROAn....5.....W} and introduced NN designations for new stars; both of them have been replaced with GJ numbers in the mean time. By now, all the previously known nearby stars thus have GJ designations.
One can still tell from where they originated by their number: Woolley numbers start at 9001 and the largest assigned number is 9850, while NN numbers range from 3001 to 4388.
The widely used Gliese-Jahrei{\ss} (GJ) numbers have been kept in the CNS5 (see \secref{sec:CNS5_numbering_scheme} for details).

\subsection{Other catalogues}
\label{sec:hist_other_catalogues}

\subsubsection{The RECONS project}
\label{sec:hist_RECONS}

The REsearch Consortium On Nearby Stars (RECONS; founded in 1994) has made an enormous effort to discover and characterise stars and ultra-cool dwarfs in the 25~pc volume, primarily by measuring their trigonometric parallaxes from the ground \citep[e.g.][]{henry18}. 
They provided updated astrometric, photometric and multiplicity information, and identified many new nearby stars since the publication of the Yale Parallax Catalog \citep{YPC_van_Altena_1995_gcts.book.....V}. Their results are published in The Solar Neighborhood series of papers in \textit{The Astronomical Journal} (49 papers as of June 2022), making the compilation of the 25~pc sample from their work rather cumbersome.
Only statistics for the 10~pc volume is given on their website.
As of 2018, the RECONS database contained 317 systems within 10~pc.
An overview of new nearby stars within 10~pc discovered by RECONS is provided in \citet{henry18}.

\subsubsection{Catalogue of stars within 50~pc based on \gdrtwo{}}
\label{sec:hist_50pc_sample}

\citet{torres_50pc_2019A&A...629A.139T} studied the dynamical evolution of the comets in the Oort cloud. In particular, they characterised stellar encounters with the Solar System by integrating their orbits based on \gdrtwo{} data. For this purpose, the catalogue of stars within 50~pc based on \gdrtwo{} data has been constructed. The catalogue lists 14~659 stars.

However, the approach used for source selection is a potential concern. Sources with high-quality astrometry were selected based on the relative uncertainty of the parallax (smaller than 20\%), flux excess factor, and unit weight error ($UWE$). The re-normalisation of $UWE$ ($RUWE$; \citealt{lindegren18b}) was not incorporated when selecting sources for the catalogue. Such re-normalisation would preserve objects with extreme colours (e.g.\ such as brown dwarfs) from being removed because, contrary to $UWE$, not only is the object's $G$ magnitude taken into account when calculating $RUWE$, but also its colour.

As the authors pointed out themselves, the catalogue is incomplete. The prime culprit here is the \gaia{} magnitude limit and, as a consequence, a significant number of faint low-mass stars is not detected.
Furthermore, some of the brightest sources are also missing due to the \gaia{} saturation limit.

\subsubsection{GCNS}
\label{sec:hist_GCNS}

The GCNS \citep{gaiaedr3_gcns} aims to identify all sources in \gedrthree{} with reliable astrometry and a non-zero probability of being located within 100~pc from the Sun. 
This is done by retrieval of sources with parallaxes larger than 8~mas, removal of spurious sources with machine-learning (using the random forest classifier) and by Bayesian distance estimation for the remaining sources.

The full catalogue lists 331\,312 sources.
The overall completeness of the GCNS is expected to be 95\% or better for spectral types up to M8 (translating into $M_G = 15.7~{\rm mag}$).

It is important to bear in mind that the prime focus of the GCNS is to provide a homogeneous census of nearby stars.  Its volume is larger than that of the CNS5, but bright and red sources could be missing since the GCNS is based only on \gedrthree{} data. As we will see, due to the inclusion of infrared surveys in the CNS5, the CNS5 is complete for much later spectral types well into the brown dwarf regime. 

\subsubsection{The 10~pc sample by Reyl\'e et al. (2021)}
\label{sec:hist_10pc}

\citet{Reyle_2021A&A...650A.201R} compiled the sample of stars, brown dwarfs, and exoplanets located within 10~pc of the Sun as a quality assurance test for the GCNS. Using this sample, it is possible to not only identify very nearby stars that are missing from the GCNS (or future \gaia{} data releases), but also infer the completeness of the GCNS by estimating the expected number of objects of different types from the number densities for the 10~pc volume.

Objects for the 10~pc sample were selected from SIMBAD using a strict cut on parallax at $\geq 100$~mas, with parallax uncertainties playing no role in the selection process. The sample was supplemented with components of unresolved binaries, brown dwarfs with recently published parallaxes from \citet{Best_2021AJ....161...42B} and \citet{kirkpatrick19, Kirkpatrick_2021ApJS..253....7K}, and confirmed exoplanets from the Extrasolar Planets Encyclop{\ae}dia\footnote{\url{http://exoplanet.eu/}} as well as from the NASA Exoplanet Archive\footnote{\url{https://exoplanetarchive.ipac.caltech.edu/}}.

The sample consists of 540 objects (including 77 exoplanets) in 339 systems. The Sun and its planets are not included.
The 10~pc sample has separate entries for every object, including the unresolved components of multiple systems and exoplanets.

Naturally, among the catalogues reviewed in this section, the 10~pc sample by \citet{Reyle_2021A&A...650A.201R} has the highest completeness due to its smaller volume. At the same time, the number of objects contained within this volume is insufficient for statistical studies of the solar neighbourhood.

Future updates of the  10~pc sample are anticipated to have only a minor impact on the stellar content of the sample; the majority of the new additions to the sample are expected to come from single stars that have been resolved into multiple components.
In the substellar regime, the expected additions will be ultra-cool dwarfs (particularly those in the Galactic plane) and, of course, new exoplanets.

\section{Construction of the CNS5 catalogue}
\label{sec:construction}

\subsection{Data selection}
\label{sec:data_selection}

Our goal for the compilation of the CNS5 is to 
 maximise completeness within the 25~pc limit. Therefore, the wealth of data from several different catalogues and surveys has to be carefully selected and combined. In the following subsections we give a full account of our selection criteria, applied corrections, and how we combined data from the different catalogues to consolidate the CNS5.

\subsubsection{\gedrthree{}}
\label{sec:eDR3}

Due to its high completeness and high accuracy \gedrthree{} 
is an excellent starting point for the preparation of a catalogue of nearby stars.
For the vast majority of objects the \gaia{} parallax clearly indicates whether the star is located within 25~pc from the Sun or not. Therefore, CNS5 is built primarily based on the data selected from \gedrthree{} and then, if necessary, supplemented with data from \hip{} and other catalogues and sources.

The primary sample for CNS5 is constructed by retrieving from \gedrthree{}  all the objects that are possibly located within 25 pc from the Sun:
\begin{equation}
    \label{eq:plx_cut}
    \varpi_{\rm EDR3}-Z_{5, 6}+3\eta\sigma_\varpi{}_{\rm EDR3}\geq 40~\text{mas}, 
\end{equation}
where \parallax{$_{\rm EDR3}$} denotes the measured parallax in the \gedrthree{} catalogue, \parallaxerror{$_{\rm EDR3}$} its standard error, $Z_{5, 6}$ is the parallax zero-point for the five and six parameter solutions in \gedrthree{},
and $\eta$ is the parallax error inflation factor. The parallax zero-point correction and the inflation factor for the parallax error will be discussed below.
The query returns 5876 sources.

\begin{figure}
    \centering
    \includegraphics[width=0.49\textwidth]{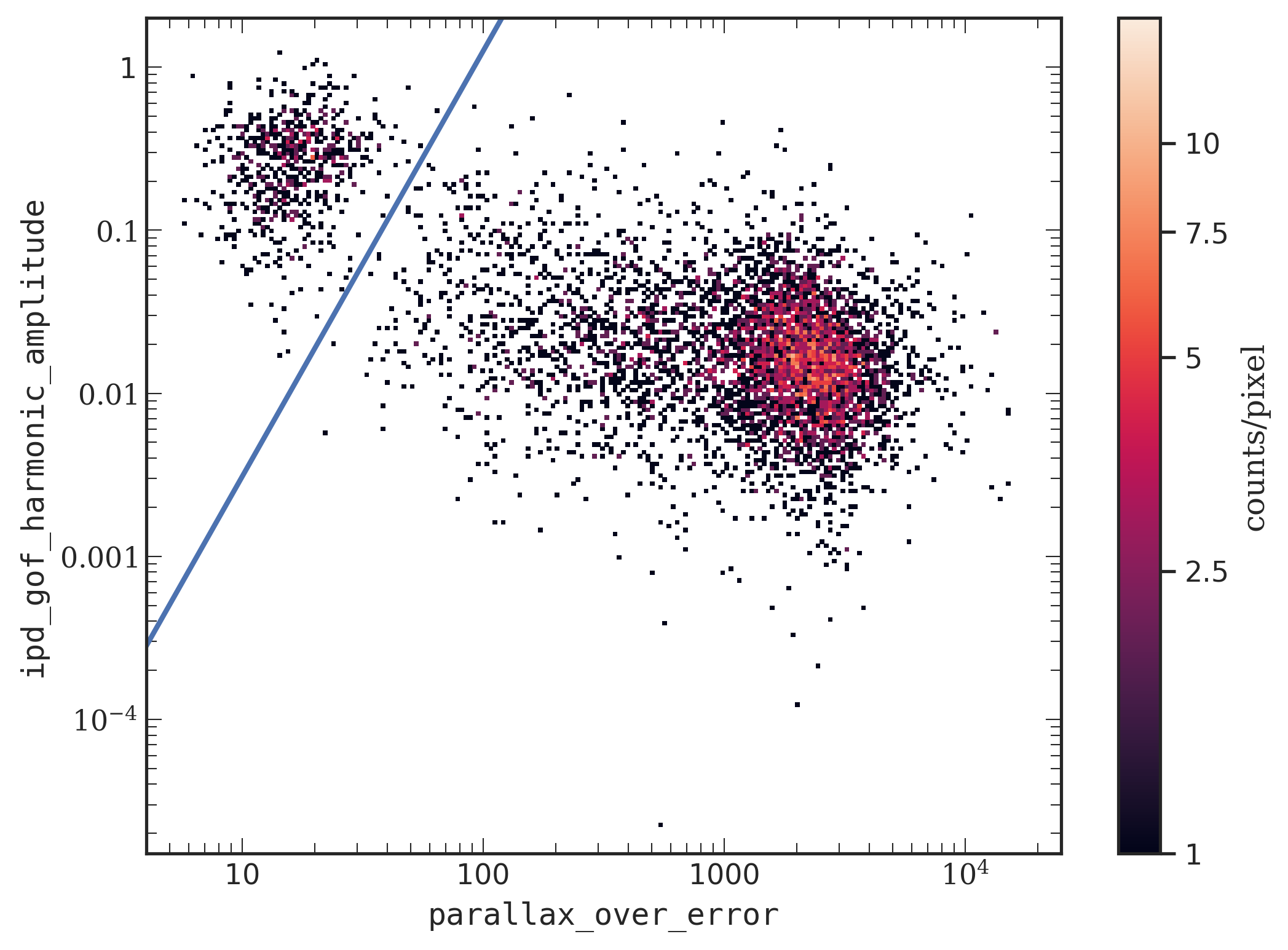}
    \caption{Dichotomy between sources with spurious and reliable astrometric solutions. The blue line corresponds to \equref{eq:ipd_gof_harm_ampl_cut}; 692 sources (11.78\%) are located above this threshold and hence their astrometric solutions are spurious.}
    \label{fig:ipd_harm_ampl_cut_25}
\end{figure}

Sources with spurious astrometric solutions are removed by applying a simple yet powerful cut on the amplitude of the Image Parameter Determination goodness-of-fit (IPD GoF; \texttt{ipd\_gof\_harmonic\_amplitude} in \gedrthree{}). Thus, we construct our sample by retaining only sources that satisfy the following condition:
\begin{equation}
    \label{eq:ipd_gof_harm_ampl_cut}
    A_{\rm GoF}<10^{-5.12}(\varpi_{\rm EDR3}/\sigma_\varpi{}_{\rm EDR3})^{2.61},
\end{equation}
where $A_{\rm GoF}$ is the amplitude of the IPD GoF.

Figure~\ref{fig:ipd_harm_ampl_cut_25} shows that sources with spurious solutions and nearby sources with good parallaxes form two distinct groups in (\texttt{ipd\_gof\_harmonic\_amplitude, parallax\_over\_error}) parameter space and are separated by \equref{eq:ipd_gof_harm_ampl_cut}.
This cut is discussed and validated in detail in Appendix~\ref{sec:harm_ampl_cut}.

One of the strengths of this cut is that the photometric parameters themselves are not involved in the selection process. This is opposed to other widely used selection criteria such as $RUWE$, where the re-normalisation factor is a function of the $G$-band magnitude and the $BP-RP$ colour \citep{lindegren18b} or the
photometric excess factor, which is based solely on photometry \citep{Evans_gaia_dr2_photom_2018A&A...616A...4E, lindegren18}.
By filtering with \equref{eq:ipd_gof_harm_ampl_cut} we avoid introducing a selection bias against, for instance, variable stars. It has been shown that an excess in $RUWE$  can be induced not only by binarity of an object but also by its photometric variability \citep{Belokurov_ruwe_2020MNRAS.496.1922B}. Such a source is observed at a broad range of magnitudes (and sometimes even colours) and consequently this affects the $RUWE$ re-normalisation coefficient, which is derived assuming constant apparent magnitudes. The photometric excess factor must be interpreted with caution too: while it is common to use it as a data quality indicator to identify issues with crowding or background subtraction, it is important to bear in mind that variable sources frequently have large values of flux excess as well \citep{gaiaedr3_photometry}.

It is important to bear in mind that our cut, in principle, may reject a few unresolved or marginally resolved binaries. However, the number of rejected objects should be negligible and probably comes down to the 12 objects listed in \tabref{tab:gcns_rejected}. Furthermore, we do not know the true parallaxes of these objects since their astrometric solution (which is based on the assumption of the single-star model) might be affected by binarity. Thus, it is not clear whether these objects even belong to the 25~pc volume.

Parallaxes published in \gedrthree{} are known to be underestimated by a few tens of microarcseconds,  and a correction of this systematic parallax offset is needed \citep{gaiaedr3_zpt}.
In this paper it has also been demonstrated that the parallax zero-point correction (parallax bias) depends on the G magnitude of the object, its colour\footnote{In the form of an effective wavenumber $\nu_{\rm eff}$ for a five-parameter solution or a pseudo-colour $\hat{\nu}_{\rm eff}$ for a six-parameter solution.} and ecliptic latitude $\beta$. 

We would like to stress that the parallax zero-point correction is not negligible for nearby stars, even though their parallaxes are several orders of magnitude larger than the parallax zero-point offset.
It has been shown by Bailer-Jones\footnote{\url{https://www2.mpia-hd.mpg.de/homes/calj/gedr3_distances/FAQ.html}} that the distances in the GCNS are overestimated on average by 0.22~pc, and the dominant cause for this difference is that GCNS does not incorporate a correction for the parallax zero-point offset.

We computed parallax zero-point values for each source in our sample from \gedrthree{} according to the procedure of \citet{gaiaedr3_zpt} and subtracted it from the parallax value published in \gedrthree{}.
16 sources would not have been part of the CNS5 sample if we had not applied the parallax zero point, so the correction is not only important for deriving the most accurate distances but also sample selection and its completeness.

Regarding the parallax uncertainties in \gedrthree{}, several studies validated them using different approaches and have shown that the uncertainties are underestimated \citep{El_Badry_inflation_2021MNRAS.506.2269E,
gaiaedr3_validation, Maiz_Apellaniz_inflation_2021A&A...649A..13M, Vasiliev_inflation_2021MNRAS.505.5978V, Zinn_inflation_2021AJ....161..214Z}.
We inflated parallax errors in our sample using an empirical function for the inflation factor derived in \citet{El_Badry_inflation_2021MNRAS.506.2269E}.
It is important to bear in mind that the inflation factor must be interpreted as a lower limit. The underestimation of the parallax uncertainty 
is even larger for sources with large $RUWE$ values, binaries with small angular separation, and for sources in the vicinity of other bright sources.
For the bright stars outside the interpolation region $7 \magrm<G<21~{\rm mag}$, we have decided to also inflate the parallax uncertainties, given that for $2 \magrm <G<7 \magrm$ the inflation function is well-behaved and flattens towards a constant value of $\sim1.15$ . By doing so, we ensure that the parallax uncertainty distribution remains continuous within the whole magnitude range $2 \magrm <G<21 \magrm$. Consequently, this will yield more accurate and consistent results in astrophysical applications of our catalogue.

Six objects in the CNS5 would not have been included if the parallax errors would not have been inflated.
These objects all have $G$ magnitudes in the range $11.03 \magrm \leq G \leq 15.35 \magrm$, so that they are well within the interpolation region $7 \magrm<G<21~\magrm$ of Eq.~(16) in \citet{El_Badry_inflation_2021MNRAS.506.2269E}.

The parallaxes from \gdrtwo{} were not considered during the data selection process, and there are several reasons that justify this decision.
Typically, the parallax uncertainties in \gedrthree{} are smaller by a factor 0.8 or better than in \gdrtwo{} due to improved calibrations and the longer time span of the observations: \gdrtwo{} is based on 22 months of observations while \gedrthree{} is based on data collected during 34 months \citep{gaiaedr3_astromertry}.
In a few cases, the opposite is true and, despite the longer time span of the observations, the parallax uncertainty is larger in \gedrthree{}.
This can often be a consequence of the object being an unresolved binary with the orbital period comparable with the \gedrthree{} temporal baseline. 
Over time, the contribution of the orbital motion accumulates and can induce a photocentre perturbation.
This leads to the significant inconsistency of the observed displacement of the source with the single-star model fit, which in turn translates into increased uncertainties of parallaxes and proper motions.

Furthermore, we have decided against adding new objects with the parallaxes from \gdrtwo{} which have just two-parameter solutions in \gedrthree{} or are missing completely. The vast majority of such sources have spurious solutions in \gdrtwo{} and did not meet the acceptance criteria during reprocessing in \gedrthree{} \citep{gaiaedr3_astromertry}. Including these sources would thus contaminate our catalogue. 
We would like to remark that sources with spurious solutions in \gdrtwo{} cannot be eliminated with our criterion in \equref{eq:ipd_gof_harm_ampl_cut}
since the IPD GoF statistics are not included in that data release. Finally, it appears not to be feasible to construct a volume-limited sample of nearby stars based on \gdrtwo{} which would be free from spurious entries and at the same time without removing too many real sources. As we outlined above, the widely used selection criteria -- namely $RUWE$ and the photometric excess factor -- are sub-optimal for this task. Alternatively, applying machine-learning will not necessarily provide a classification which is robust enough. As reported in \citet{gaiaedr3_gcns}, when the same random forest classifier as the one which was used to construct the GCNS has been applied to \gdrtwo{} (but with the predictive variables adapted to those available in \gdrtwo{}), the resulting sample still contained 15 sources with $\varpi > 500\ \mathrm{mas}$, whereas in \gedrthree{} there are only two such sources -- \object{Barnard's star} and \object{Proxima Centauri}.

\subsubsection{\hip{}}
\label{sec:HIP}
While \gedrthree{} includes the majority of known nearby stars,
it is certainly not complete, as in particular the very brightest stars are missing.
As the next step, we thus supplemented our \gedrthree{} sample with entries from the \hip{} Catalogue \citep{Perryman_HIP_1997}, which is complete for 
$V$ magnitudes brighter than 7.3--9.0~mag.
By doing so, we add objects which are too bright for \gaia{} and thus missing, or which do not have a full astrometric
solution in \gedrthree{}. Furthermore, \gedrthree{} is also incomplete for high proper motion stars: $\sim8\%$ of known stars with proper motions $\gtrsim 0.6~\mbox{arcsec\,yr}^{-1}$ are missing in \gedrthree{} \citep{gaiaedr3_validation}.
Another major advantage of including data from \hip{} is that for very bright objects it usually provides astrometry of higher quality than the one from \gedrthree{}.

Similar as for the selection of stars from \gedrthree{}, objects from the \hip{} Catalogue were selected based on their parallax. In addition, we impose an upper limit on parallax error to avoid adding objects with extremely large parallax uncertainties, which often result from photocentre wobble of unresolved or marginally resolved binaries:
\begin{equation}
\label{eq:hip_plx}
\left.\begin{aligned}
\text{(i)}\quad& \varpi+3\sigma_\varpi{} \geq 40~\text{mas}\\
\text{(ii)}\quad& \sigma_\varpi{} \leq 10~\text{mas}
\end{aligned} \quad \right\},
\end{equation}
where \parallax{} is the parallax from the \hip{} Catalogue and \parallaxerror{} its formal error.
In this paper we use the second reduction of the \hip{} astrometric catalogue with improved parallaxes and their formal errors \citep{vanleeuwen07, vanL07pap}.
The selected sample contains 2\,007 sources.
We adopted the cross-match with \gedrthree{} provided by \citet{gaiaedr3_marrese}, described in \citet{gaiaedr3_documentation_ch9_marrese}.

\begin{figure}
        \centering
        \includegraphics[width=0.49\textwidth]{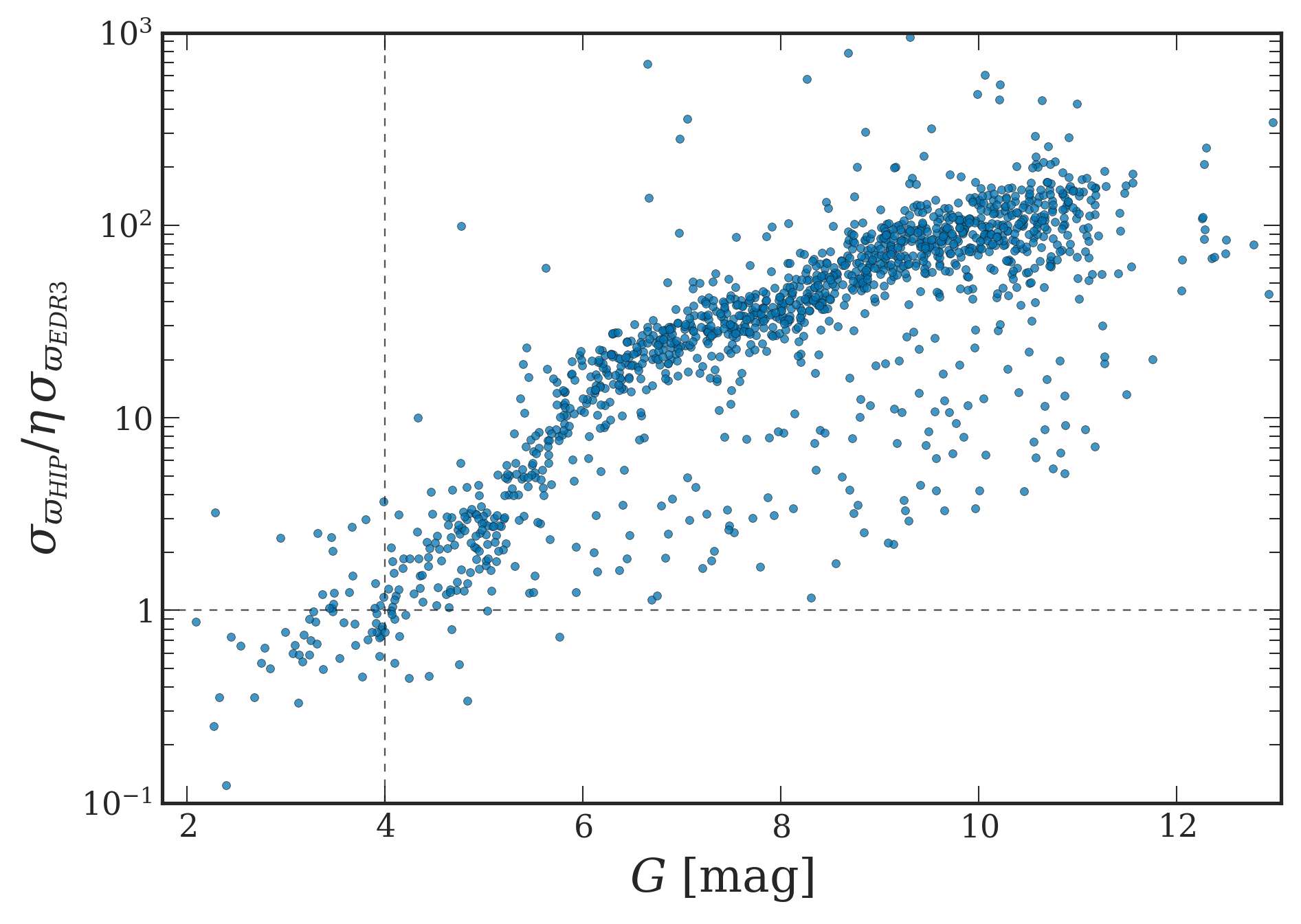}
        \caption{Ratio of parallax errors in the \hip{} and \gedrthree{} catalogues as a function of $G$ magnitude. Parallax errors from \gedrthree{} are the inflated ones. For the objects brighter than $G \approx 4$\,mag (vertical dashed line) parallax uncertainties in \hip{} are typically smaller than those from \gedrthree{}.}
        \label{fig:err_hip_gaia}
    \end{figure}

A comparison of parallax errors from \hip{} and \gedrthree{} shows that for objects fainter than $G\,{\approx}\,6$\,mag the \gedrthree{} parallaxes typically have one to two orders of magnitude higher precision
than those measured by \hip{}.
However, parallaxes from \hip{} for very bright objects have smaller standard errors, due to the saturation limit of \gaia{}.
Figure~\ref{fig:err_hip_gaia} shows the ratio of parallax errors from \hip{} and inflated errors from \gedrthree{} (\parallaxerror{$_{\rm HIP}$}/$\eta$\parallaxerror{$_{\rm EDR3}$}) as a function of the 
\gaia{} $G$-band magnitude.
Here we found that for the objects brighter than $G \approx 4$\,mag \hip{} parallaxes are typically better (i.e.\  \parallaxerror{$_{\rm HIP}$}/$\eta$\parallaxerror{$_{\rm EDR3}$}~$< 1$).
Therefore, the parallax values
with the smaller uncertainties (\parallaxerror{$_{\rm HIP}$} or $\eta$\parallaxerror{$_{\rm EDR3}$}) will be adopted as a resulting parallax in the CNS5 for those objects which are present in both catalogues, \gedrthree{} and \hip{}.

\subsubsection{Ultra-cool dwarfs}
\label{sec:ultracooldwarfs}

Objects fainter than the \gaia{} magnitude limit of 20--21~mag in $G$ band are so far missing from our sample. 
This concerns many of the L and T dwarfs in the 25~pc volume. 
Thus we decided to add nearby ultra-cool dwarfs from the compilation by 
\citet{Best_2021AJ....161...42B}, who recently provided a sample which is complete 
within 25~pc for spectral types from L0 to T8.
All stars in their sample have parallaxes,
many of them measured as part of the Hawaii Infrared Parallax Program \citep[][and later papers in the series]{dupuy2012}. As the next step, we updated and supplemented the resulting sample with \textit{Spitzer} parallaxes from \citet{Kirkpatrick_2021ApJS..253....7K}.

As usual, we included only objects that are possibly within the 25~pc volume according to Eq.~\ref{eq:hip_plx} and 
discarded those objects with parallax errors larger than 10~mas. 
Again, only trigonometric parallaxes were considered. 
This resulted in a sample of 906 objects.
Typical parallax errors in the resulting sample of ultra-cool dwarfs range from 1 to 10~mas, larger than those from \hip{} and \gaia{}, as can be seen in \figref{fig:e_plx_cns3_cns5}. 
Nevertheless, adding the ultra-cool dwarfs helps tremendously with completeness 
towards the smallest masses. In total, 462 ultra-cool dwarfs without counterparts in \gedrthree{} were added to the CNS5 from the sample of \citet{Best_2021AJ....161...42B} and 79 from the sample of \citet{Kirkpatrick_2021ApJS..253....7K}.

\subsubsection{CNS4}
\label{sec:CNS4}

We add  sources from the internal version of the CNS4 which are missing so far in our sample from \gedrthree{}, \hip{} and \citet{Best_2021AJ....161...42B}.
This is done by identifying and discarding those objects in CNS4 that (i) are already included in our sample, (ii) now have higher-precision parallax measurements that move them outside the 25~pc limit, (iii) were classified as a nearby star in the past, whereas newer studies showed that they actually have non-stellar nature such as, for instance, a plate flaw (\object{GJ~2087}) or even BL~Lac objects (\object{GJ~3750} and \object{GJ~3848}), (iv) are sources listed as a component of a nearby star, whereas newer studies could not confirm them, concluded that they are non-existing or showed that they are not physically bound and therefore are background stars, (v) have trigonometric parallaxes with errors exceeding an upper limit of $10~\text{mas}$, (vi) have only photometric parallax measured.

Cross-matching CNS4 with our sample was often compounded by inconsistencies in the CNS4 between listed designations and coordinates, which were not always up to date. 
Such cases had to be handled manually and frequently required significant amounts of detective work.

For every object retained in the resulting sample, we collected updated astrometry from the literature and supplemented it with optical, NIR and MIR photometry whenever possible.
Our attempts to identify a counterpart in \gedrthree{} with a two parameter solution were successful for objects predominantly of intermediate brightness, whereas an unambiguous identification of \gedrthree{} counterparts for fainter objects was often stymied by the presence of two or more sources of similar brightness (and with only a two parameter solution in \gedrthree{} for them) in the vicinity.
Confident cross-matching with \gedrthree{}, even when only a two parameter solution was available, provides not only highly accurate coordinates but also photometry in the \gaia{} bands.
When combined with parallaxes from the literature, this allows to reliably place these sources in the CMD.

The objects added from the CNS4 are typically M dwarfs of intermediate brightness with only a two parameter solutions in \gedrthree{}. 
The additional 27 objects from CNS4 are crucial to maximise completeness of the CNS5.

\begin{table*}
\caption{\object{GJ~559} and \object{GJ~551} as an example for indexing binary and multiple systems in the CNS5.}
\label{tab:binsystGJ}
    \centering
    \begin{tabular}{llccl}
        \hline
        \hline
         \texttt{cns5\_id} & \texttt{gj\_id} & \texttt{component\_id} & \texttt{n\_components} & \texttt{gj\_system\_primary}   \\
         \hline
         CNS5 3591 & 551 & C & 3 & 559 \\
         CNS5 3628 & 559 & AB & 3 & - \\
         \hline
    \end{tabular}
\end{table*}

\subsection{The Sun}
\label{sec:individual_objects}

As was pointed out in the introduction, the aim of the CNS5 is to provide the most complete census of nearby stars possible, so that it can be used for statistical studies of the solar neighbourhood. Thus, there is no strong reason not to include the nearest star - the Sun. On the other hand, due to its proximity, its  treatment is different from that of other stars in the catalogue, and a few notes have to be made here. 

In the catalogue entry for the Sun we list only apparent magnitudes in different photometric bands. Consistent with previous versions of the CNS catalogues, we do not assign any GJ designation to this entry.

Traditionally, the distance between the Sun and Earth (and thus absolute magnitudes of the Sun) was estimated from the solar parallax\footnote{Defined as the angle subtended by the equatorial radius of the Earth at the mean distance of the Sun.}.
The determination of the solar parallax has been a fundamental problem in observational astronomy over the centuries \citep{solplx_1857AJ......5...53B, solplx_1878MNRAS..38..429T, solplx_1909MNRAS..69..544H, solplx_1912PASP...24..211T, solplx_1943ASPL....4..144W, solplx_1961Natur.190..519T}. For an overview of principles of the determination of the solar parallax with \gaia{} please refer to \citet{solplx_gaia}.
However, since 2012 the Astronomical Unit (au) is a conventional unit of length and equals exactly 149\,597\,870\,700~m (translating to the solar parallax of $\varpi_\odot=8.794143836\arcsec$), as defined by \citet{iau_2012_b2}.
Given that the stellar parallax is non-definable for the Sun and that our catalogue only lists trigonometric stellar parallaxes, no parallax value is given for the Sun.

Turning now to photometry, the solar magnitudes and colours serve as an important reference point in many areas of astrophysics. However, we do not have a measured value of neither $G_\sun$, nor its colours $(G-RP)_\sun$ or $(BP-RP)_\sun$. The approach to be taken here is to use magnitudes and colours estimated from flux calibrated solar reference spectra. \citet{casagrande_2018MNRAS.479L.102C} reported magnitudes of the Sun in the \gaia{} bands in the \texttt{Vegamag} system derived from absolute flux measurements, both ground-based and from space, combined with model spectra. Hence, we adopt $G_\sun=-26.90$\,mag (this corresponds to an absolute magnitude of $M_{G_\sun}=4.67$\,mag), $(G-RP)_\sun = 0.49$\,mag and $(BP-RP)_\sun=0.82$\,mag for the Sun.

Similarly, \citet{Willmer_2018ApJS..236...47W} estimated the apparent and absolute magnitudes of the Sun in a large number of other broad-band filters used in various surveys and observatories. Thus we tabulate in the CNS5 the NIR and MIR magnitudes of the Sun, that is, in 2MASS ($J, H, K_s$) and WISE ($W_1, W_2, W_3, W_4$) filters. We remark that the magnitudes of the Sun listed in the CNS5 are in the \texttt{Vegamag} system.

\section{Description of the catalogue}
\label{sec:description}

\subsection{Numbering scheme}
\label{sec:CNS5_numbering_scheme}

All objects in the catalogue are designated in the format $\textbf{CNS5}~NNNN$, where $NNNN$ represents the sequence number assigned consecutively when entries are ordered by right ascension. This running number is unique for every object within the CNS5. Objects in the next releases of the catalogue will be renumbered entirely (with the corresponding acronym that specifies the release number, e.g.\ CNS6).

In addition, the widely used Gliese-Jahrei{\ss} (GJ) numbers have been kept in the CNS5. 3\,461 new objects in the catalogue were given GJ numbers for the first time. GJ numbers assigned in the CNS5 range from GJ~10001 to GJ~13461. Contrary to CNS5 numbers, GJ numbers are not ordered by right ascension and do not have to be unique for every object -- different components of binary or multiple systems may have the same GJ number.

Frequently, information on binarity of an object is encoded in a designation by appending the suffixes A, B  etc. to the system's designation. This approach could be confusing and sometimes could even lead to erroneous notations in publications.

For example, Proxima Centauri ($= \object{\text{GJ}~551}=\alpha~\text{Cen~C}$) is occasionally referred to as GJ~551~C. However, neither GJ~551~A nor GJ~551~B do exist. $\alpha~\text{Cen~A}$ and $\alpha~\text{Cen~B}$ are \object{GJ~559~A} and \object{GJ~559~B}, respectively.

To avoid further confusion in cases when a primary and a secondary component have different designations or new information on binarity of an object has become available, we decided to keep indexing of binary and multiple systems in the CNS5 separately from the designations.
This is done by introducing additional columns in the CNS5 where we list the GJ number for the system (\texttt{gj\_id}), suffixes for a component (\texttt{component\_id}), the total number of components in a system (\texttt{n\_components}), and, if the GJ number for a component differs from the GJ number of the primary of the system, we list the GJ number of the primary (\texttt{gj\_system\_primary}). This is shown in Table~\ref{tab:binsystGJ} for the example of the triple system $\alpha~\text{Cen}$.
In cases where we assigned new GJ numbers
to binary or multiple systems, we assigned separate GJ numbers for each component listed
in the CNS5.

Components are designated in the CNS5 with capital letters in descending order of brightness in the $G$ band.
Thus, in binary systems containing a white dwarf, the secondary will have a higher mass than the primary in the majority of the cases.
We opted not to designate components in the order of their separation from the primary, as it is frequently done, since the separation, especially that of nearby multiple systems, is changing with time.

\subsection{CNS5 content}
\label{sec:data_structure}
The data structure of the CNS5 is described in detail in \tabref{tab:cat_discription}. Columns (1-4) contain object identifiers.
In addition to CNS5 and GJ designations, we provide, if available, object identifiers in \hip{} and \gedrthree{} catalogues too. This facilitates cross-matching of our catalogue with other catalogues by using one of these identifiers.

Positions in our catalogue are not propagated to the common reference epoch, but provided at the mean epoch of the original catalogue. This implies that the epoch is not constant throughout the CNS5 catalogue. The epoch to which the position refers is listed for each object in the \texttt{epoch} column.
References for coordinates, as well as for all other parameters listed in the CNS5, are provided in the form of bibcodes as assigned by the SAO/NASA Astrophysics Data System.

Parallaxes listed in CNS5 are always trigonometric; no photometric or spectroscopic parallaxes were considered.
Parallaxes, when selected from \gedrthree{}, are corrected for the parallax zero-point and the corresponding errors are the inflated ones. Proper motions for all stars are listed as well and for all but 82 sources come from the same reference as the parallax. 

The $G$-band photometry of \gedrthree{} sources with 2-parameter or 6-parameter astrometric solutions is corrected as described in \citet{gaiaedr3_photometry,gaiaedr3_summary}.
Magnitudes in \gedrthree{} are listed without uncertainties. The error values given in CNS5 are calculated from the electron flux and its relative error as:
\begin{equation}
\sigma_{X} = \frac{2.5}{\ln(10)\,I_X}\sigma_{I_X},
\end{equation}
where $X$ is the corresponding \gaia{} band ($G, BP$ or $RP$), $I_X$ is the flux in this band. For simplicity, here we neglect the asymmetry of the error distribution for sources with a low flux-over-error ratio ($I/\sigma_I \lesssim 10$).
The assumption of symmetric errors is reasonable: all sources with $G$ magnitudes in the CNS5 have $I_G/\sigma_{I_G} > 10$ and the sources with a low flux-over-error ratio in $RP$ or $BP$ bands represent a tiny minority in the CNS5 (13 sources with $I/\sigma_I < 10$ in $RP$ band and 286 such sources in $BP$ band).

\begin{figure}
    \centering
    \includegraphics[width=0.48\textwidth]{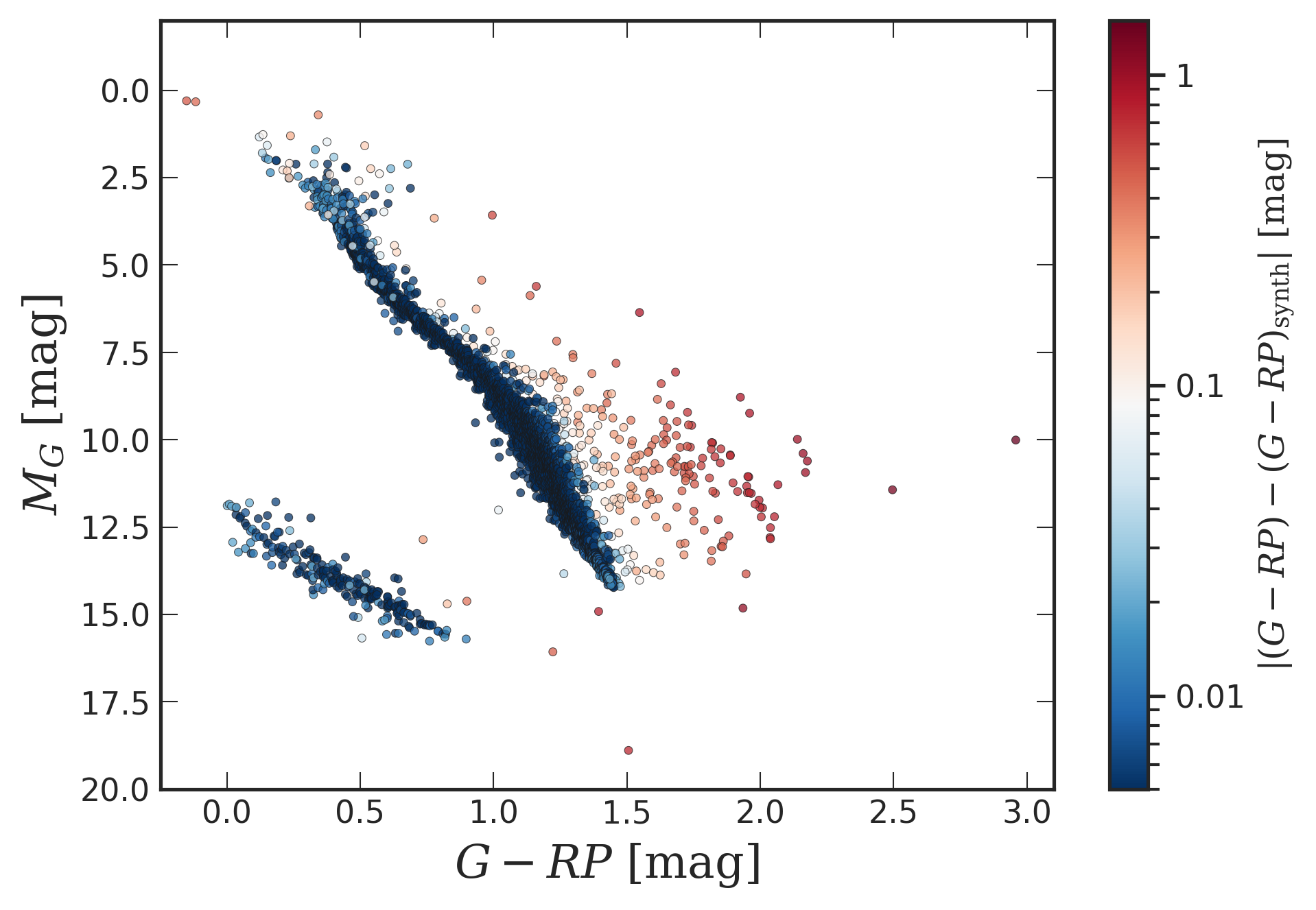} \\
    \includegraphics[width=0.48\textwidth]{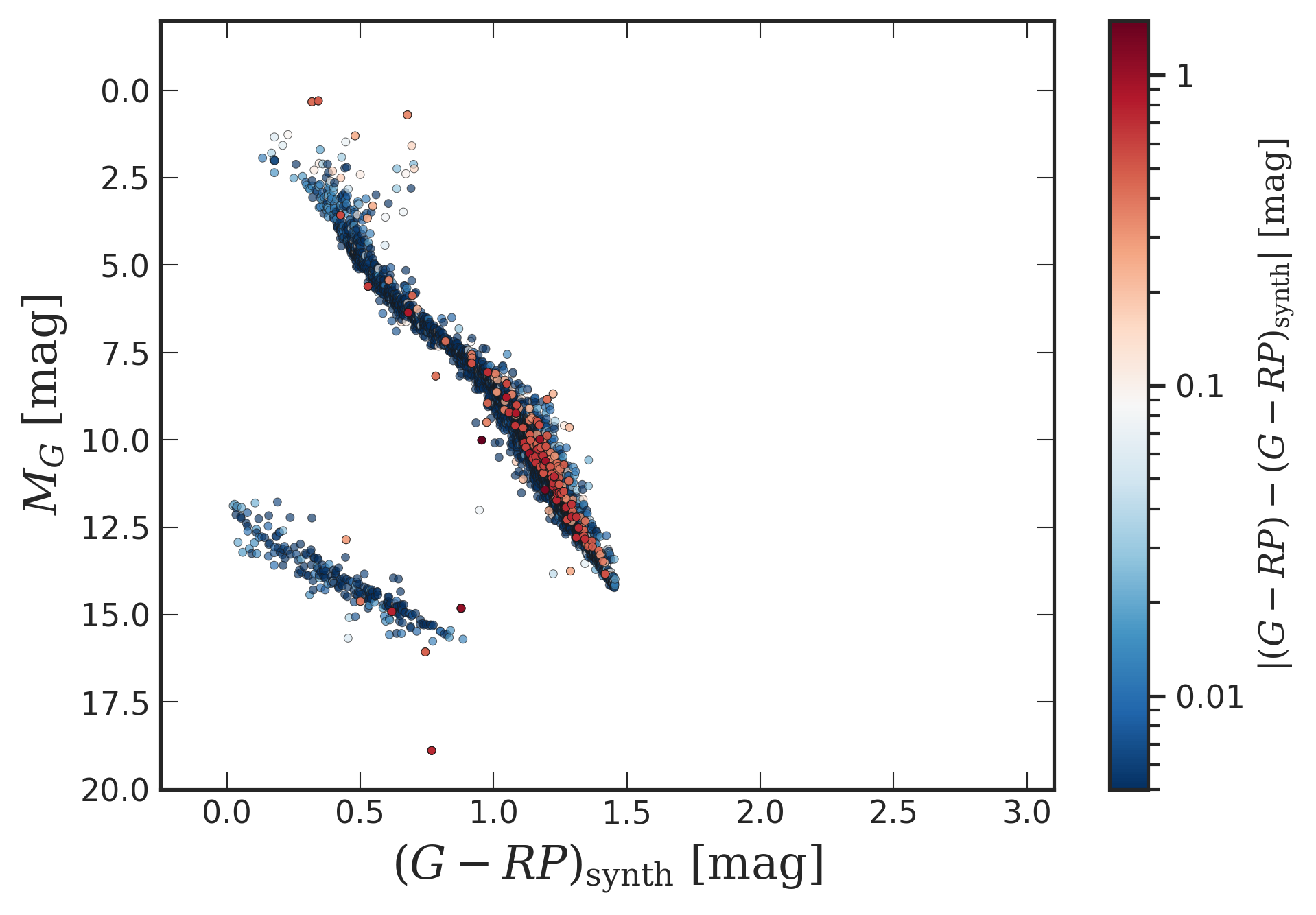}
    \caption{\textit{(Top)}: CMD for objects in the CNS5 which have a counterpart in \gedrthree{} using the published $G-RP$ colour.\\ \textit{(Bottom)}: CMD for the same sample, but using synthetic (deblended) $G-RP$ colours. The colour bar shows the absolute difference between the measured and synthetic $G-RP$ colours. The stars coded in red are the ones with the largest differences between the two colours.}
    \label{fig:hrd_g_rp_synth_25pc}
\end{figure}

\begin{figure}
    \centering
    \includegraphics[width=0.48\textwidth]{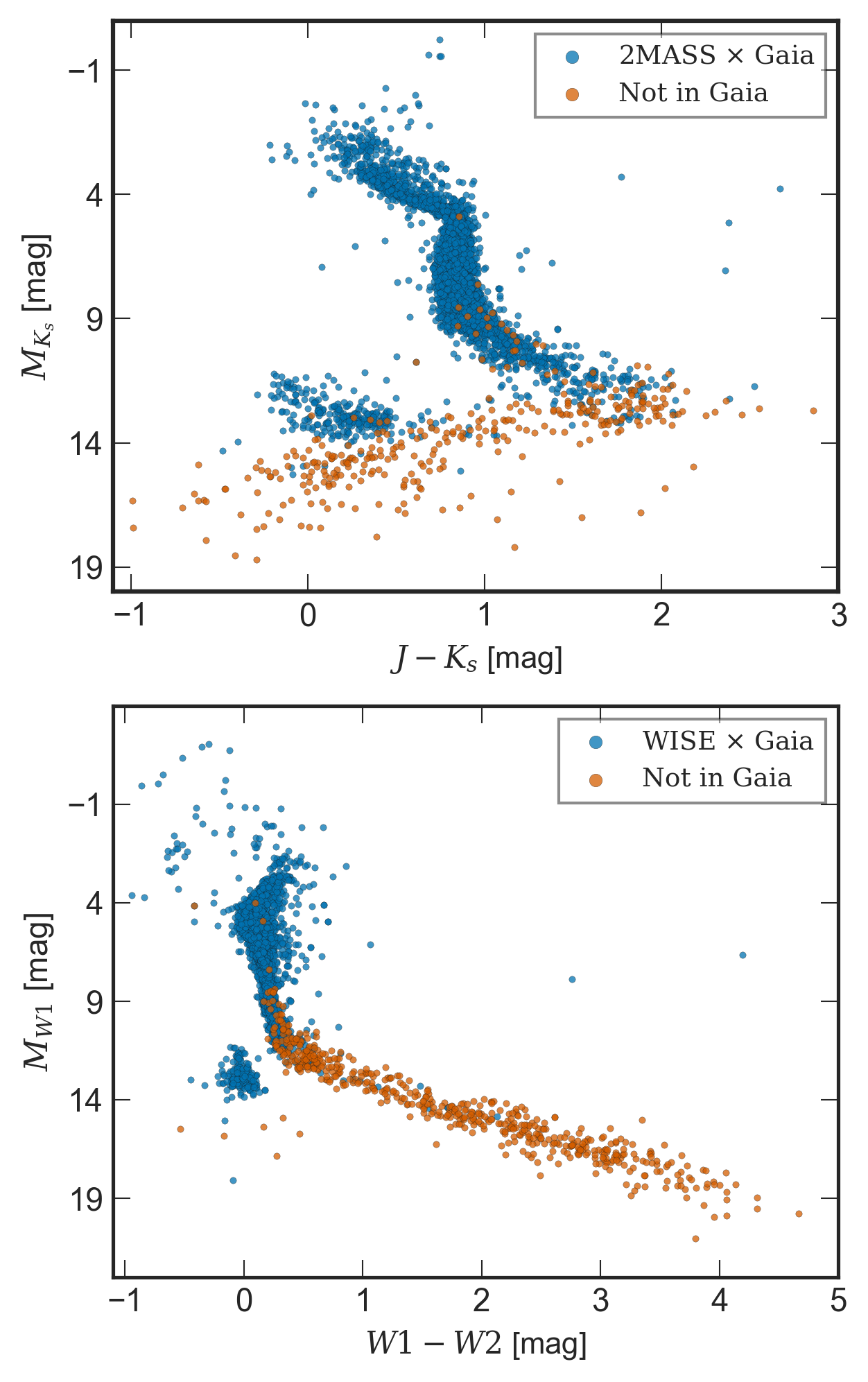} 
    \caption{\textit{(Top):} CMD of the \twomass{} sample. The blue points represent \twomass{} sources cross-matched with \gedrthree{} while the orange ones are not present in \gaia{}. Six extreme outliers are out of this frame. \\
    \textit{(Bottom):} Similarly, a CMD of the \wise{} sample, with the blue points representing \wise{} sources cross-matched with \gedrthree{} while the orange ones are not present in \gaia{}. Two extreme outliers are out of this frame.}
    \label{fig:CMD_2mass_wise}
\end{figure}

For objects with photometry in \gedrthree{} we also list the synthetic (deblended) $G$ and $G-RP$ magnitudes. These are calculated from our model fit to the \gaia{} data for a sample of objects with high-quality photometry and is fully described in Appendix~\ref{sec:synthetic_magnitudes}.
The deblended magnitudes in our catalogue are provided only for sources with a sufficiently large flux-over-error ratio in $BP$ and $RP$ bands ($I_{BP,\, RP}/\sigma_{I_{BP,\, RP}} > 20$) and with a colour within the applicability range of $0\magrm<BP-RP<4.25$\,mag. We supply the deblended photometry for all sources which match these criteria, irrespective of whether the correction is significant or not. In case it is available, the deblended photometry is always given as the resulting photometry.

For all objects selected from the \hip{} catalogue, which have no counterpart in \gedrthree{}, we provide converted values of $G$
and $RP$ magnitudes. 
Converted magnitudes were calculated using the photometric relationships $G-H_p = f(V-I)$ from \citet{gaiaedr3_photometry}.
We opted for this parameter space because here, contrary to ($G-H_p, B-V$), red giants and dwarfs display the same behaviour and no distinction between them is needed.
However, the reader should be aware that $V-I$ colours in \hip{} are a compilation from the literature and originate from a variety of ground-based measurements obtained more than 30 years ago. As demonstrated by \citet{Koen_2002MNRAS.334...20K} and by \citet{Platais_2003A&A...397..997P}, $V-I$ colours listed in \hip{} in some cases can be grossly incorrect and should be treated with caution.

\begin{figure*}
    \centering
    \includegraphics[width=0.90\textwidth]{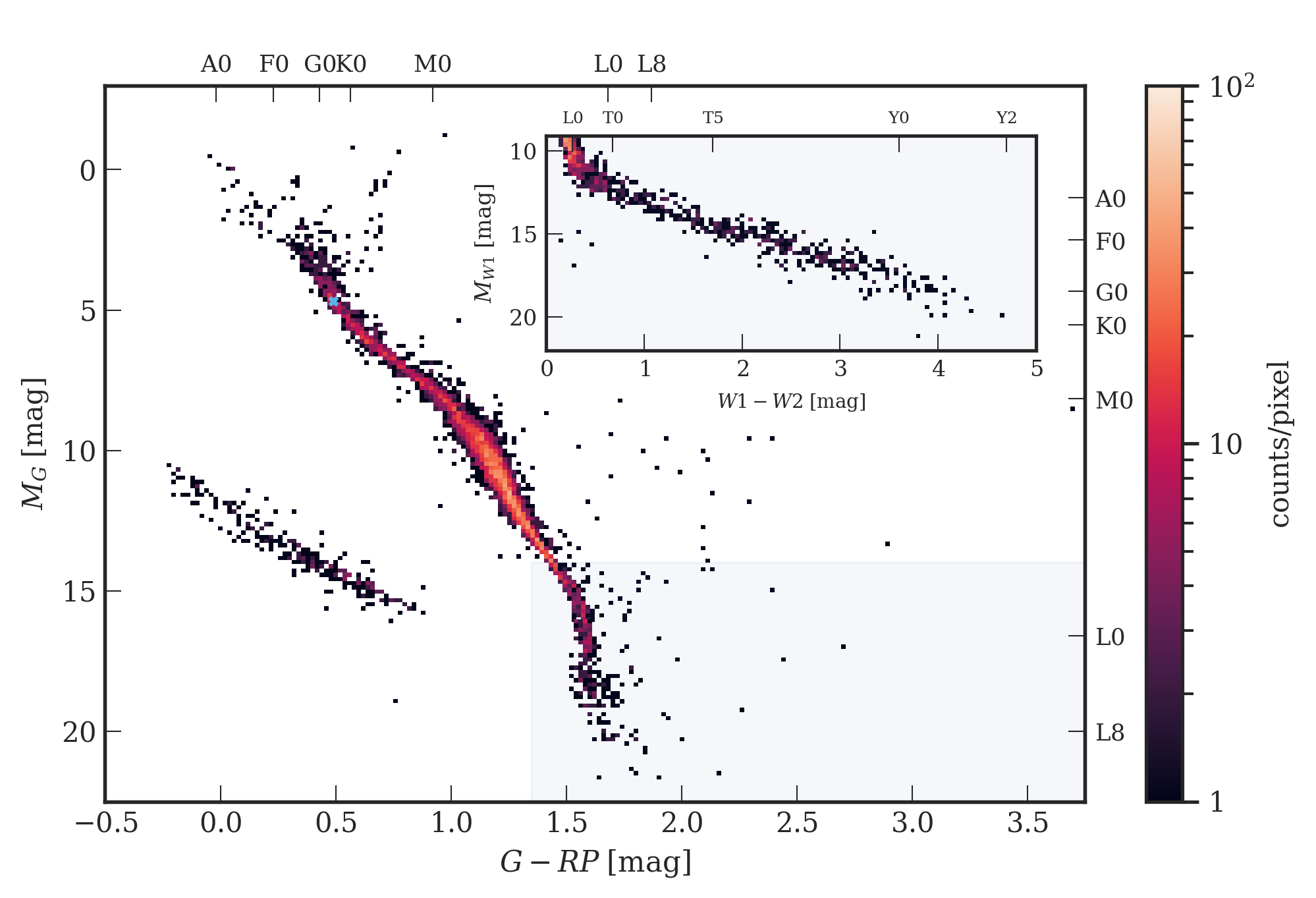}
    \caption{CMD of CNS5 objects in \gedrthree{} and \wise{} bands. The main plot includes \gedrthree{} objects with  resulting $G -RP$ colours. This includes objects with synthetic colours and also objects which fall out of its applicability range. It also includes objects from \hip{} which do not have a counterpart in \gedrthree{}, with converted $G$ and $RP$ magnitudes. The top and right axis show the spectral type and the blue cross represents the Sun's location on this CMD.The grey shaded area corresponds to the region in which the number of MS objects not present in \gaia{} start to increase in \wise{} bands. Thus, this region is also represented as an inset plot in \wise{} bands with the spectral type on the top axis.
    The colour bar shows the density in terms of star counts per pixel. On the main plot, each pixel is $[0.020 \times 0.15] \magrm^2 $ while on the inset plot it is $[0.045 \times 0.25] \magrm^2 $. One extreme outlier falls outside of the main plot region.}
    \label{fig:CNS_&_wise_cmd}
\end{figure*}

The remaining columns referring to photometry contain NIR and MIR magnitudes in the 2MASS ($J, H$ and $K_s$) and WISE ($W1, W2, W3$ and $W4$) passbands with their corresponding uncertainties. We used the cross match between \gaia{} and 2MASS and \gaia{} and WISE, respectively, provided with \gedrthree{} for the identification of objects. 

Further, we list the number of components 
in each system in the CNS5. 
We made an effort to identify at least obvious common proper motion pairs in the CNS5 by selecting pairs with projected separation less than 1~pc and proper motions consistent with a Keplerian orbit \citep[see][Equation 1 and Equations 3--6 in their paper]{El_Badry_inflation_2021MNRAS.506.2269E}.

The limit on the projected separation of 1~pc corresponds to the separation where the Galactic tidal field typically starts to dominate the gravitational attraction of two stars (e.g.\ \citealt{Binney_2008gady.book.....B, El_Badry_inflation_2021MNRAS.506.2269E}). The chosen limit seems to be appropriate since we found no pairs with a projected separation above 0.85~pc, and only two pairs ($0.3\%$) with a projected separation larger than 0.33~pc. All separations are thus well below the adopted limit, so that these systems have a large probability of being physically associated. The mean projected separation of the identified common proper motion pairs is 1857~AU, whereas the median is 185~AU (corresponding to an orbital period of $\sim80\,000$~years and $\sim2500$~years, respectively, for a typical binary with a total mass of $1 M_\sun$).

We consider a pair of stars to be a common proper motion pair if its observed scalar proper motion difference does not exceed the maximum expected proper motion difference for a circular orbit of a binary with a total mass of $5 M_\sun$ at the $2\sigma$ level. For $86.6\%$ of the identified pairs, we found that the observed proper motion difference is at least two times smaller than the expected difference due to orbital motion. Only $0.6\%$ (four pairs) have a proper motion difference larger than the expected difference due to orbital motion but still consistent at the $2\sigma$ level with a Keplerian orbit. Loosening this selection criterion further would result in contamination from chance alignments with background stars.

When it comes to identifying common proper motion pairs in the CNS5, we decided to neglect the parallax discrepancy. In this way we can include also those nearby systems whose physical separation between components is large enough that their parallaxes differ significantly. 
This choice is justified by the low projected source density in the CNS5 catalogue and large parallaxes and proper motions of objects in the solar neighbourhood.
We found that parallax discrepancy, $\Delta\varpi/\sigma_{\Delta\varpi} = |\varpi_1 - \varpi_2|/\sqrt{\sigma_{\varpi_1}^2+\sigma_{\varpi_2}^2}$, for nearby binaries in \gedrthree{} could be as large as $11.9\sigma_{\Delta\varpi}$ (\object{Gaia~EDR3~5479222240596469632} and \object{Gaia~EDR3~5287046368477870848}).

Finally, we would like to remark that the common proper motion pairs identified in the CNS5, by construction, include all the nearby binaries from \citet{El_Badry_inflation_2021MNRAS.506.2269E} as well as systems with three or more components and pairs where only one component has parallax and proper motion given in \gedrthree{} and the second has them from other surveys, for instance from \textit{Spitzer}.
This results in 696 common proper motion pairs in the CNS5.

So far, we consider for the number of components given in the catalogue only
visual binary or multiple components, i.e.\ those which have a separate entry in the
CNS5 catalogue. 
Spectroscopic companions or otherwise detected companions from the literature will be added in the next version of the CNS catalogue.

Finally, we provide radial velocities from \gedrthree{}
or from the compilations by
\citet{Best_2021AJ....161...42B} where available.

\begin{table*}
\caption{Description of the CNS5 content. The entry for \object{Proxima Centauri} (\object{GJ~551}) is shown as an example.}
\label{tab:cat_discription}
    \centering
    \begin{tabular}{l|lll}
        \hline
        \hline
         Column name & Unit & Description & Example  \\
         \hline
         \texttt{cns5\_id} & & CNS5 designation & 3591\\
         \texttt{gj\_id} & & Gliese-Jahrei{\ss} number& 551\\
         \texttt{component\_id} & & Suffix for a component of binary or multiple system& C\\
         \texttt{n\_components} & & Total number of components in the system & 3\\
         \texttt{primary\_flag} & & Flag indicating the primary of a multiple system & False\\
         \texttt{gj\_system\_primary} & & GJ number of the primary component of the system & 559\\
         \texttt{gaia\_edr3\_id} & & Source identifier in \gedrthree{}& 5853498713190525696\\
         \texttt{hip\_id} &  & \hip{} identifier & 70890\\
         \texttt{ra} & deg & Right ascension & 217.39232147200883\\
         \texttt{dec} & deg & Declination& -62.67607511676666\\
         \texttt{epoch} & a & Reference epoch for coordinates & 2016.0\\
         \texttt{coordinates\_bibcode} & & Reference for \texttt{ra, dec} & 2020yCat.1350....0G\\
         \texttt{parallax} & mas & Absolute trigonometric parallax & 768.0665391873573\\
         \texttt{parallax\_error} & mas & Error of \texttt{parallax} & 0.056201234\\
         \texttt{parallax\_bibcode} & & Reference for \texttt{parallax} & 2020yCat.1350....0G\\
         \texttt{pmra} & mas a$^{-1}$ & Proper motion in right ascension ($d(\alpha \cos\delta)/dt$) & -3781.741008265163\\
         \texttt{pmra\_error} & mas a$^{-1}$ & Error of \texttt{pmra}& 0.03138607740402222\\
         \texttt{pmdec} & mas a$^{-1}$ & Proper motion in declination & 769.4650146478623\\
         \texttt{pmdec\_error} & mas a$^{-1}$ & Error of \texttt{pmdec}& 0.05052453279495239\\
         \texttt{pm\_bibcode} & & Reference for proper motion& 2020yCat.1350....0G\\
         \texttt{rv} & km s$^{-1}$ & Radial velocity (spectroscopic) & -22.4\\
         \texttt{rv\_error} & km s$^{-1}$ & Error of \texttt{rv}& 0.5\\
         \texttt{rv\_bibcode} & & Reference for the radial velocity & 2006A\&A...460..695T\\
         \texttt{g\_mag} & mag & $G$ band mean magnitude (corrected) & 8.984749\\
         \texttt{g\_mag\_error} & mag & Error of \texttt{g\_mag}& 0.0007106\\
         \texttt{bp\_mag} & mag & \gedrthree{} integrated $BP$ mean magnitude & 11.373116\\
         \texttt{bp\_mag\_error} & mag & Error of \texttt{bp\_mag}& 0.0025825\\
         \texttt{rp\_mag} & mag & \gedrthree{} integrated $RP$ mean magnitude& 7.5685353\\
         \texttt{rp\_mag\_error} & mag & Error of \texttt{rp\_mag}& 0.0017553\\
         \texttt{g\_mag\_from\_hip} & mag & Converted $G$ band magnitude from \hip{}& \\
         \texttt{g\_mag\_from\_hip\_error} & mag & Error of \texttt{g\_mag\_from\_hip}& \\
         \texttt{g\_rp\_from\_hip} & mag & Converted $G-RP$ colour from \hip{} & \\
         \texttt{g\_rp\_from\_hip\_error} & mag & Error of \texttt{g\_rp\_from\_hip}& \\
         \texttt{g\_mag\_resulting} & mag & Resulting $G$ band magnitude& 8.984749\\
         \texttt{g\_mag\_resulting\_error} & mag & Error of \texttt{g\_mag\_resulting}& 0.0007106\\
         \texttt{g\_rp\_resulting} & mag & Resulting $G-RP$ colour & 1.3984906\\
         \texttt{g\_rp\_resulting\_error} & mag & Error of \texttt{g\_rp\_resulting}& 0.0033824\\
         \texttt{g\_rp\_resulting\_flag} &  & Flag related to \texttt{g\_rp\_resulting} & 0\\
         &  &  and \texttt{g\_rp\_resulting\_error}\tablefootmark{a} & \\
         \texttt{j\_mag} & mag & 2MASS $J$ band magnitude& 5.357\\
         \texttt{j\_mag\_error} & mag & Error of \texttt{j\_mag}& 0.023\\
         \texttt{h\_mag} & mag & 2MASS $H$ band magnitude& 4.835\\
         \texttt{h\_mag\_error} & mag & Error of \texttt{h\_mag}& 0.057\\
         \texttt{k\_mag} & mag & 2MASS $K_s$ band magnitude& 4.384\\
         \texttt{k\_mag\_error} & mag & Error of \texttt{k\_mag}& 0.033\\
         \texttt{jhk\_mag\_bibcode} & & Reference for NIR magnitudes& 2003tmc..book.....C\\
         \texttt{w1\_mag} & mag & WISE $W1$ band magnitude& 4.195\\
         \texttt{w1\_mag\_error} & mag & Error of \texttt{w1\_mag}& 0.086\\
         \texttt{w2\_mag} & mag & WISE $W2$ band magnitude& 3.571\\
         \texttt{w2\_mag\_error} & mag & Error of \texttt{w2\_mag}& 0.031\\
         \texttt{w3\_mag} & mag & WISE $W3$ band magnitude& 3.826\\
         \texttt{w3\_mag\_error} & mag & Error of \texttt{w3\_mag}& 0.015\\
         \texttt{w4\_mag} & mag & WISE $W4$ band magnitude& 3.664\\
         \texttt{w4\_mag\_error} & mag & Error of \texttt{w4\_mag}& 0.024\\
         \texttt{wise\_mag\_bibcode} & & Reference for MIR magnitudes& 2014yCat.2328....0C\\
         \hline
    \end{tabular}
    \tablefoot{
    \tablefoottext{a}{{The definitions of the flag values are: 0 = resulting $G-RP$ is deblended, 1 = resulting $G-RP$ is uncorrected (i.e.\ \gedrthree{} catalogue value is listed), 2 = $G-RP$ is converted from \hip{}.}}
    }
\end{table*}

\begin{table}
\caption{Description of the photometric content of the CNS5.}
\label{tab:cat_discription_photometry}
    \centering
    \begin{tabular}{ll}
        \hline
        \hline
         Parameter & \# of objects   \\
         \hline
         \texttt{g\_mag} & 5234 \\
         \texttt{bp\_mag}  & 5148 \\
         \texttt{rp\_mag}  & 5157 \\
         \texttt{g\_mag\_from\_hip}  & 137 \\
         \texttt{g\_rp\_from\_hip}  & 137 \\ 
         \texttt{g\_mag\_resulting}  & 5372 \\
         \texttt{g\_rp\_resulting}  & 5261 \\
         \texttt{j\_mag}  & 5348 \\
         \texttt{h\_mag}  & 5347 \\
         \texttt{k\_mag}  & 5334 \\
         \texttt{w1\_mag}  & 4812\\
         \texttt{w2\_mag}  & 4814\\
         \texttt{w3\_mag}  & 4812\\
         \texttt{w4\_mag}  & 877\\
         Has \texttt{k\_mag} but no \texttt{g\_mag\_resulting} & 334\\
         Has \texttt{g\_mag\_resulting} but no \texttt{k\_mag} & 372\\
         Has \texttt{w1\_mag} but no \texttt{g\_mag\_resulting} & 537\\
         Has \texttt{g\_mag\_resulting} but no \texttt{w1\_mag} & 1097\\
         \hline
    \end{tabular}
\end{table}

\subsection{Availability of the CNS5 catalogue}
\label{sec:access}

The CNS5 catalogue is hosted by the German Astrophysical Virtual Observatory (GAVO)\footnote{\url{https://dc.g-vo.org/CNS5}}
and is also available through the VizieR Catalogue Service\footnote{\url{https://vizier.cds.unistra.fr/viz-bin/VizieR}}.

\section{Catalogue properties}
\label{sec:content}

\subsection{Colour-magnitude diagrams}
\label{sec:CMD}

Figure~\ref{fig:hrd_g_rp_synth_25pc} (top) shows the CMD for the selection of \gedrthree{} objects in the CNS5 with published (i.e.\  unadjusted) colours.
Here, we see that some objects have a large offset towards the red of the main sequence (MS), which is unphysical and is caused by blending or contamination from nearby sources.

\begin{figure}
    \centering
    \includegraphics[width=0.49\textwidth]{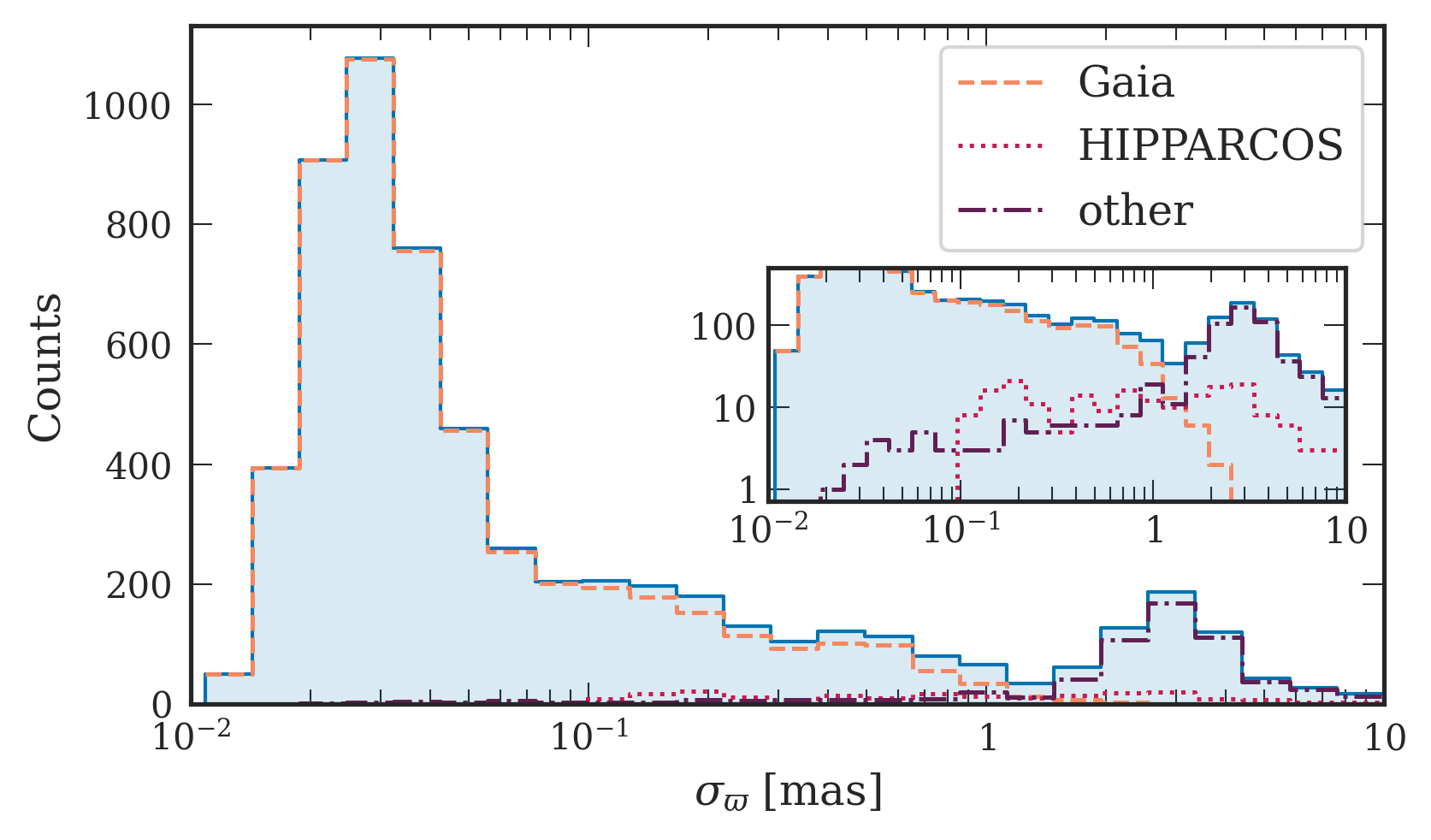}
    \caption{Parallax uncertainties in the CNS5. The inset represents the zoom-up of the vertical axis and with logarithmic scale being used on each of the axes. Distributions of the parallax uncertainties in the subsamples from \gaia{}, \hip{} and other references are overplotted separately.}
    \label{fig:e_plx_cns3_cns5}
\end{figure}

In order to correct for it, we derive synthetic $G-RP$ magnitudes (see \appref{sec:synthetic_magnitudes} for details) in the applicability range of $0\magrm <BP-RP<4.25 \magrm$ and if  the flux over error is greater than 20 for both $BP$ and $RP$ magnitudes.
Figure~\ref{fig:hrd_g_rp_synth_25pc} (bottom) shows the CMD using these synthetic $G-RP$ magnitudes for all stars, although the corrections are small for many of them (coded in blue in the figure). The stars coded in red colours have large corrections, and they now all fall right on the main-sequence where one would expect them, illustrating that our correction works extremely well. 

In \figref{fig:CMD_2mass_wise} we show the infrared CMDs using \twomass{} (top) and \wise{} (bottom) photometry. 
In both diagrams the blue points are sources with a counterpart in  \gedrthree{},  
while the objects in orange are absent in \gedrthree{}. 
In the \twomass{} CMD, the lowest mass objects turn left towards the blue. This effect is associated with the 1.6 and 2.2~\textmu m methane absorption bands \citep{Burgasser_1999ApJ...522L..65B} and with the dissipation of clouds across the L/T transition of brown dwarfs \citep{saumon08}.
As the absorption in cooler T~dwarfs gets stronger, making their $J-K_s$ colour bluer, these low-mass brown dwarfs approach the location of the white dwarfs (WDs) in the CMD. In contrast, in \wise{} $W1-W2$ colour the ultra-cool dwarfs continue the trend towards the red for lower masses. 
As a result, the lower end of the MS in \wise{} is well separated from the WDs.
Photometric errors for the fainter objects in particular are also smaller in \wise{}, and
as seen in  \secref{sec:data_structure}, \wise{} goes deeper and is more complete for ultra cool dwarfs than \twomass{}, thus making \wise{} ideal to study the faint end of the main-sequence.

In \figref{fig:CNS_&_wise_cmd}, we present the CMD of the CNS5 objects in \gedrthree{} and \wise{} bands combined. 
We use synthetic \gaia{} colours (Appendix~\ref{sec:synthetic_magnitudes}) where applicable, and have converted the magnitudes and colours of \hip{} stars into the \textit{Gaia} system to also include those stars in the plot. 
There are still a few stars located redwards of the main-sequence; these are all stars for which we could not provide deblended magnitudes, because they fall outside of the colour range where a reliable correction could be derived, or they have S/N smaller than 20. Their colours in the CMD are still dominated by blending or contamination, whereas their $G$ band absolute magnitude is reliable. 
We note that for objects in our catalogue the uncertainty of position in the CMD is dominated by colour uncertainty and not by that of the parallax. 
The mode of the underlying parallax error distribution in the CNS5 is $0.025~\text{mas}$ (see \figref{fig:e_plx_cns3_cns5}). This means that the parallax uncertainty typically contributes only about $0.0014\magrm$ to the uncertainty of the absolute magnitude for an object at the 25~pc distance, whereas the mode of $G-RP$ uncertainty distribution is $0.0018\magrm$.
From \figref{fig:CMD_2mass_wise} (bottom) we know that the number of MS objects which do not have a counterpart in \gedrthree{} starts to increase from $M_{W1} \gtrsim 9 \:\magrm $ or $ M_G \gtrsim 14 \: \magrm$ (cf.\ Fig.~\ref{fig:g_w1}). This region is indicated by the grey shaded area in the main plot. We show this lower-main sequence region as an inset plot in the \wise{} bands, where completeness is much higher than in \gaia{}.

\subsection{Completeness of the CNS5}
\label{sec:completeness}

A precise assessment of the completeness of the CNS5 catalogue is essential not only to know until which absolute magnitude or spectral type we are complete, but it is also a crucial component for deriving a luminosity function (see \secref{sec:luminosity_function}) and for many other similar applications.
Using a number density simply computed for the volume with 25~pc radius would yield a luminosity function which is heavily biased due to incompleteness at the faint end. Consequently, number densities at faint magnitudes would be underestimated. In contrast, when number densities are derived within the completeness limit, this bias is avoided.

Assuming that stars in the solar neighbourhood have a constant number density,
we define the completeness limit of the CNS5 as the largest distance $r_c$ at which the distribution of objects in the catalogue is still consistent with being spatially uniform. 
Beyond this distance the observed number density starts to drop because more and more objects are too faint to be observed (or to have reliable parallaxes) and hence are missed in our volume-limited sample.
Furthermore, it is natural to expect that the distance $r_c$ will be smaller for objects with fainter absolute magnitudes.

A uniform space density prior is reasonable for our 25~pc sample also because there are no known clusters within this distance. 
The nearest cluster, the Hyades, is located at a distance of $47.50\pm0.15$~pc of the Sun \citep{gaiadr2_hrd}, has half-mass radius of 4.1~pc and tidal radius of 9~pc \citep{roeser_2011}. 
Using \gdrtwo{}, the present-day tidal tails of the Hyades were mapped within 200~pc of the Sun and there are only 18 probable members and 3 contaminants located within 25~pc of the Sun \citep{roeser_2019}. 
According to \citet{jerabkova_2021A&A...647A.137J}, the number of candidate members of the Hyades tidal tails located within 25~pc is even lower: 
they confirm only 7 members using \gdrtwo{} and 5 members when using more precise astrometry from \gedrthree{} in their analysis.
These numbers are not large enough to yield any statistically significant overdensity within our 25~pc volume or to contaminate the luminosity function of the solar neighbourhood.

We assessed the completeness of our catalogue using a Kolmogorov-Smirnov test. 
Here, we select a subsample of stars for each absolute magnitude bin for which a completeness limit should be estimated. 
Then for each such subsample the cumulative distribution function (CDF) of the distance $r$ is derived and compared with the expected CDF of a uniform distribution.
A note should be made here that we do this on the component level and not on the system level. For simplicity, stars with distances close to the 25~pc limit were treated according to their nominal parallax values. 

By applying the Kolmogorov-Smirnov test, we find the largest distance $r_c$ within 25~pc
at which both, the empirical and the uniform CDF, are statistically indistinguishable at the 5\% level. 
Within the derived distance $r_c$ our catalogue can be regarded as statistically complete for the probed magnitude interval.

\begin{figure*}
    \centering
    \includegraphics[width=0.49\textwidth]{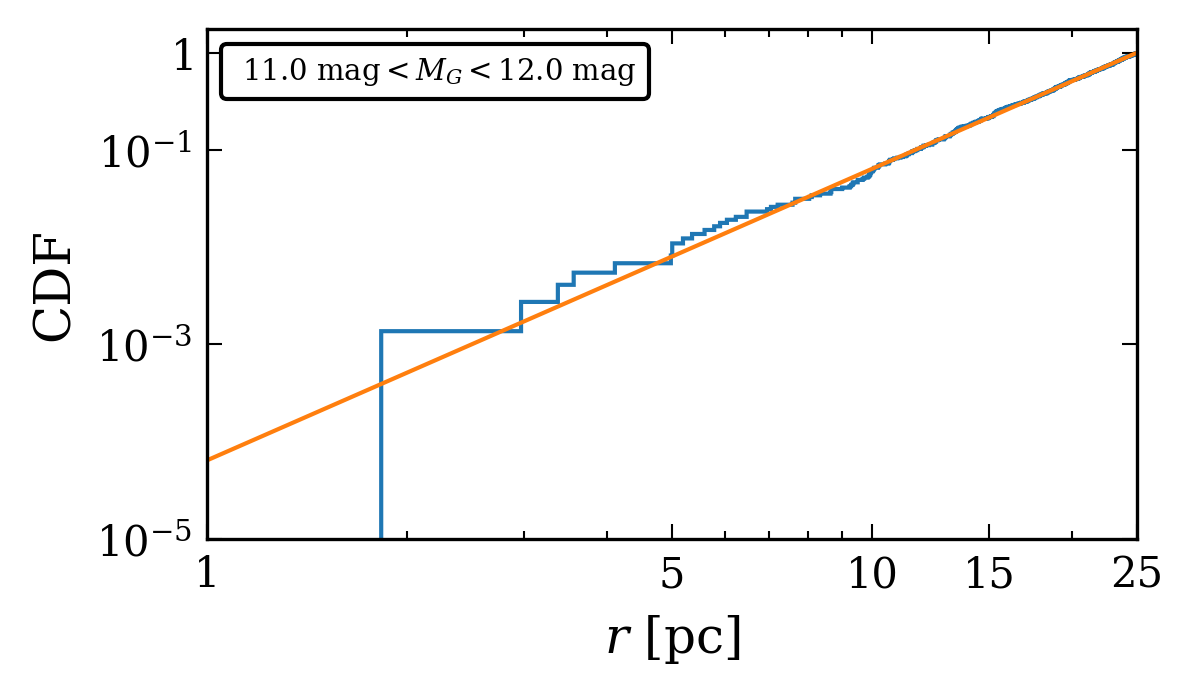}
    \includegraphics[width=0.49\textwidth]{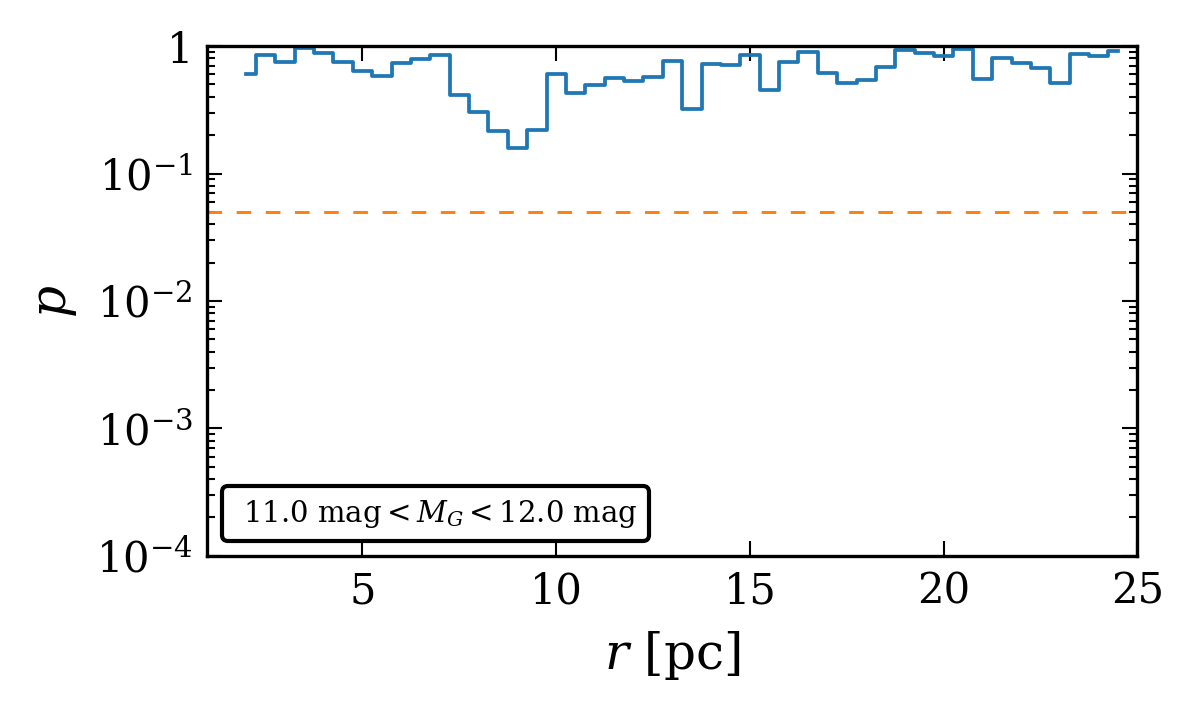}
    \includegraphics[width=0.49\textwidth]{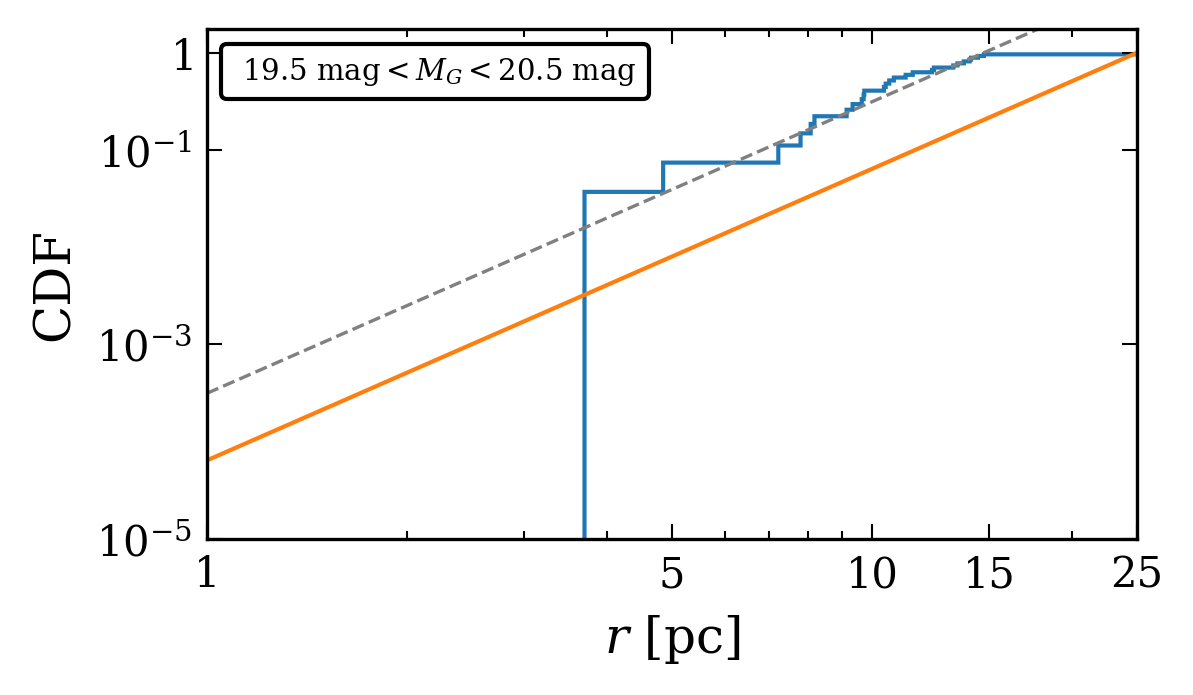}
    \includegraphics[width=0.49\textwidth]{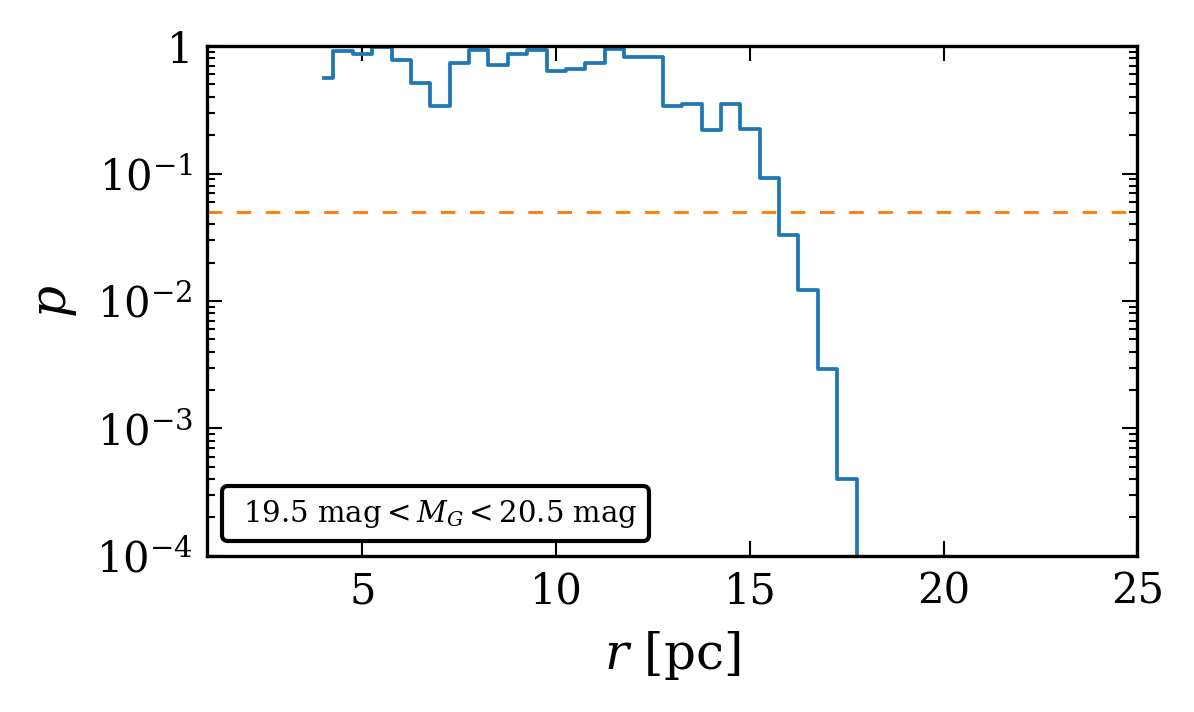}
    \caption{Here we illustrate how we estimate the completeness limit with the help of the Kolmogorov-Smirnov test. The \textit{left} column shows the empirical (shown in blue) and analytical (shown in orange) CDFs. The \textit{right} column shows the $p$-value from the Kolmogorov-Smirnov test as a function of probed distance limit. The dashed line indicates the $p=0.05$ threshold. Values below the threshold indicate that the empirical and analytical CDFs are significantly different, which means that the spatial distribution of objects in the subsample is not uniform anymore. Thus, the subsample cannot be regarded as statistically complete beyond the completeness radius. Two magnitude ranges are shown here as an example. The dashed line in the \textit{lower left} panel denotes the re-normalised CDF using the derived completeness limit for this magnitude bin.}
    \label{fig:cdf_ks}
\end{figure*}

Here we recall that the analytical CDF of the cumulative number of stars  as a function of distance $n(r)$ for a sphere of radius $R_0=25~{\rm pc}$ and uniform space density can be derived from the normalised probability density function (PDF):
\begin{equation}
n(r) = 3\frac{r^2}{R_0^3}.
\end{equation}
Thus, the CDF of $n(r)$ is given as:
    \begin{equation}
    \left.\begin{aligned}
    CDF(r) = \frac{r^3}{R_0^3}, &\qquad {0 \leq r \leq R_0}.
    \end{aligned} \right.
    \end{equation}

For each empirical sample we construct a corresponding test sample of $10^4$ objects. 
The distance of an object in the test sample is given by 
\begin{equation}
r= R_0 \cdot y^\frac{1}{3},
\end{equation}
where $y$ is 
a random number between zero and one.

\figref{fig:cdf_ks} examplifies CDFs and results of the Kolmogorov-Smirnov test for two cases.
While the distribution of the sources in the magnitude bin centered at $G=11.5\magrm$ (\textit{upper row}) is consistent with being uniform over all probed distances up to 25~pc, 
it is evident that the empirical and analytical CDFs of the sources in the bin at $G=20\magrm$ are considerably different (\textit{lower left panel}) and the sample is statistically complete only up to 16~pc (\textit{lower right panel}).
For illustrative purposes, we show that the empirical CDF is represented by the analytical CDF quite well when the latter is normalised at the derived completeness limit (dashed line in the \textit{lower left panel}).

\begin{figure*}
    \centering
    \includegraphics[width=0.49\textwidth]{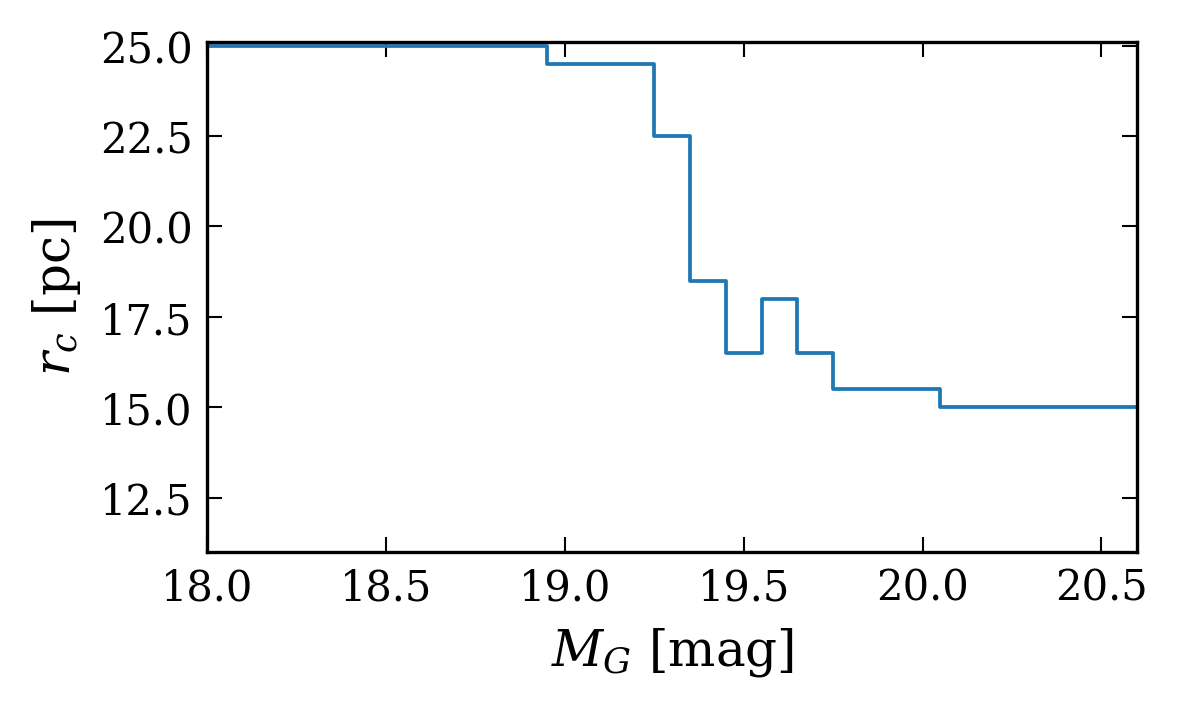}
    \includegraphics[width=0.49\textwidth]{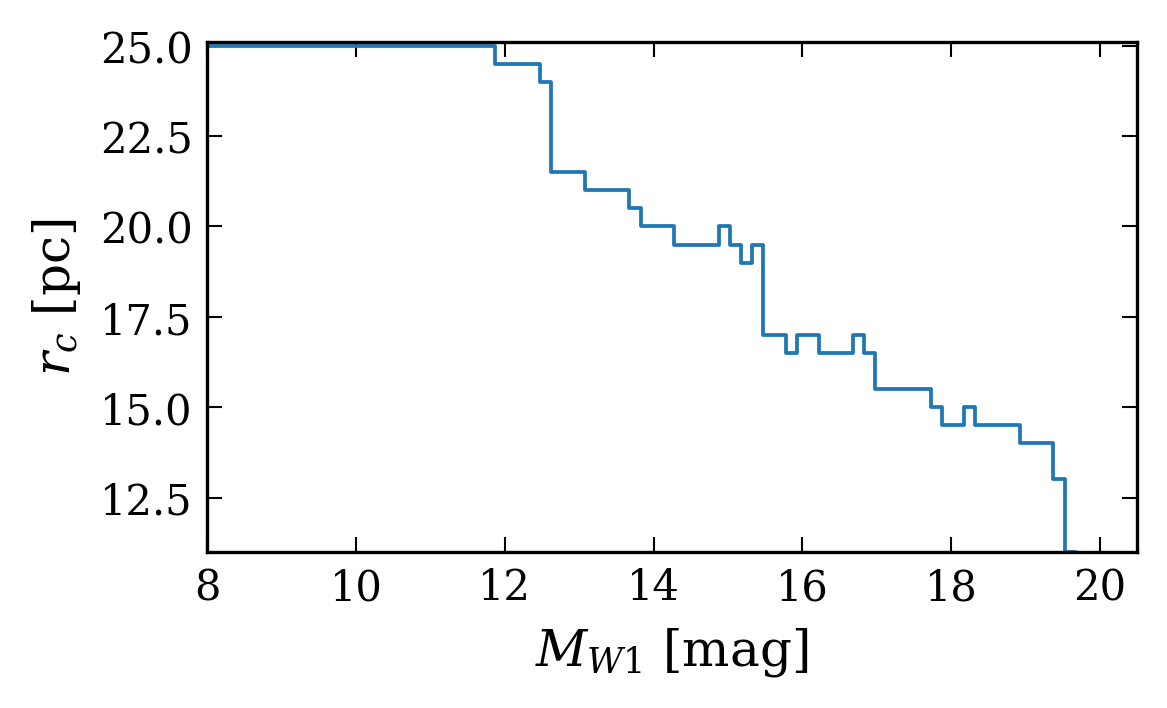}
    \caption{\textit{Left: }Completeness limit as a function of absolute magnitude in the $G$-band. The limit is computed for a sliding bin with 0.1~mag step and 1~mag bin width. \textit{Right:} Completeness limit as a function of absolute magnitude in the $W1$-band. Here a 0.15~mag step and a 2~mag bin width was used. The completeness limits are 18.9~mag in G~band and 11.8~mag in W1.}
    \label{fig:rc_g}
\end{figure*}

Following the approach outlined above, we derive a completeness limit of 18.9~mag in G-band and of 
11.8~mag in W1-band absolute magnitudes
(\figref{fig:rc_g}). 

\begin{figure}
    \centering
    \includegraphics[width=0.49\textwidth]{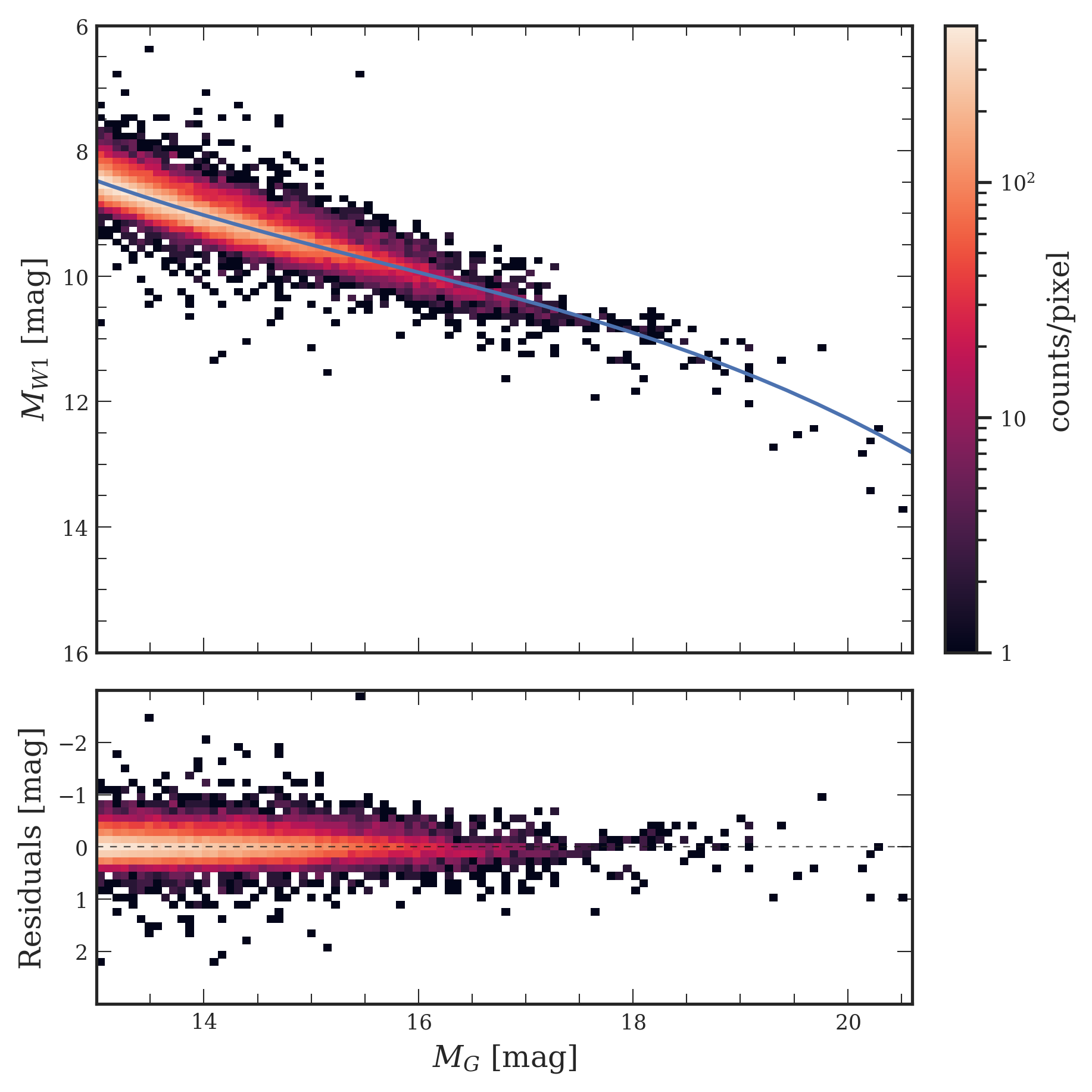}
    \caption{Relation between the absolute magnitude in $G$ and $W1$ bands for the main-sequence stars fainter than $M_{G} = 13.0 \magrm$. The solid blue line is the approximate relation in \equref{eq:g_w1}.}
    \label{fig:g_w1}
\end{figure}

When comparing the completeness limits or the luminosity functions in optical and mid-infrared wavelengths, an approximate relation between the absolute magnitudes in these wavelengths is useful. In order to derive the transformation coefficients, we used the sample of the main-sequence stars from the GCNS with counterparts in the CATWISE catalogue \citep{CatWISE_2020ApJS..247...69E}, in order to make the sample larger but comparable to the CNS5. 
We used the following criteria:
\begin{equation}
    \label{eq:gcns_w1_g}
    \begin{aligned}\setcounter{mysubequations}{0}
    \text{\mysubnumber}\quad& M_{G} > 13.0 \magrm, \\
    \text{\mysubnumber}\quad&\texttt{gcns\_prob} > 0.99, \\
    \text{\mysubnumber}\quad&\texttt{wd\_prob} < 0.05, \\
    \text{\mysubnumber}\quad& \text{\equref{eq:ipd_gof_harm_ampl_cut}:}\ A_{\rm GoF} < 10^{-5.12}~(\varpi/\sigma_\varpi)^{2.61} . \\
    \end{aligned} \quad
\end{equation}
The resulting sample contains 31\,875 sources.

An approximate transformation between $M_{W1}$ and $M_G$ magnitudes can be obtained with a cubic polynomial fit in this sample, for which we obtain
\begin{equation}
\label{eq:g_w1}
M_{W1} = 0.0073714\,{M_G}^3-0.3474\,{M_G}^2+5.896\,M_G-25.6\quad,
\end{equation}
where all absolute magnitudes are in the unit of mag. 
The resulting relation for the main-sequence stars fainter than $M_{G} = 13.0 \magrm$ is shown in \figref{fig:g_w1}.

\begin{figure}
    \centering
    \includegraphics[width=0.49\textwidth]{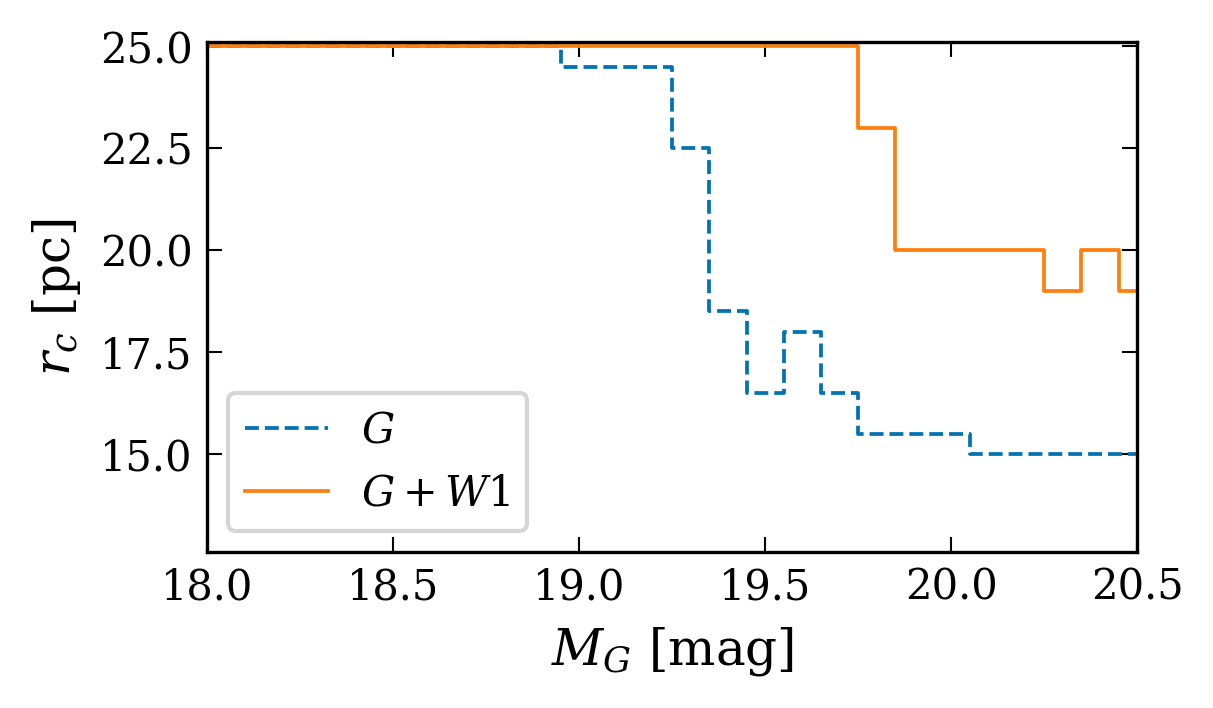}
    \caption{Completeness limit of the CNS5 in $G$-band when ultra-cool dwarfs with converted $M_G$ magnitudes are included (solid orange line). The completeness limit in $G$-band from \figref{fig:rc_g} is shown for comparison (dashed blue line).}
    \label{fig:g_w1_compl}
\end{figure}

Applying this transformation to sources without a counterpart in \gedrthree{} but with $W1$ magnitudes, so that we obtain approximate $G$ magnitudes for those ultracool dwarfs as well, we estimate that the CNS5 is complete down to $G\approx19.7\magrm$ (\figref{fig:g_w1_compl}), 0.8~mag fainter than based on \gaia{} sources alone.

\begin{figure}
    \centering
    \includegraphics[width=0.48\textwidth]{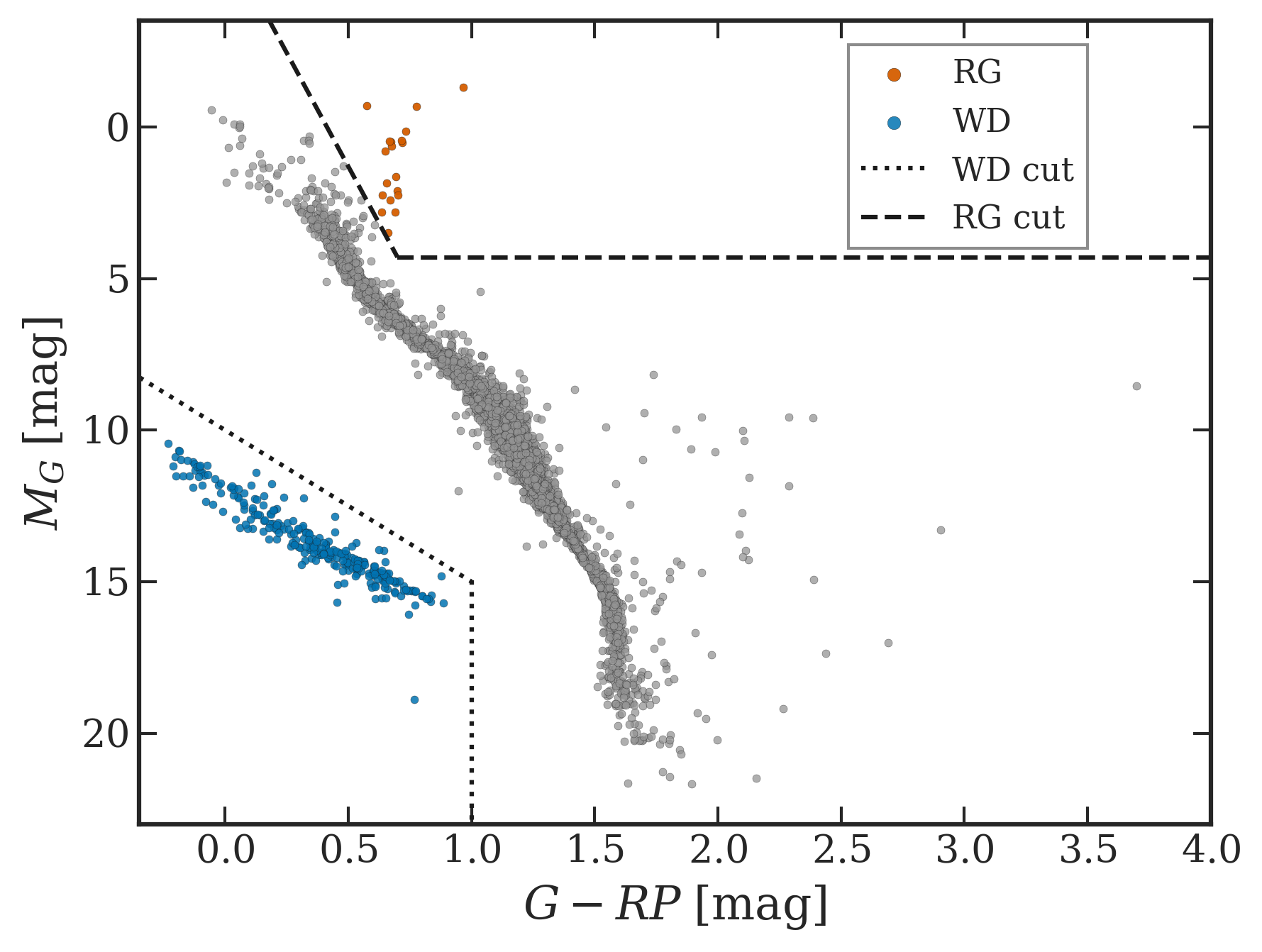} 
    \caption{CMD for the \gedrthree{} and \hip{} sample in the CNS5. The dashed line separates red giants from the main-sequence stars, while all objects below the dotted line are considered as white dwarfs. We classify the remaining sample as MS stars.}
    \label{fig:CMD_cuts}
\end{figure}

Given that white dwarfs (WD) are quite rare and, as shown later, represent only $5\%$
of the stellar content in the solar neighbourhood, we separately assessed the WD completeness.
We selected objects located in the WD region of the CMD by defining the following cut (see \figref{fig:CMD_cuts}):
\begin{equation}
    \label{eq:wd_cut}
    \left.\begin{aligned}
    \text{(i)}\quad&M_G>10.0 \magrm +5(G-RP)\\
    \text{(ii)}\quad&G-RP<1.0 \magrm
    \end{aligned} \quad \right\}
\end{equation}

Having applied the Kolmogorov-Smirnov test to the whole WD sample as well as to subsamples of different magnitude bins, we found that the distribution of white dwarfs in the solar neighbourhood is consistent with being uniform and that the 25~pc white dwarf sample can be regarded as statistically complete.
Therefore, this eliminates the possible concern that the shape and the location of the cut-off of the observational white dwarf luminosity function are affected by Malmquist bias (e.g.\ \citealt{Liebert_1979ApJ...233..226L, Iben_1984ApJ...282..615I, Garcia_Berro_2016NewAR..72....1G}).
We emphasise that in our completeness assessment we refer to the sample of single or resolved white dwarfs. As for unresolved companions, it plausible to speculate that we are more incomplete, which affects the derived number density of white dwarfs in the solar neighbourhood.

We also assessed the sky distribution of all stars in the CNS5 within the completeness limit and observe that their distribution is not only homogeneous but also isotropic.
However, the distribution of ultracool dwarfs in our catalogue with distances beyond the completeness limit has a smaller density in the Galactic plane than outside of it.
This indicates that the majority of missing brown dwarfs are probably hiding in the Galactic plane regions.
Recently, \citet{Best_2021AJ....161...42B} have performed a detailed analysis of anisotropies of brown dwarfs over the sky. Similarly to what we observe, they concluded that their sample shows a deficiency of brown dwarfs at low Galactic latitudes.

\subsection{Luminosity functions}
\label{sec:luminosity_function}

\subsubsection{Main-sequence luminosity functions}

In this section we illustrate that the CNS5 provides an excellent sample to derive the observational luminosity functions in optical as well as in MIR wavelengths.
As a first step, we want to  separate our sample into WDs and MS stars. WDs were selected using criteria defined in \equref{eq:wd_cut}.
The following limit was used to separate red giants (RG) from the main-sequence stars (MS):
\begin{equation}
    \label{eq:ms_rg_cut}
    \left.\begin{aligned}
    \text{(i)}\quad&M_G<-6.2 \magrm+15\,(G-RP)\\
    \text{(ii)}\quad&M_G<4.3 \magrm
    \end{aligned} \quad \right\} 
\end{equation}

As a result of these cuts (see \figref{fig:CMD_cuts}) we get a total of 20 RGs and 264 WDs.
All the remaining objects are assumed to belong to the main-sequence.

\begin{figure}
    \centering
    \includegraphics[width=0.49\textwidth]{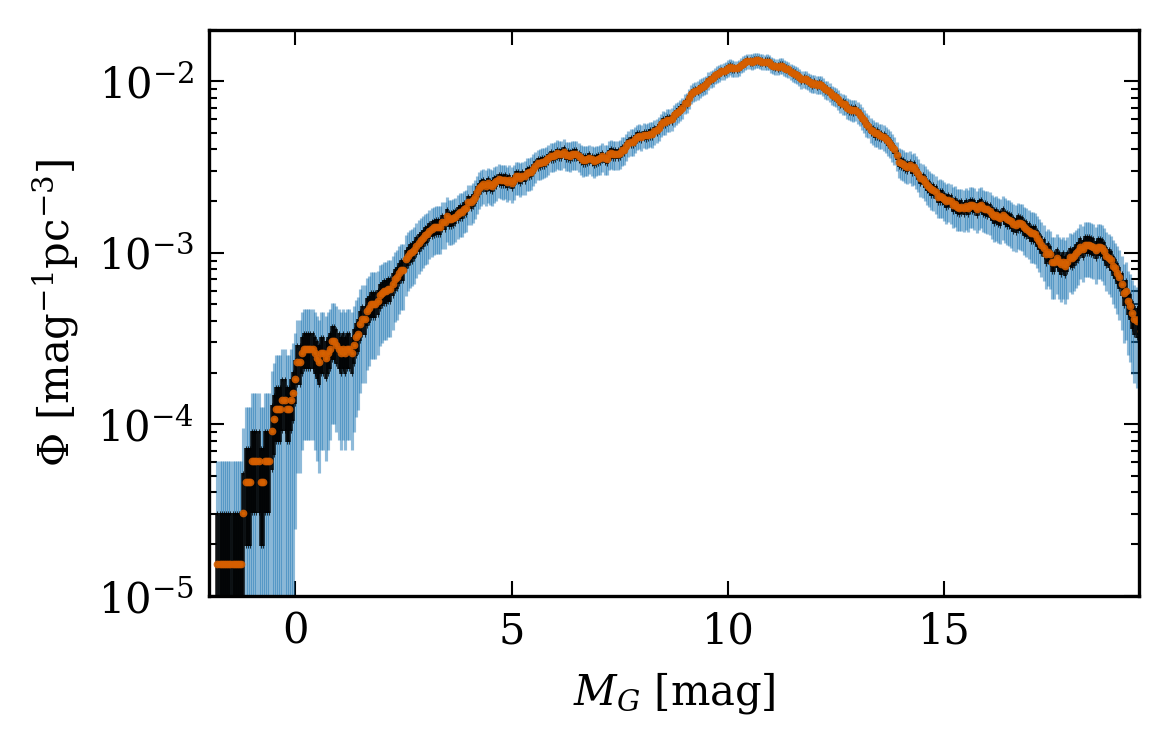}
    \caption{Luminosity function in $G$-band of the main-sequence stars in the solar neighbourhood. Black and blue error bars correspond respectively to $1\sigma~(68.3\%)$ and $3\sigma~(99.7\%)$ confidence intervals.}
    \label{fig:lf_ms_g}
\end{figure}

\begin{figure}
    \centering
    \includegraphics[width=0.49\textwidth]{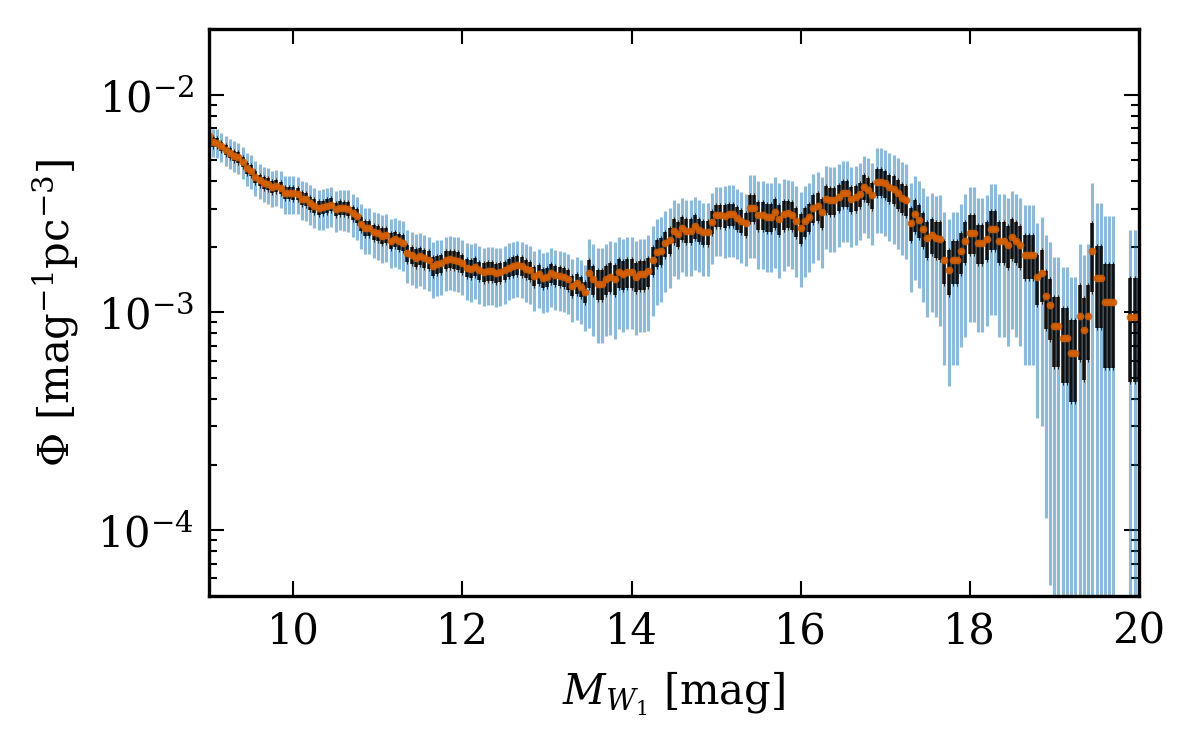}
    \caption{MIR luminosity function of the faint end of the main-sequence. Black and blue error bars correspond respectively to $1\sigma~(68.3\%)$ and $3\sigma~(99.7\%)$ confidence intervals. The shown magnitude range corresponds to the range where the majority of objects with no counterpart in \gedrthree{} are located (see \figref{fig:CMD_2mass_wise}). $M_{W_1}=9 \magrm$ translates approximately to $M_{G}=14 \magrm$.}
    \label{fig:lf_ms_w1}
\end{figure}

As it has been demonstrated in the previous section, the CNS5 catalogue is complete down to 19.7~mag in $G$ band. Therefore, the completeness correction was not required there. However, to derive the local observational luminosity function in the mid-infrared, we applied the classical $V_{max}$ technique \citep{Schmidt_1968ApJ...151..393S} using the derived distance limits from a Kolmogorov-Smirnov test to compute the effective limiting volume.
The derived luminosity functions for the main-sequence stars are shown in \figref{fig:lf_ms_g} and \ref{fig:lf_ms_w1}.
It is interesting to note that the presence of the dip at $M_G = 17.7\magrm$, where the stellar to substellar boundary is located \citep{gaiaedr3_gcns}, can be claimed with confidence. Also the increase in the luminosity function after this dip is physical and can be seen in both, optical and MIR luminosity functions.

\subsubsection{White dwarf luminosity function}

The white dwarf luminosity function (WDLF) is a crucial ingredient for the characterisation of the stellar content of the solar neighbourhood \citep{Weidemann_1967ZA.....67..286W}. In particular, the faint end of the WDLF offers the possibility to infer the age of the local stellar population \citep{Schmidt_1959ApJ...129..243S}.
The finite age of the population and thus limited cooling time of WD progenitors implies that there is an absolute magnitude fainter which no WD can be found in the local population. This is observed as an abrupt cut-off in the WDLF whose position is determined by age of the oldest white dwarfs in the solar neighbourhood \citep{Liebert_1979ApJ...233..226L, Liebert_1988ApJ...332..891L, Yuan_1992A&A...261..105Y, Harris_2006AJ....131..571H, Catalan_2008MNRAS.387.1693C}.

Furthermore, the WDLF is sensitive to the star formation rate (SFR). Thus, inverting the observational WDLF allows to reveal the star formation history of the solar neighbourhood and even to probe star formation bursts occurred in the past \citep{Yuan_1992A&A...261..105Y, Rowell_2013MNRAS.434.1549R}.

However, estimates of both, the SFR as a function of look-back time and of the age of the local population, are dependent not only on the adopted white dwarf evolutionary models, but also sensitive to the properties of the observational dataset which is used to derive the WDLF.
Therefore, the observational WDLF will better correspond to reality when derived from a volume-limited sample with maximised completeness and where the absolute magnitudes are estimated using parallaxes with exquisite precision and accuracy and not from reduced proper motions as it was done in the past (e.g.\ \citealt{Harris_2006AJ....131..571H}).

\begin{figure}
    \centering
    \includegraphics[width=0.49\textwidth]{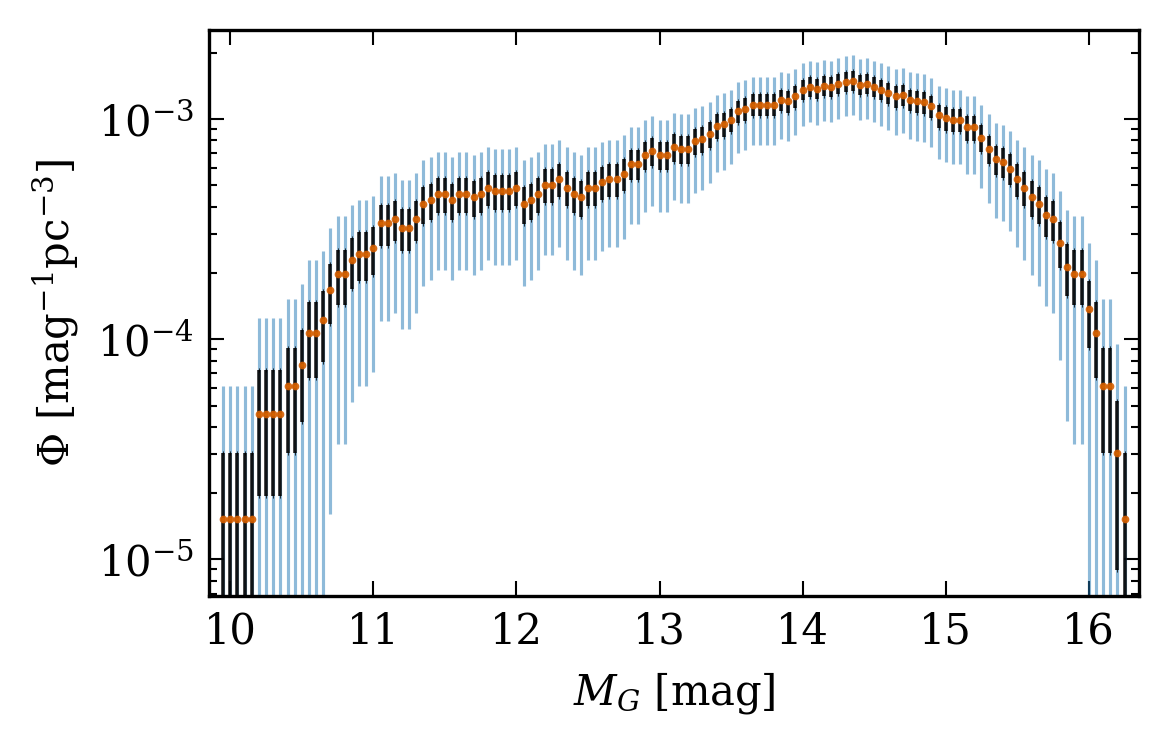}
    \caption{Luminosity function of the white dwarfs in the solar neighbourhood. Black and blue error bars correspond respectively to $1\sigma~(68.3\%)$ and $3\sigma~(99.7\%)$ confidence intervals.}
    \label{fig:lf_wd}
\end{figure}

Such a sample of white dwarfs is provided by the CNS5. Thus, we derive the observational WDLF in \gaia's $G$-band using the same approach as outlined above for MS stars.
The derived luminosity function is shown on \figref{fig:lf_wd}.

The total number density is another important parameter for describing a local WD population and is also used for the normalisation of the theoretical WDLFs.
The CNS5 contains 264 WDs within the 25~pc volume. This translates into a total number density of $(4.03\pm0.25)\times10^{-3}\,\text{stars pc}^{-3}$. Our result is
consistent with previous findings \citep{Liebert_2005ApJS..156...47L, gaiaedr3_gcns}.

\section{Discussion and Outlook}
\label{sec:discussion}

We first assess the overall completeness of CNS5 based on the completeness limits
derived in 
Sect.~\ref{sec:completeness}.
From Fig.~\ref{fig:g_w1_compl}, we estimate
that we statistically include all systems in the 25~pc volume with absolute $G$ magnitudes brighter than about 19.7~mag (or absolute $W1$ magnitudes brighter than 11.8~mag), corresponding to a spectral type of about L8 
(\citealt{dupuy2012,pecaut13}
\footnote{updates available at \url{https://www.pas.rochester.edu/~emamajek/EEM_dwarf_UBVIJHK_colors_Teff.txt}}).
Thus, the CNS5 includes virtually all main-sequence stars as well as virtually all brown dwarfs up to spectral type L8.
We stress that our completeness assessment refers to well-resolved components in case of binary or multiple stars; we are certainly more incomplete when it comes to the identification of companions, especially those of low mass or
of close separation.

Since we derived the completeness limit in a statistical sense, namely by judging from the density of objects in smaller and closer volumes to larger and more distant volumes, we do not claim that every single system is indeed listed, but the number of missing systems should be rather small compared to the number of objects listed in the catalogue. 

This is a major step forward; previous completeness limits for brown dwarfs in particular, but also for M dwarfs, were much closer than the 25 pc distance. 
The larger volume and larger number of objects that comes with an increased completeness radius enables statistical studies with much smaller error bars. 

Previously, RECONS \citep{henry16} had listed 366 objects within the 10~pc sample, which they estimated was 90\% complete. CNS5 has added 26 objects within the 10~pc volume, which now contains 392 objects with parallaxes larger or equal to 100~mas. Even when compared to the GCNS and considering the $3\sigma$ confidence interval of the parallaxes, the CNS5 lists 108 more objects within the 10~pc and 739 more within the 25~pc volume because it includes not only \textit{Gaia} data, but also \hip{} for the bright stars otherwise missing in \textit{Gaia} as well as the infrared ground-based parallax survey from \citet{Best_2021AJ....161...42B, Kirkpatrick_2021ApJS..253....7K},
which helps tremendously with the completeness of substellar objects.  

When comparing the CNS5 with the 10~pc sample from \citet{Reyle_2021A&A...650A.201R}, we found that
the CNS5 adds only two objects not listed in the 10 pc sample, presumably due to the different selection approaches used. 
Conversely, 148 sources from their 10~pc sample are missing in the CNS5 because they are either an exoplanet, a component of an unresolved binary (these will be part of the CNS6), or if they have a parallax uncertainty $\sigma_\varpi > 10$~mas or if their parallax was just an estimate without uncertainty (such as \object{CWISE J061741.79+194512.8} AB).

The CNS5 contains 5931 objects in total. Among them, there are 5230 stars (including the Sun, 20 giant stars and 264 white dwarfs) and 701 brown dwarfs. Here, we assume that the approximate stellar/substellar boundary lies at $80~M_{Jup} = 0.076~M_\sun$, corresponding to spectral type L1 ($M_G=16.90\magrm$ or $W1-W2=0.28\magrm$). 
This yields a stellar number density of $(7.99\pm0.11)\times10^{-2}\,\text{stars pc}^{-3}$, while the number density for brown dwarfs is $(1.07\pm0.04)\times10^{-2}\,\text{objects pc}^{-3}$.

About 72\% of stars (or $\sim63\%$ of stars and brown dwarfs) in the CNS5
are M~dwarfs (there are 3760 main-sequence stars in the CNS5 with absolute magnitudes in the range $8.16\magrm\leq~M_G\leq16.60\magrm$ as appropriate for M~dwarfs, out of a total of 5230 stars).
For comparison, the fraction of M-dwarfs in the 10~pc sample was found to be $\sim61\%$ \citep{Reyle_2021A&A...650A.201R}.
The number of M-dwarfs within the 25~pc volume could still change in the future, in particular due to unresolved multiple systems in the CNS5.

An interesting quantity is also the fraction of stars relative to the fraction of brown dwarfs, which is directly related to the star formation process. Previous estimates included a factor of
6 times more stars than brown dwarfs in the 8~pc volume by \citet{kirkpatrick2012}, 
10 times more stars than brown dwarfs in the 10~pc volume from the RECONS project \citet{henry16} and a factor of 5.2 from \citet{bihain2016} within 6.5~pc, while newer estimates seem to converge at a factor around 6--7
\citet{kirkpatrick19,henry19}. There are 4.4 stars per each brown dwarf listed in the 10~pc sample from \citet{Reyle_2021A&A...650A.201R}. In the CNS5, the fraction of stars to brown dwarfs is $4.6\pm0.4$ for the 15~pc volume and it is consistent with the values derived for smaller volumes (see \tabref{tab:stars_to_bd_ratio}). Within the total volume of 25~pc, the factor is $7.5\pm0.3$, indicating that  more brown dwarfs than stars are missing at larger distances.

\begin{table}
\caption{Star to brown dwarf ratio in the CNS5 for different distance limits. The ratio increases significantly at distances larger than 15~pc, indicating that more brown dwarfs than stars are missing at larger distances.}
\label{tab:stars_to_bd_ratio}
    \centering
    \begin{tabular}{l|llll}
        \hline
        \hline
         $r$ & \# of stars & \# of BDs & Total & Stars/BDs  \\
         \hline
         6.5~pc & 116 & 22 & 138 & $5.3\pm1.4$ \\
         8.0~pc & 164 & 40 & 204 & $4.1\pm0.8$ \\
         10~pc & 319 & 74 & 393 & $4.3\pm0.6$ \\
         15~pc & 1066 & 231 & 1297 & $4.6\pm0.4$ \\
         20~pc & 2590 & 413 & 3003 & $6.3\pm0.3$ \\
         25~pc & 5230 & 701 & 5931 & $7.5\pm0.3$ \\
         \hline
    \end{tabular}
\end{table}

The brown dwarf census is still somewhat limited. Assuming the star to brown dwarf ratio is about 5, we can estimate an approximate number of brown dwarfs currently missing. Knowing that there are 5230 stars in the 25\,pc volume, one would expect 1046 brown dwarfs in the same volume. As mentioned above, 701 of them are listed in the CNS5, so that about 345 brown dwarfs are missing.
Therefore, we  conclude that about one third of brown dwarfs, corresponding to about 6\% of all objects within 25~pc, are still to be discovered. 

In \secref{sec:completeness}, we have shown that the white dwarf sample in the CNS5 is statistically complete within 25~pc. 
We derive a number density of white dwarfs of $(4.03\pm0.25)\times10^{-3}\,\text{stars pc}^{-3}$. 
Thus, white
dwarfs represent 5\% of the local stellar
population. 
We do not see a single faint blue white dwarf as seen by \citet{Scholz_FBWD_2022RNAAS...6...36S}, but there are only 59 such objects in the GCNS, so our volume of 25~pc might be too small to include one of these rare objects.

Similarly, we do not clearly see the Jao
gap \citep{jao18} at absolute $G$ magnitudes of 10.14~mag, although one could notice a small feature if looking closely. The gap becomes more prominent only if looking at larger volumes; it is visible in the GCNS \citep{gaiaedr3_gcns}.

We plan to compile another version of the CNS, termed CNS6, based on \gdrthree{}
in the future. In particular, we will
include more information on multiplicity,
including spectroscopic companions from the literature and homogeneously derived stellar parameters for all objects in the catalogue.

The CNS5 catalogue also provides the
opportunity to investigate the
exoplanet inventory in a sample
which is unbiased with regard
to spectral type or apparent
magnitude, in contrast to other typical planet search samples. 
The planet population within
25~pc might be more representative of the overall planet population in the Milky Way, at least in those
environments that resemble the
solar neighbourhood, than just
the overall known planet population which is highly biased by the sensitivities of the various planet search methods. 
Furthermore,
the astrometric method is most
sensitive around the most nearby stars and thus ideally complements our knowledge about exoplanets in
the nearby stars sample, which so far mainly comes from the transit
and the radial velocity methods. Since astrometry is most sensitive at intermediate periods or
semi-major axes, it ideally complements the transit and radial velocity methods which are most sensitive to the shortest periods, and direct imaging detections at the widest separations. 
Significant contributions regarding intermediate period/semi-major axis planets in the nearby stars sample are expected especially from \textit{Gaia} DR4, which will also contain epoch astrometry.

\begin{acknowledgements}
The authors thank the referee, Dr.~C\'eline Reyl\'e, for constructive and valuable comments on the manuscript and insightful suggestions that led to a significant improvement of this paper.
We kindly thank Christian Dettbarn for his work on an early version
of the catalogue and Markus Demleitner for providing the catalogue via the Virtual Observatory.

Part of this work was supported by the International Max Planck Research School for Astronomy and Cosmic Physics at the University of Heidelberg, IMPRS-HD, Germany.
A.G. and A.J. gratefully acknowledge funding from the Deutsche Forschungsgemeinschaft (DFG, German Research Foundation) -- Project-ID 138713538 -- SFB 881 (``The Milky Way System'', subproject A06).

This work has made use of:
TOPCAT \citep{topcat_2005ASPC..347...29T, topcat_2019ASPC..523...43T}, a GUI analysis package for working with tabular data in astronomy;
Astropy, a community-developed core Python package for Astronomy \citep{astropy2018};
Scipy, a set of open source scientific and numerical tools for Python \citep{scipy_2020NatMe..17..261V};
the VizieR catalogue access tool and the SIMBAD database operated at CDS, Strasbourg, France;
the National Aeronautics and Space Administration (NASA) Astrophysics Data System (ADS).

This work has made use of
data from the European Space Agency (ESA) mission {\it Gaia} (\url{https://www.cosmos.esa.int/gaia}), processed by the {\it Gaia}
Data Processing and Analysis Consortium (DPAC, \url{https://www.cosmos.esa.int/web/gaia/dpac/consortium}). Funding for the DPAC has been provided by national institutions, in particular the institutions participating in the {\it Gaia} Multilateral Agreement.
\end{acknowledgements}

\bibliographystyle{aa}
\bibliography{refs}

\begin{appendix}
\section{Selecting astrometrically clean datasets from \gedrthree{}}
\label{sec:harm_ampl_cut}

Data selection from \gedrthree{} is compromised by the presence of sources with spurious astrometric solutions in the published catalogue. In this section, we discuss our quality cut which eliminates these sources. The validation of this approach will follow in the next subsection.

There is no unique selection criterion (or combination of criteria) which would differentiate between sources with spurious and reliable astrometric solutions in \gedrthree{} for every use case. For each specific task one should choose individually appropriate selection criteria.
Our aim is to define a cut which is conservative enough to retrieve only objects with reliable astrometry (parallaxes in particular). At the same time, the adopted cut should not lessen the completeness of our sample by removing objects which do have reliable parallaxes but with inconsistencies in photometry or even with null values of $G, BP$ or $RP$ fluxes.

$RUWE$ (re-normalised unit weight error; \citealt{lindegren18b}) and photometric excess factor (\citealt{Evans_gaia_dr2_photom_2018A&A...616A...4E, lindegren18}; \texttt{phot\_bp\_rp\_excess\_factor} in the \gedrthree{} catalogue) are the two most widely used parameters in various selection criteria. Deriving both of these parameters involves photometric measurements: the re-normalisation factor of $RUWE$ depends on the magnitude and colour of the source, and photometric excess is defined as the flux ratio $(I_{BP}+I_{RP})/I_G$. A deviation from the expected flux excess indicates inconsistency between $G$, $BP$ and $RP$ photometry and therefore often used to exclude such sources. However, one should keep in mind that this metric does not reveal the origin of the inconsistency: it is not possible to distinguish between objects where only $BP$ and $RP$ fluxes from the \gaia{} photometric instrument are affected and objects where the $G$ flux from the astrometric field is problematic too.

It was shown that both parameters can reach high values for sources with peculiarities in their photometry: an excess in $RUWE$ can be induced by photometric variability or by the presence of an unresolved companion \citep{Belokurov_ruwe_2020MNRAS.496.1922B}, while the photometric excess factor tends to be larger for objects with peculiar SEDs as well as for various types of variable stars \citep{gaiaedr3_photometry}. Therefore, making  the decision whether to include an object in the catalogue or not based on photometric excess or $RUWE$ would bias our catalogue against such types of objects. 

\begin{figure}
    \centering
    \includegraphics[width=0.48\textwidth]{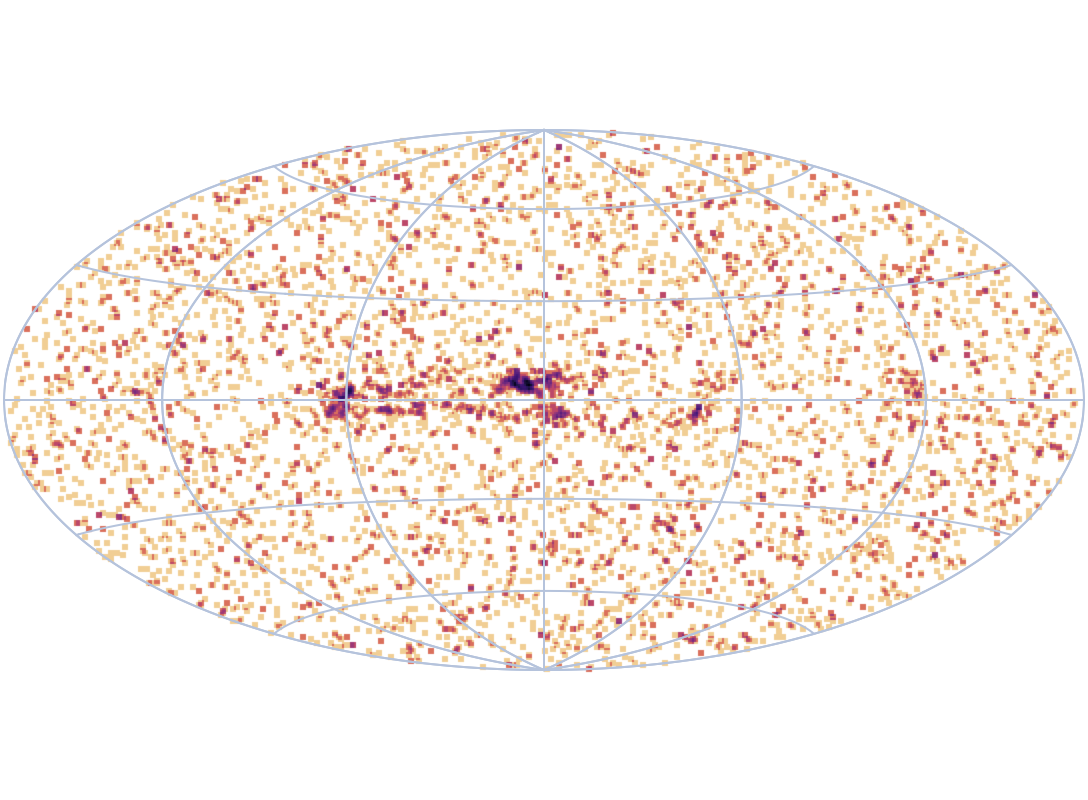}
    \includegraphics[width=0.48\textwidth]{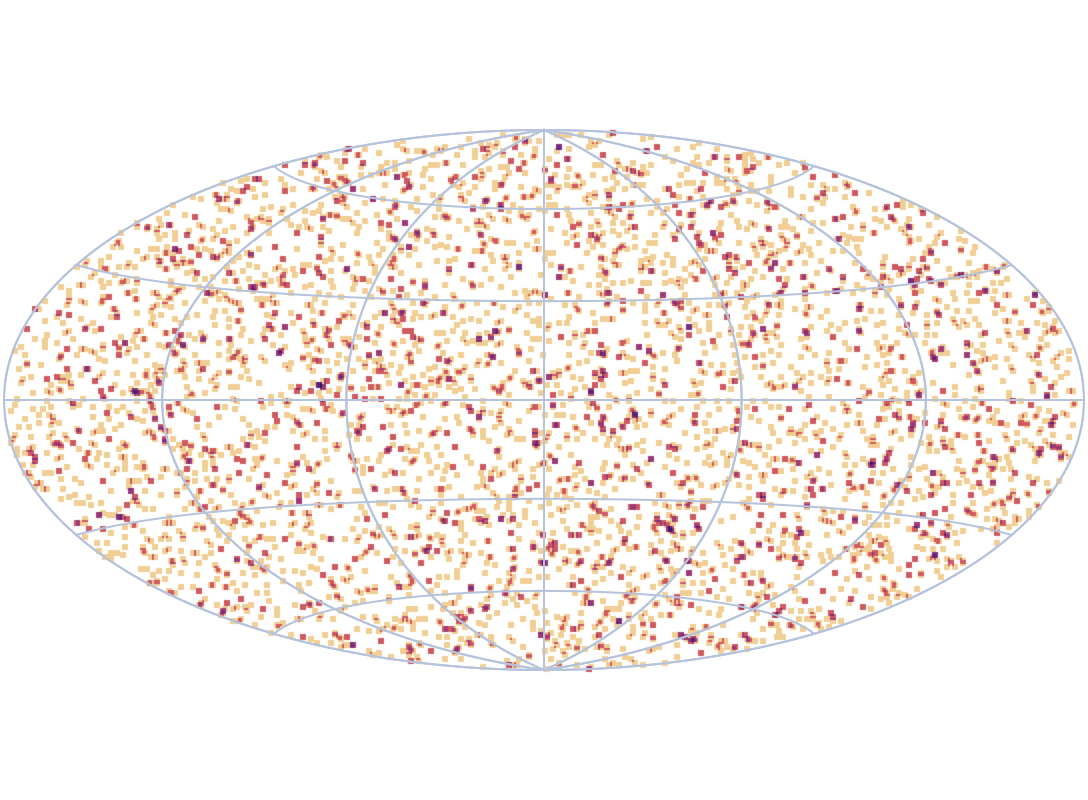} 
    \caption{\textit{Top:} Map of the 25 pc dataset from \gedrthree{} in Galactic coordinates with $l = b = 0$ at the centre. The overdensity along the Galactic plane is an artefact from spurious solutions. \textit{Bottom:} sources from the 25 pc dataset which remained after applying the cut defined in \equref{eq:A_ipd_gof_harm_ampl_cut}.}
    \label{fig:skymap_all}
\end{figure}

We recall that spurious solutions are often caused by mistakes in cross-matching of individual measurements when assigning the \texttt{source\_id}.
Measurements of two (or more) different objects associated to the same \texttt{source\_id} will result in a spurious catalogue entry: the pipeline interprets the measured displacement as an apparent motion of the object and describes it through a non-physical astrometric solution. This leads to a spuriously large parallax, yet with relatively large formal uncertainties (as a consequence of large centroid dispersion).
Obviously, such confusion occurs mostly in crowded fields, for instance, close to the Galactic plane and the Magellanic clouds. This is illustrated by \figref{fig:skymap_all} (\textit{top panel}): an overdensity along the Galactic plane is clearly visible in the sky distribution of the sample (hereafter, 25~pc sample), retrieved by selecting all objects from \gedrthree{} satisfying \equref{eq:plx_cut}.
Furthermore, the mismatching probability for a source also depends on the distribution of scan directions relative to the position angle of confused objects.
The necessary information can be obtained from the Image Parameter Determination goodness-of-fit \citep[IPD GoF;][]{gaiaedr3_documentation_ch13_hambly}, which is parameterised as
\begin{equation}
    \ln(\mathrm{GoF}) = c_0 + c_2 \cos(2\psi) + s_2\sin(2\psi),
\end{equation}
where $\psi$ is the position angle of the scan direction and $c_0, c_2, s_2$ are Fourier coefficients.
If an object is a binary, this will be indicated by a large amplitude of the IPD GoF (\texttt{ipd\_gof\_harmonic\_amplitude} in \gedrthree{}), which indicates the level of asymmetry in the image and is computed as
\begin{equation}
    A_{\rm GoF} = \sqrt{c_2^2+s_2^2}.
\end{equation}

\begin{figure}
    \centering
    \includegraphics[width=0.49\textwidth]{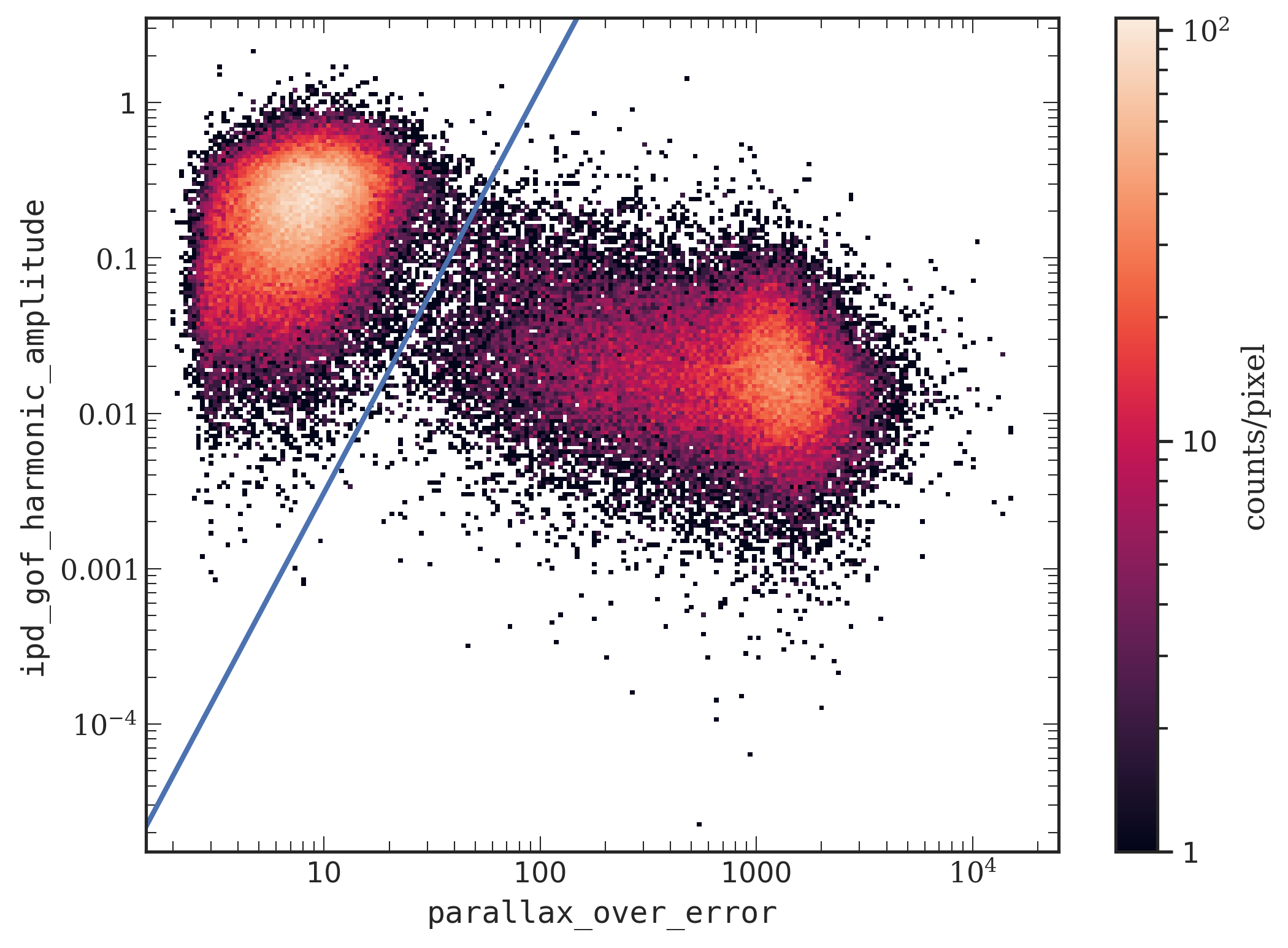} 
    \caption{Same as \figref{fig:ipd_harm_ampl_cut_25} but for 50~pc sample. As opposed to 25~pc sample, the 50~pc sample is dominated by sources with spurious astrometric solutions: 42~607 sources (50.84\%) are located above the blue line -- the threshold defined in \equref{eq:A_ipd_gof_harm_ampl_cut}.}
    \label{fig:ipd_harm_ampl_cut}
\end{figure}

Given the above, we can expect a dichotomy between spurious and reliable astrometric solutions in the ($A_{\rm GoF}$, $\varpi/\sigma_\varpi$) parameter space. This dichotomy is shown for the 50~pc sample in \figref{fig:ipd_harm_ampl_cut}: sources with spurious solutions (the cluster of points in the top left corner) are well separated from the sources with reliable astrometric solutions.

For an astrometric solution to be considered reliable, we require the source to satisfy \equref{eq:ipd_gof_harm_ampl_cut}, reproduced here for convenience:
\begin{equation}
    \label{eq:A_ipd_gof_harm_ampl_cut}
    A_{\rm GoF} < 10^{-5.12}~(\varpi/\sigma_\varpi)^{2.61},
\end{equation}
where $\varpi/\sigma_\varpi$ corresponds to \texttt{parallax\_over\_error} as published in the \gedrthree{} catalogue; that is, parallax zero-point corrections or error inflation are not taken into account for this selection. Here, the slope of the line is chosen so that it follows beneath the cluster of sources with spurious solutions.

However, this approach comes with its limitation: it works best for nearby stars and it is only applicable in the high-S/N regime. In other words, the cut is tailored to objects with significant parallaxes, where $\varpi/\sigma_\varpi$ is larger than that of sources with spurious astrometric solution.
For more distant sources (\figref{fig:ipd_harm_ampl_cut}, cf.~\figref{fig:ipd_harm_ampl_cut_25}), the parallax-over-error ratio decreases, the valley between the clusters of spurious and reliable solutions becomes less clean, and the slope of the cut has to be adjusted in each individual case according to demands for completeness and purity.

\begin{figure}
    \centering
    \includegraphics[width=0.48\textwidth]{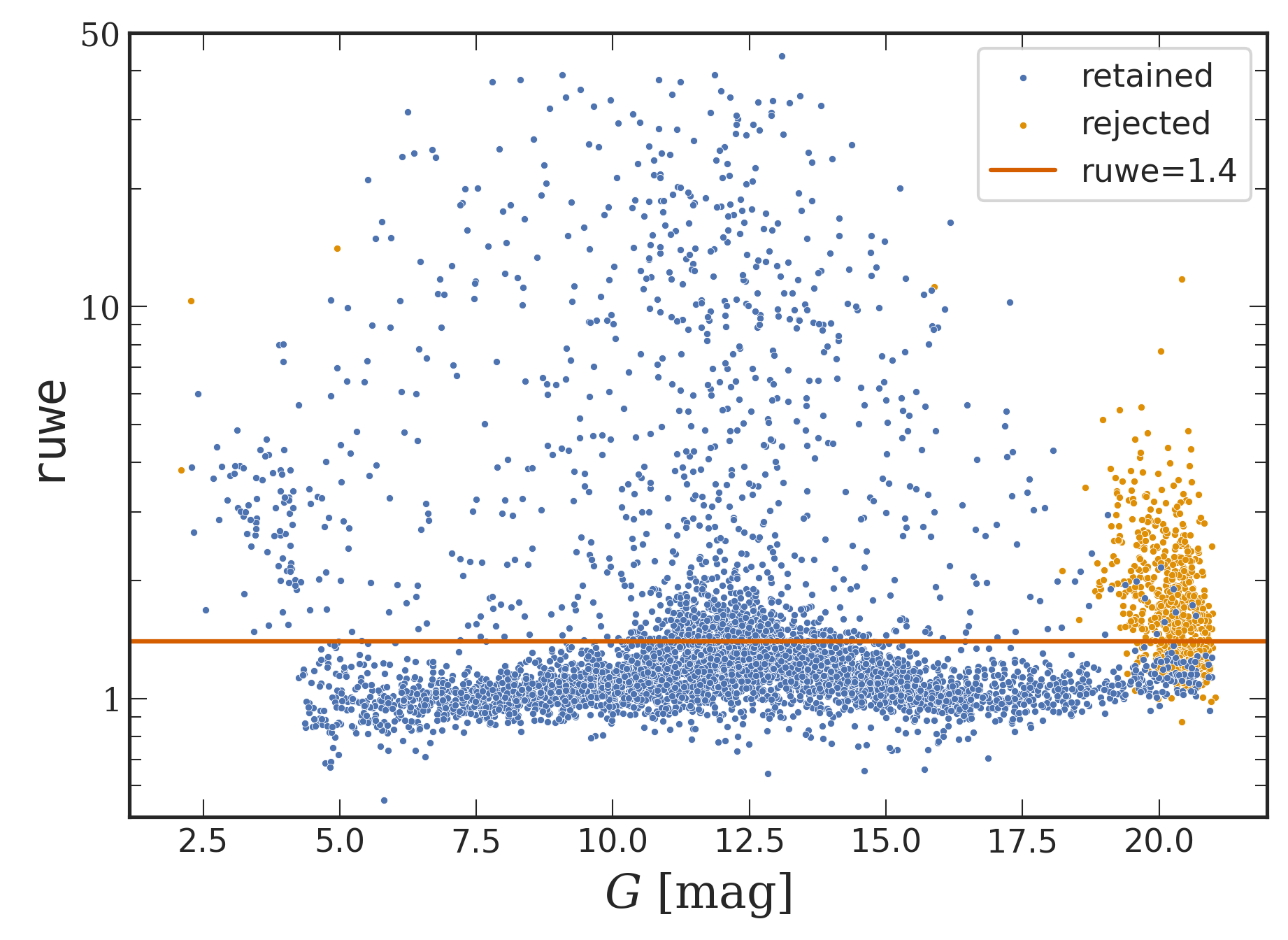} 
    \caption{Distribution of $RUWE$ for sources retained (shown in blue) and rejected (shown in yellow) by criterion defined in \equref{eq:A_ipd_gof_harm_ampl_cut} as a function of $G$ magnitude. The solid horizontal line shows the $RUWE=1.4$ threshold. Note the presence of rejected sources with $RUWE$ bellow the threshold.}
    \label{fig:ipd_harm_ampl_ruwe}
\end{figure}

Figure~\ref{fig:ipd_harm_ampl_ruwe} illustrates that the widely used cut $RUWE < 1.4$ is not an optimal approach when completeness of a sample is important. A large number of objects, especially at magnitudes around $G\approx12~{\rm mag}$, have small amplitudes of the IPD GoF (below the threshold defined by \equref{eq:A_ipd_gof_harm_ampl_cut}), but would otherwise have been removed by the cut on $RUWE$. On the other hand, we should note the presence of spurious objects with $G\approx20~{\rm mag}$ with $RUWE<1.4$. Including them in the sample would contaminate CNS5 and yield an overestimated completeness and distorted luminosity function.

\subsection{Selection criterion validation}
\label{sec:cut_validation}

In the following, we examine whether \texttt{ipd\_gof\_harmonic\_amplitude} and \texttt{parallax\_over\_error} are sufficient to define a rigorous selection criterion for objects with significant parallaxes to eliminate spurious astrometric solutions in \gedrthree{} and whether such selection criterion yields results as good as with other classifiers which are in use.

\subsubsection{`Fidelity' from Rybizki et al. (2022)}
\label{sec:cut_validation_fidelity}

\begin{figure}
    \centering
    \includegraphics[width=0.49\textwidth]{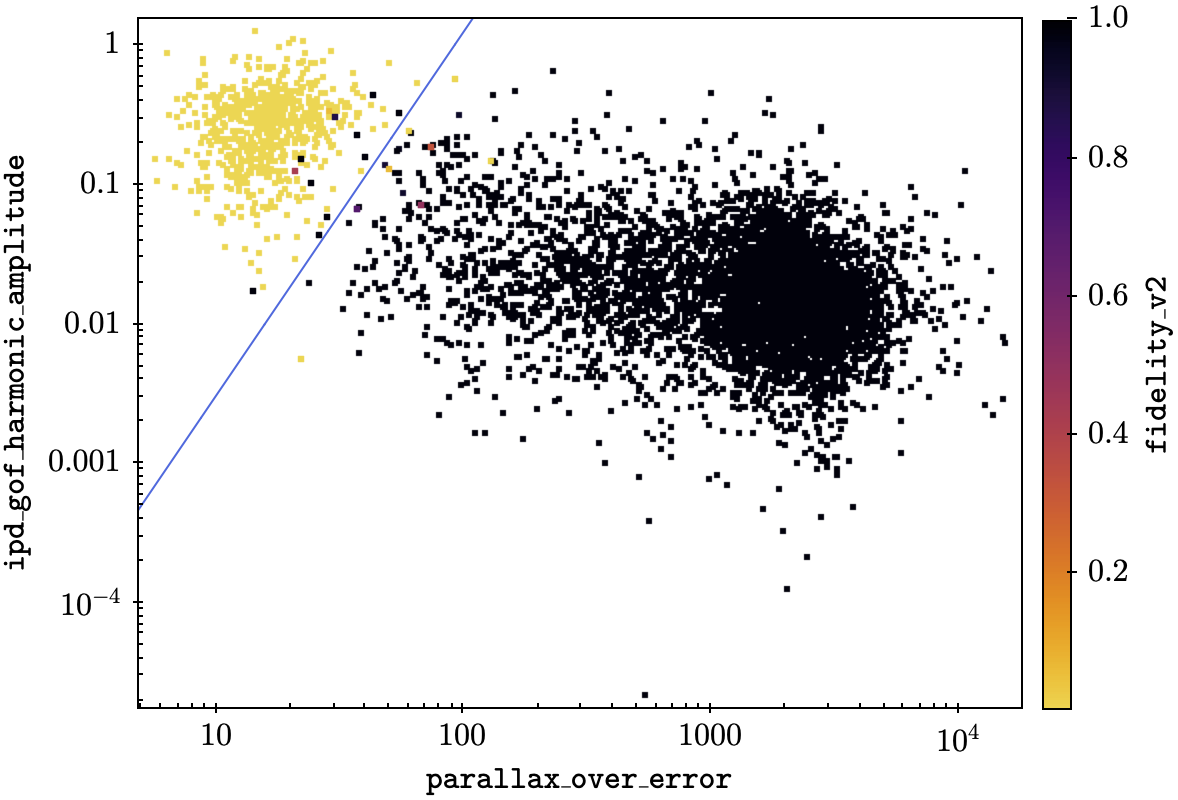}
    \caption{`Astrometric fidelity' from \citet{Rybizki_fidelity_2021arXiv210111641R} as a function of position in ($A_{GoF}, \varpi/\sigma_{\varpi}$) space for all the sources from \gedrthree{} that are possibly located within 25~pc of the Sun. The blue solid line corresponds to the cut described by \equref{eq:A_ipd_gof_harm_ampl_cut}.}
    \label{fig:fidelity}
\end{figure}

\citet{Rybizki_fidelity_2021arXiv210111641R} used machine learning to classify astrometric solutions in \gedrthree{} and to eliminate spurious solutions. They trained a neural network using 17 parameters from \gedrthree{} (including \texttt{ipd\_gof\_harmonic\_amplitude} and \texttt{parallax\_over\_error}) and provided an `astrometric fidelity' for each source  - a parameter having values in the range between 0 (spurious solution) and 1 (reliable solution).
This approach is somewhat similar to what has been done for the GCNS, now extended to the entire data release.

Here, we validate our cut on harmonic amplitude against the updated version of the classifier (\texttt{fidelity\_v2} column in the \texttt{gedr3spur.main} table\footnote{Hosted at the German Astrophysical Virtual Observatory (GAVO; \url{https://dc.zah.uni-heidelberg.de/})}). The principal difference between the original (\texttt{v1}) and the updated (\texttt{v2}) classifier is the new high-density training data and crossmatching of the training sample to Pan-STARRS1 instead of to 2MASS in order to avoid bias against faint blue objects \citep{Rybizki_fidelity_2021arXiv210111641R}.

Figure~\ref{fig:fidelity} illustrates the remarkable agreement between our cut and the fidelity parameter.
Of the 5184 objects classified as good by us, only 8 have fidelity < 0.9, while there are only 9 sources with fidelity > 0.9 rejected by our cut. Hence we conclude that in $99.84\%$ of the cases the cut on harmonic amplitude and the neural network classifier agree on whether the astrometric solution of an object is spurious or not.

\begin{table*}
\caption{Sources in the GCNS within 25~pc rejected by our selection criterion defined in \equref{eq:A_ipd_gof_harm_ampl_cut}.}
\label{tab:gcns_rejected}
\centering
\begin{tabular}{lclllllll}
\hline \hline
  \multicolumn{1}{l}{\gedrthree{}} &
  \multicolumn{1}{c}{Known} &
  \multicolumn{1}{c}{\parallax} &
  \multicolumn{1}{c}{\parallaxerror} &
  \multicolumn{1}{c}{\gmag} &
  \multicolumn{1}{l}{2MASS designation} &
  \multicolumn{1}{c}{$K$} &
  \multicolumn{1}{c}{$\sigma_K$} \\
  \texttt{source\_id} & binary & [mas] & [mas] & [mag] && [mag] &  [mag]  \\
\hline
2361294885296927488 &  N & 38.05 & 1.58 & 20.80 & 00135779-2235200 & 14.04 & 0.05\\
4993479684438433792 &  Y & 39.92 & 0.73 & 2.09 & 00261699-4218216 & -0.17 & 0.38\\
2800081221135505536 &  N & 42.89 & 1.65 & 20.83 & 00282091+2249050 & 13.78 & 0.06\\
5151358868307074432 &  Y & 49.35 & 1.63 & 20.23 & 02052940-1159296 & 13.00 & 0.03\\
251476382498509952 &  N & 36.20 & 1.74 & 20.84 & 03454743+5137159 & 13.79 & 0.06\\
3056233482992410496 &  N & 43.35 & 1.52 & 19.69 & 07414279-0506464 & 12.39 & 0.03\\
761918578311083264 &  Y & 42.24 & 1.06 & 19.98 & 11122567+3548131 & 12.72 & 0.03\\
3591011474602344320 &  N & 43.65 & 1.99 & 20.86 & 11181292-0856106 & 14.18 & 0.08\\
3896357089270247168 &  N & 37.32 & 2.66 & 20.70 & 11582077+0435014 & 14.44 & 0.06\\
1222646935698492160 &  Y & 42.24 & 0.98 & 2.27 & 15344125+2642529 & 2.21 & 0.36\\
5952826016600852352 &  N & 36.13 & 1.29 & 20.49 & 17224991-4449366 & >13.81 & \\
2411855072801158656 &  Y & 39.40 & 1.06 & 4.95 & 23190668-1327310 & 3.35 & 0.26\\
\hline\end{tabular}
\end{table*}

\begin{figure}
    \centering
    \includegraphics[width=0.49\textwidth]{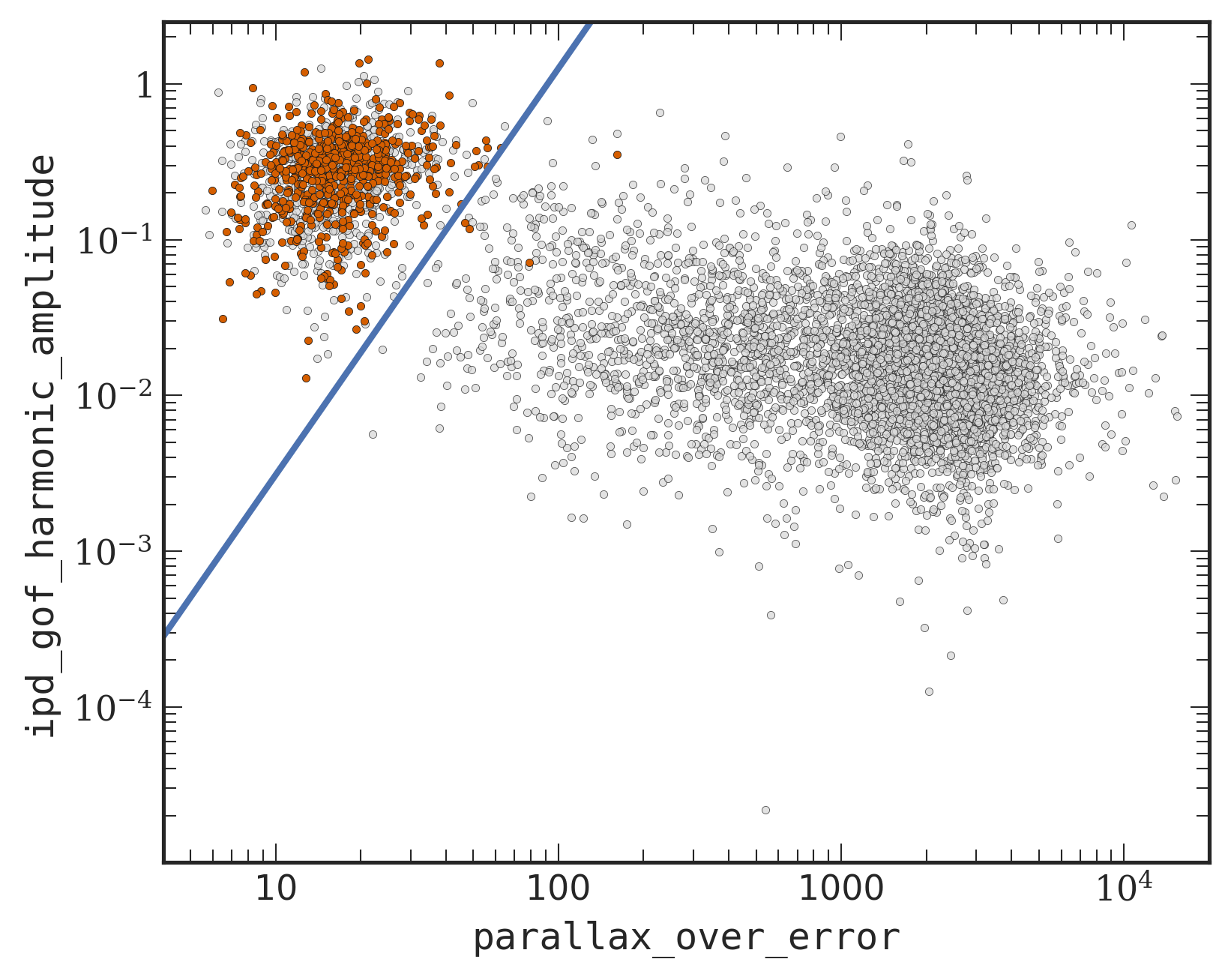} 
    \caption{Amplitude of the IPD GoF for the control sample of sources with spurious astrometric solutions as a function of $|\varpi/\sigma_\varpi|$ (red markers) overplotted on the 25 pc sample (grey markers). The blue solid line corresponds to the cut described by \equref{eq:A_ipd_gof_harm_ampl_cut}.}
    \label{fig:ipd_harm_ampl_cut_negative}
\end{figure}

\subsubsection{Sample with negative parallaxes}
\label{sec:cut_validation_negative}
Performance of a selection criterion for a particular sample can be assessed through a control sample of sources with spurious astrometric solutions. Such control sample can be constructed under the fundamental assumption that the probability of positive or negative parallax in a spurious solution is approximately equal; that is, the probability density function is symmetric around $\varpi=0$. Assuming this symmetry and the fact that any astrometric solution with a highly negative parallax (i.e. a negative parallax where the deviation from zero is statistically significant) is spurious by definition, we can estimate the number of sources with spurious solutions in our sample of objects with positive parallaxes \citep{gaiaedr3_validation}.
Additional support for this approach comes from \citet{gaiaedr3_gcns}: using the large negative parallax sample, they have constructed the training set for a random forest classifier and utilised it to generate the GCNS.

That being said, we select the control sample symmetrically to \equref{eq:plx_cut}:
\begin{equation}
    \label{eq:A_neg_parallax}
    \varpi-3\sigma_\varpi \leq -40~\text{mas}.
\end{equation}
The resulting sample contains 630 objects.

Consequently, when we apply our cut on harmonic amplitude (\equref{eq:A_ipd_gof_harm_ampl_cut}) to the control sample we may infer the fraction of sources with spurious solutions, which are not filtered out by the cut and evaluate whether this fraction is statistically significant. If the cut was perfect it would reject all the objects from the control sample.

In \figref{fig:ipd_harm_ampl_cut_negative} we show the control sample of sources with spurious solutions in the (\texttt{ipd\_gof\_harmonic\_amplitude}, $|\varpi/\sigma_\varpi|$) parameter space (red markers), whilst the 25 pc sample is shown in the background (grey markers). As expected, the vast majority of the sources in the control sample have large amplitude of the IPD~GoF and low (absolute) values of the \texttt{parallax\_over\_error}. Therefore, these sources do not satisfy the condition in \equref{eq:A_ipd_gof_harm_ampl_cut} and thus are filtered out by our cut (indicated by the blue solid line).

Here, only 4 sources are erroneously retained by our cut.
This corresponds to $0.63\%$ of the control sample and is well below the Poisson noise $\sqrt{N}=25.1$. These results allow  us to conclude that the cut works as intended.

\subsubsection{GCNS}
\label{sec:cut_validation_gcns}

Here we compare our classifier with the classifier which was used to generate the GCNS. As outlined in \secref{sec:hist_GCNS}, sources in the GCNS were selected from \gedrthree{} using supervised machine-learning. Specifically, removal of sources with spurious solutions was done with a random forest classifier which included 14 predictive variables \citep{gaiaedr3_gcns}.

To validate our selection criterion in \equref{eq:A_ipd_gof_harm_ampl_cut}, we use the sample of sources in the GCNS that are possibly located within 25~pc of the Sun. These are the sources that satisfy our \equref{eq:plx_cut}. There are 5191 such sources in the GCNS.

When comparing both datasets, we find that there are five sources included in our sample which are rejected in the GCNS. 
What these sources have in common is relatively low flux-over-error ratio (especially in $RP$ band) and, in addition, either $100\%$ of their transits flagged as blended in the $BP$ and $RP$ bands (four sources) or there is no $BP$ and $G$ photometry at all, while 15 out of 20 $RP$ transits are flagged as blended (one source).
However, as we discuss in Appendix~\ref{sec:synthetic_magnitudes}, issues with $BP$ and $RP$ photometry, including blending, do not necessarily imply that the astrometric solution is spurious since the astrometric solution in \gaia{} is obtained using data from the CCDs in the astrometric field, whereas $BP$ and $RP$ fluxes are derived from the integration of the BP and RP low-resolution spectra.

On the other hand, 12 sources which are included in the GCNS (0.2\% of the GCNS 25~pc sample) are rejected by our classifier as they have high harmonic amplitude and low parallax-over-error ratio. 
Possible causes include extreme magnitudes (three sources have magnitudes in the range $2.09\leq G\leq 4.95$~mag, whereas the remaining nine sources are fainter than $G\geq19.69$~mag) and the probable binarity of these sources. 
The list of these sources and their properties is shown in \tabref{tab:gcns_rejected}.
All sources in the sample have a counterpart in 2MASS. 
Five of the sources are indeed known binaries.
Given the colour of the remaining seven sources and their excessive IPD GoF harmonic amplitude, these sources are promising candidates when hunting for nearby brown dwarf binaries, especially with high-resolution imaging surveys.

Overall, the selection with the cut on harmonic amplitude is consistent with the GCNS classifier in $99.8\%$ of the cases when limited to the 25~pc volume.

\FloatBarrier

\section{Deblended \textit{G-RP} colour}
\label{sec:synthetic_magnitudes}

 The $BP$ fluxes of faint sources published in \gedrthree{} are often overestimated due to the low flux threshold problem \citep{gaiaedr3_photometry, gaiaedr3_summary}. Consequently, such objects appear to be bluer in $BP-RP$ or $BP-G$ colours. Another impact of this issue is the reduced measured scatter in the $BP$-band for these sources. 

\begin{figure}
    \centering
    \includegraphics[width=0.49\textwidth]{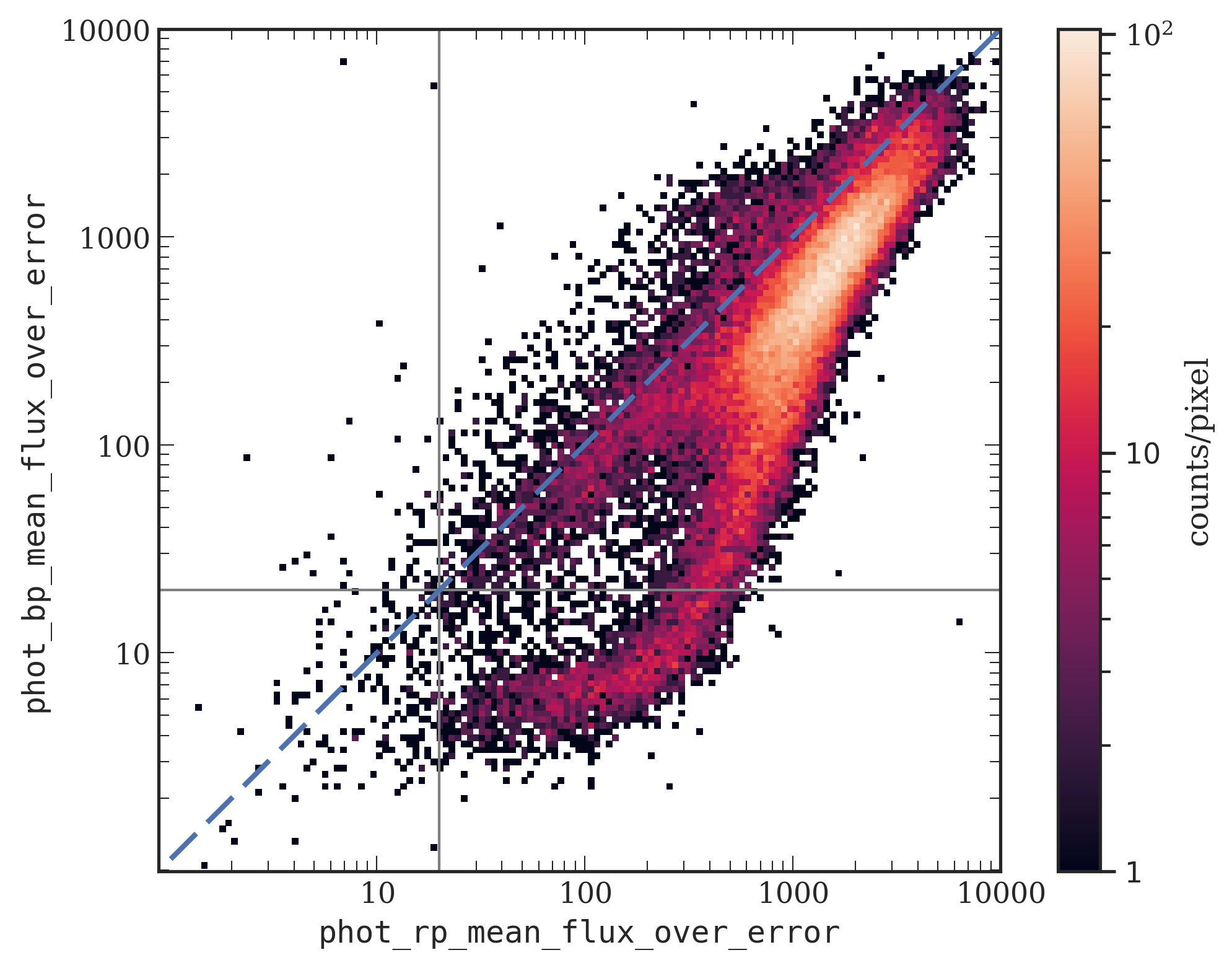}
    \caption{$BP$ and $RP$ flux-over-error ratios for 39\,873 sources in a 50 pc sample retained by \equref{eq:A_ipd_gof_harm_ampl_cut}. The vast majority of objects is located below the 1:1 line (blue dashed line), i.e. the S/N for these objects is higher in the $RP$ band than in the $BP$ band. Grey lines correspond to the S/N threshold applied in \equref{eq:synth_snr}. It is noteworthy that there is a significant fraction of objects with high-S/N $RP$ photometry, but low-S/N $BP$ photometry ($S/N_{BP}<20\ \&\ S/N_{RP}>20$).}
    \label{fig:bp_rp_flux_over_error}
\end{figure}

Furthermore, the signal-to-noise ratio for the vast majority of objects is significantly higher in the $RP$-band than that in the $BP$-band (\figref{fig:bp_rp_flux_over_error}).
Thus, the use of the $G-RP$ colour is often favoured over the $BP-RP$
(e.g.\ \citealt{Scholz_2020A&A...637A..45S, Kaltenegger_2021Natur.594..505K} to name a few).

However, when a colour-magnitude diagram is plotted against the $G-RP$ colour, there is a substantial number of objects showing an offset towards the red from the main-sequence.
Neither the number of such outliers nor their position on a CMD can be explained by the object's intrinsic properties, such as age, metallicity or binarity, or a combination thereof. 
These offsets are also not consistent with expected numbers and positions of pre-main sequence stars.

Here we recall that the $BP$ and $RP$ fluxes are the integrated mean fluxes from the \gaia{} photometric instrument, and these fluxes are obtained from CCD windows of $3.5^{\prime\prime}\times2.1^{\prime\prime}$ size downlinked to Earth, whereas the broad-band $G$ flux is determined from PSF or LSF fitting to an object window from the astrometric field (AF) CCDs \citep{gaiaedr3_documentation_ch5_2021gdr3.reptE...5B}.

If there is an additional object in the window of the source (due to, e.g., a crowded field or a binary), the measured flux will be affected by blending and the flux value will be overestimated. This primarily affects $BP$ and $RP$ bands; contamination in the $G$-band is rather infrequent due to the smaller AF window size ($0.72^{\prime\prime}\times2.1^{\prime\prime}$; \citealt{gaia_mission_2016A&A...595A...1G}).

With this in mind, when using the $G-RP$ colour -- that is, combining unbiased $G$ flux with the overestimated $RP$ flux -- blended sources or poorly resolved binaries should appear redder than well-behaved single sources, just as observed in a CMD using $G-RP$ on the abscissa.

Affected sources can be identified without prior knowledge of their position in a CMD.
For instance, the impact of blending can be assessed with a metric defined as a simple ratio between the total number of blended transits in both $BP$ and $RP$ bands and the total number of observations in these bands \citep{gaiaedr3_photometry}:
\begin{equation}
    \beta = \frac{N_{BP_{\rm blended}}+N_{RP_{\rm blended}}}{N_{BP}+N_{RP}},
\end{equation}
where $N_{BP_{\rm blended}}$ and $N_{RP_{\rm blended}}$ correspond to the number of transits flagged as blended in the $BP$ and $RP$ bands, respectively (\texttt{phot\_bp\_n\_blended\_transits} and \texttt{phot\_rp\_n\_blended\_transits} columns in the \gedrthree{} \texttt{gaia\_source} table), while $N_{BP}$ and $N_{RP}$ are the total number of observations that were used to compute the integrated mean flux and its error in the $BP$ and $RP$ bands, respectively (\texttt{phot\_bp\_n\_obs} and \texttt{phot\_rp\_n\_obs} columns in \gedrthree{}).

Furthermore, \gedrthree{} contains various quality indicators which hint at the possibility that an object is a semi-resolved binary (optical or physical). In particular, \texttt{ipd\_frac\_multi\_peak} corresponds to the percentage of windows where more than one peak was detected by the Image Parameter Determination (IPD) algorithm \citep{gaiaedr3_astromertry}.

\begin{figure}
    \centering
    \includegraphics[width=0.48\textwidth]{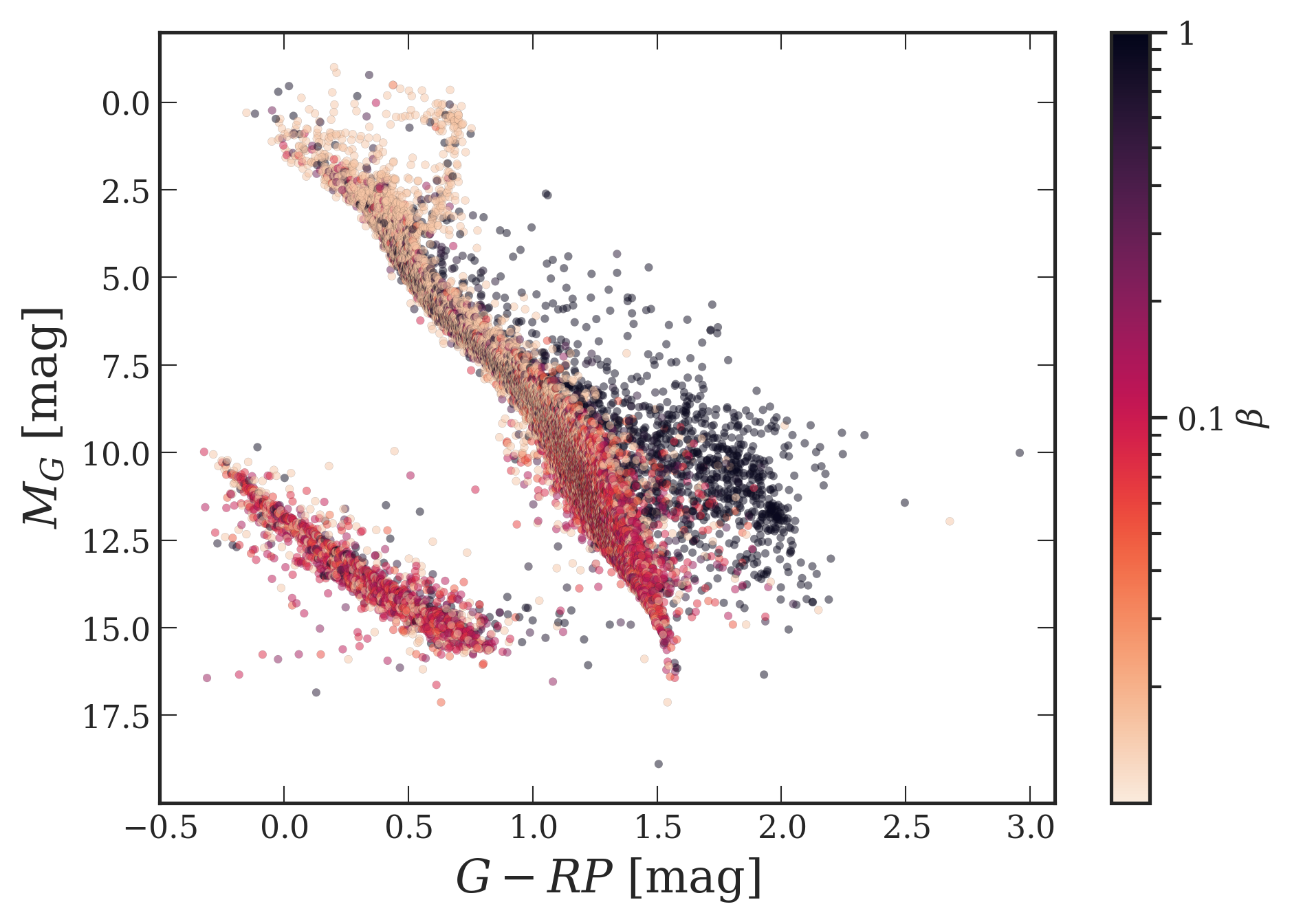}
    \includegraphics[width=0.48\textwidth]{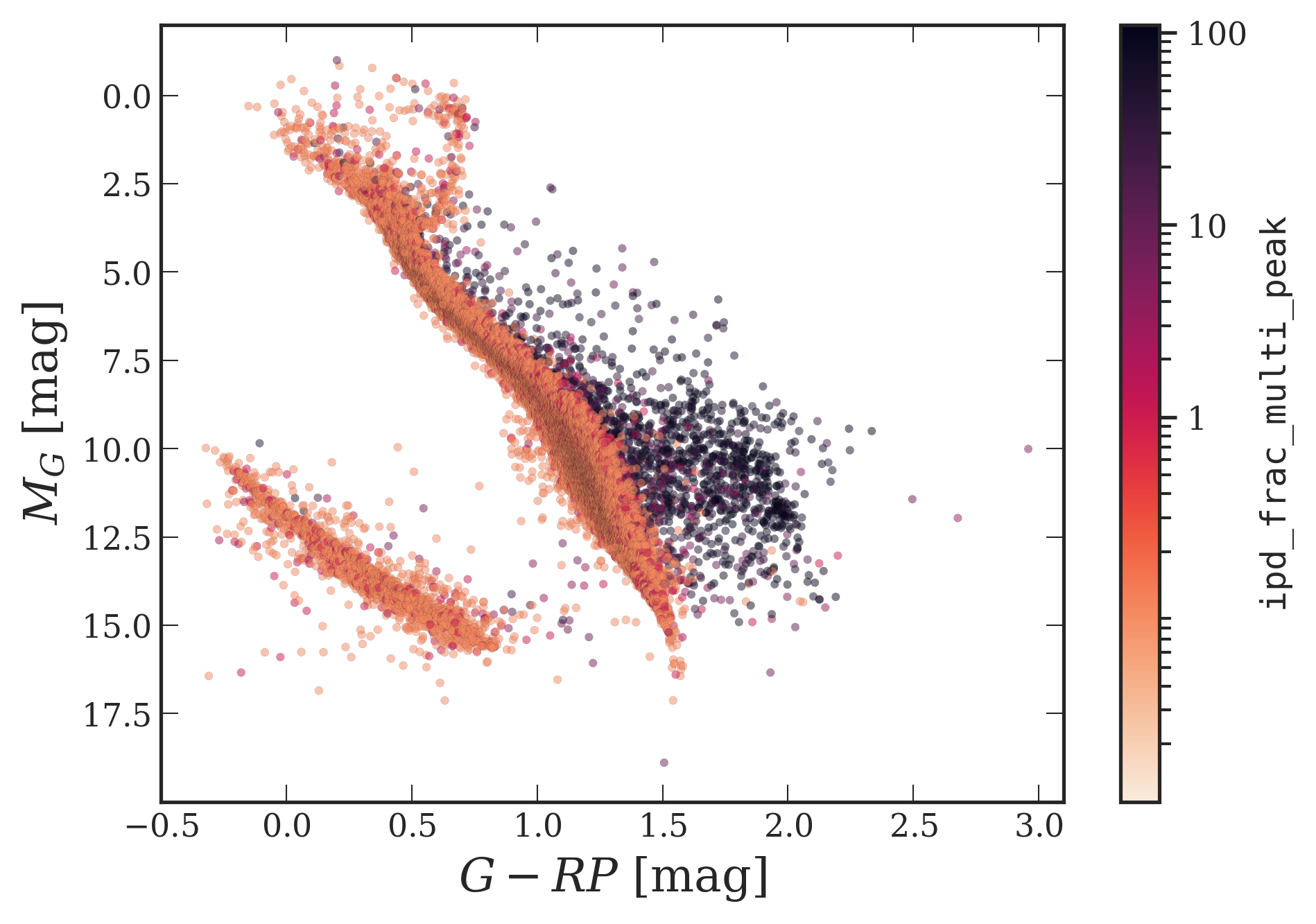}
    \caption{Colour-magnitude diagram for the 50 pc sample colour-coded with the blending metric $\beta$ (\textit{top panel}) and \texttt{ipd\_frac\_multi\_peak} (\textit{bottom panel}). Objects with large values of these two metrics exhibit an anomalous location on the diagram. A few extreme outliers are outside of the frame.}
    \label{fig:HRD_beta_multi_peak}
\end{figure}

In \figref{fig:HRD_beta_multi_peak} we demonstrate that the outliers discussed above can indeed be attributed to  inconsistencies in the photometry of blended objects (\textit{top panel}) or poorly resolved binaries (\textit{bottom panel}), or both.

Even though deblending and decontamination is not implemented in \gedrthree{}, it is already now possible to significantly improve the $G-RP$ photometry for blended and contaminated objects.

We start with the following fundamental assumption: while blending\footnote{Hereafter, we do not distinguish blending (the angular distance between the sources is so small that they fall into the same CCD window or two or more truncated windows in  the  vast  majority of the  transits) from contamination (flux of the source is altered by a bright source which is well outside of the window) and the terms are used interchangeably.}
biases the measured values of $BP$ and $RP$ fluxes, it results only in negligible changes in their ratio (i.e. $BP-RP$ colour). This holds especially true if the colour of both, the contaminating source and the contaminated source, is similar.

Therefore, the effect of blending can be dramatically reduced by defining a synthetic $G-RP$ colour, which can be derived from an empirical relationship between the $G-RP$ and the $BP-RP$ colour as described in the following.

In order to derive an empirical relationship between the $G-RP$ and $BP-RP$ colours, we construct a sample of well-behaved objects in \gedrthree{} with distances up to 100\,pc.
Here, one wants 
to use the largest possible sample from \gedrthree{}. However, at larger distances reddening starts to become significant and can bias our calibration.
Therefore, we selected only objects with parallaxes larger than 10\,mas (corresponds to the distance limit of 100\,pc) where reddening is negligible.

For this calibration, it is of fundamental importance to select only objects with excellent photometry, whereas the completeness of the sample is not crucial. Therefore, we apply additional stricter cuts on the astrometrically clean dataset.

We select only sources with high signal-to-noise ratio in both, $BP$ and $RP$ bands ($S/N>20$; the threshold is denoted by the grey lines in \figref{fig:bp_rp_flux_over_error}).
Furthermore, we exclude bright objects severely affected by saturation; namely, we consider only sources with $G>6~{\rm mag}$ \citep{gaiaedr3_astromertry, gaiaedr3_photometry}.

To ensure that we have a sample of excellent data, we select only sources with five-parameter astrometric solutions in \gedrthree{}.
In this way we also avoid systematic effects due to the uncertainty of the $G$-band corrections for the sources with six-parameter astrometric solutions \citep{gaiaedr3_photometry}.

Blended sources and (poorly) resolved binaries were removed by applying cuts on the blending metric $\beta$ and the  \texttt{ipd\_frac\_multi\_peak} statistic.

Furthermore, we excluded white dwarfs from our sample due to their larger scatter in colour-colour diagram, and limited our sample to colours in the range $0.0\magrm \leq (BP-RP) \leq 4.25~{\rm mag}$. Outside of this range the data density drops significantly and, consequently, this would yield an unreliable fit for sources with extreme colours.

To summarise, we constructed the initial sample by applying the following selection criteria:

\begin{equation}
    \label{eq:synth_snr}
    \begin{aligned}\setcounter{mysubequations}{0}
    \text{\mysubnumber}\quad& \varpi > 10\ \mathrm{mas}, \\
    \text{\mysubnumber}\quad& \text{\equref{eq:ipd_gof_harm_ampl_cut}:}\ A_{\rm GoF} < 10^{-5.12}~(\varpi/\sigma_\varpi)^{2.61}, \\
    \text{\mysubnumber}\quad&\texttt{phot\_bp\_mean\_flux\_over\_error} >20, \\
    \text{\mysubnumber}\quad&\texttt{phot\_rp\_mean\_flux\_over\_error} >20, \\
    \text{\mysubnumber}\quad& G > 6.0~{\rm mag}, \\
    \text{\mysubnumber}\quad& \texttt{astrometric\_params\_solved} = 31, \\
    \text{\mysubnumber}\quad& \beta < 0.1, \\
    \text{\mysubnumber}\quad& \texttt{ipd\_frac\_multi\_peak} \leq 2.0, \\
    \text{\mysubnumber}\quad&  G+5\log_{10}(\varpi/100)\magrm > 10.0~{\rm mag}+2.5 (BP-RP), \\
    \text{\mysubnumber}\quad& 0.0 \magrm \leq (BP-RP) \leq 4.25~{\rm mag}.
    \end{aligned} \quad
\end{equation}

\begin{figure}
    \centering
    \includegraphics[width=0.49\textwidth]{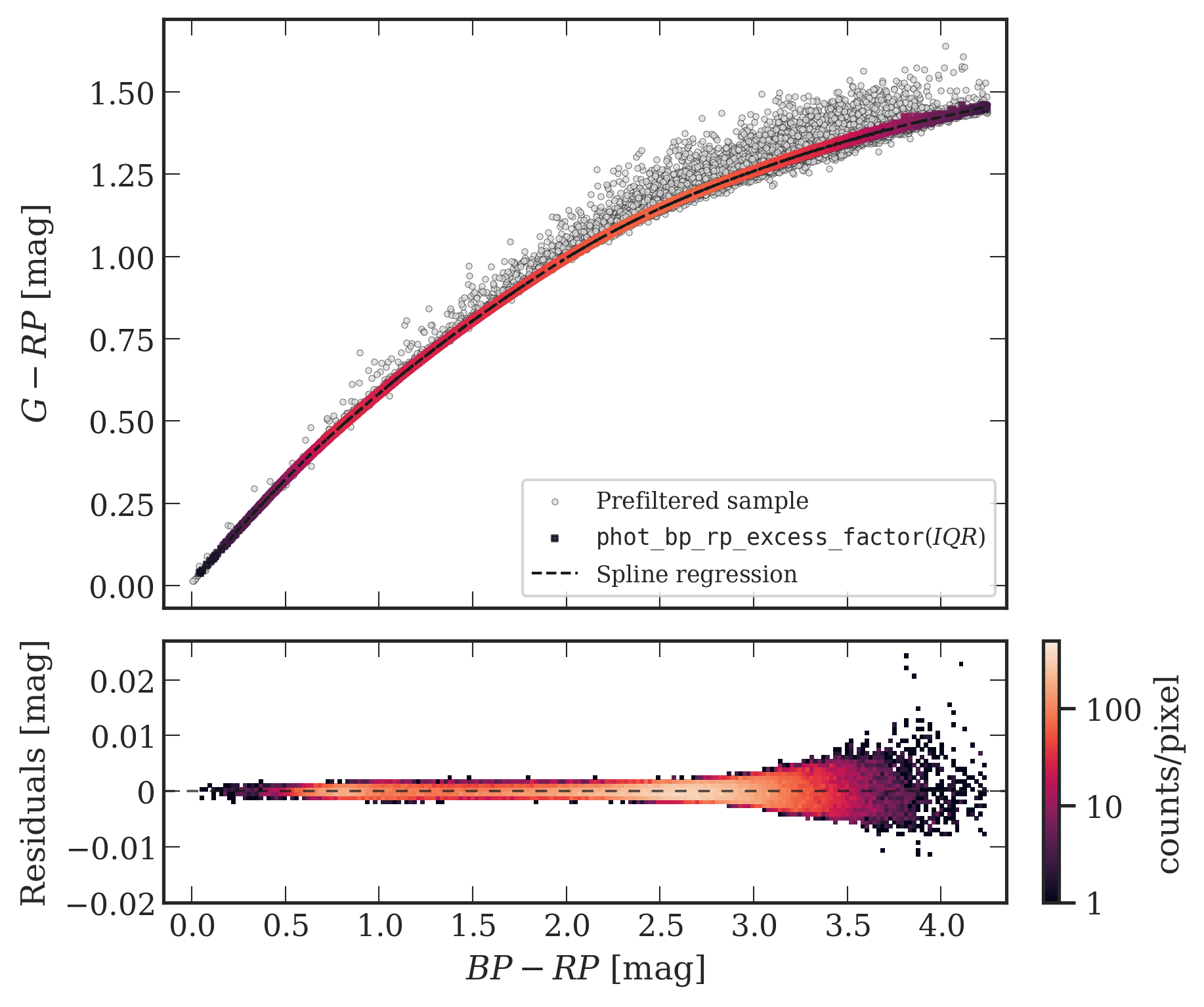}
    \caption{Colour-colour diagram for the prefiltered sample (shown in grey) and the truncated sample, with the colour indicating the number density. The dashed line depicts the model derived using spline regression. The residuals after subtracting the derived relationship are shown in the lower panel.}
    \label{fig:fit_g_rp_synth}
\end{figure}

Finally, in order to obtain a more robust statistics, we construct a truncated sample by using the $BP$ and $RP$ excess factor (\texttt{phot\_bp\_rp\_excess\_factor} in the \gedrthree{} catalogue) for deriving a trimmed estimator. 
This is done by splitting the initial sample (grey points on \figref{fig:fit_g_rp_synth}, \textit{top panel}) into bins of 0.005\,mag in the $BP-RP$ colour and retaining only objects with $BP$ and $RP$ excess within the interquartile range (IQR) values of \texttt{phot\_bp\_rp\_excess\_factor} estimated for each of these bins. There are 91\,285 sources in the resulting sample (the coloured part of \figref{fig:fit_g_rp_synth}).
The sample contains only sources with high-quality photometry. 

In the next step, we derive an empirical model for the synthetic $G-RP$ colour, where input is $BP-RP$ colour. The aim is to mitigate the inconsistencies between $G$ and $RP$ photometry caused by blending or marginally resolved binaries.

To derive the fit of $G-RP$ for isolated well-behaved sources, we use cubic splines with 500 equally spaced knots. Even though 500 knots seems like a large number, this is still an appropriate choice given the large size of our sample (each knot contains on average $\sim180$ data points). Furthermore, we wish to emphasise that in our case reducing the interpolation error and thus getting smaller residuals has higher priority than avoiding the problem of overfitting.

The residuals of the derived relationship are shown in \figref{fig:fit_g_rp_synth} (\textit{bottom panel}). 
They are very small, do not show any systematics, and illustrate the high quality of the derived relationship. The RMS error is 0.0013 mag.

The derived relationship is applicable in the range $0.0\magrm<BP-RP<4.25 \, \mathrm{mag}$. 
Another constraint to be considered when applying the relationship is the S/N of $BP$ and $RP$ fluxes in the input.
Including sources with a low S/N will lead to unreliable $G-RP$ estimates. Thus, to avoid this, we adopted $S/N=20$ as the optimal threshold value.

\begin{figure}
    \centering
    \includegraphics[width=0.49\textwidth]{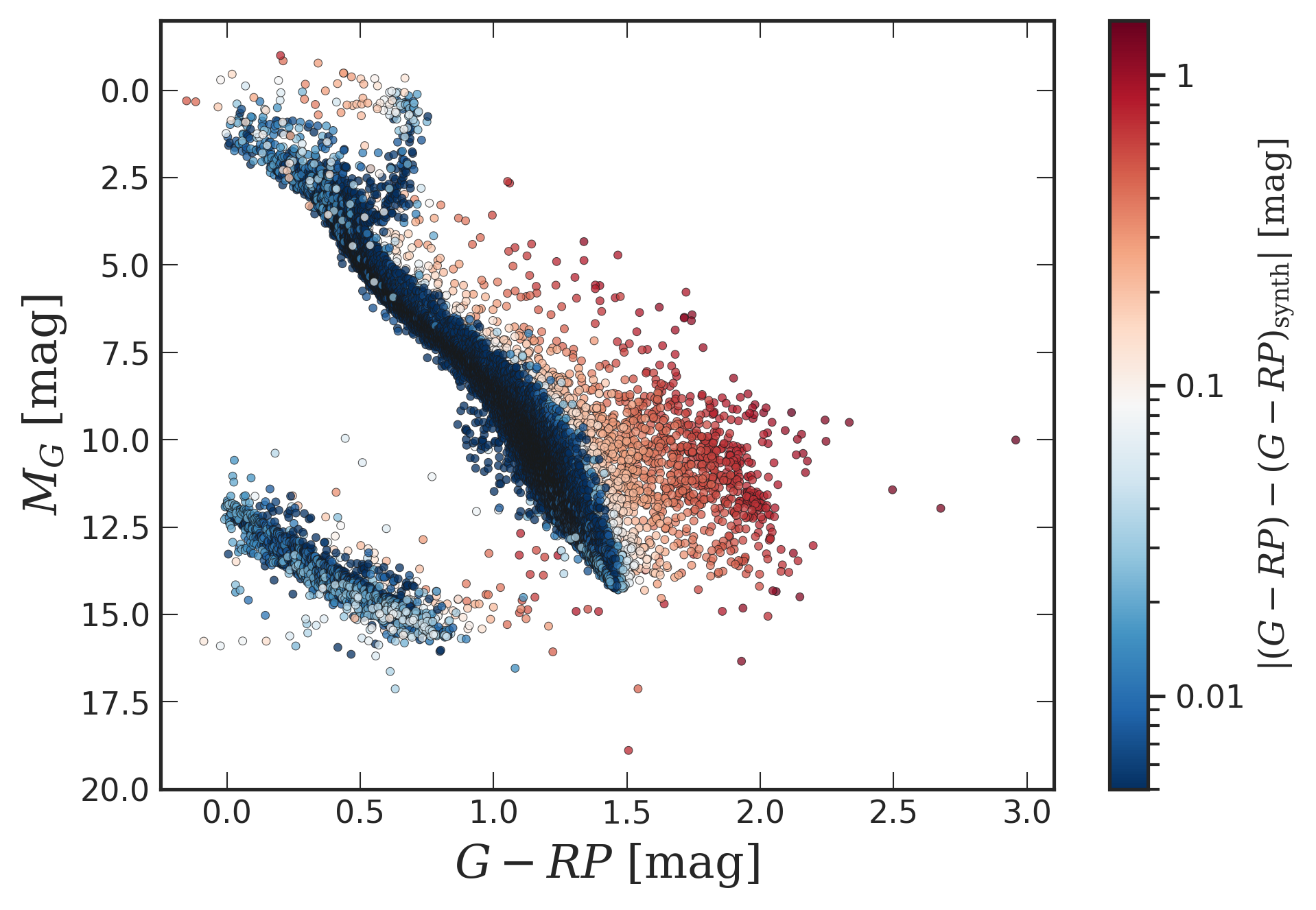}
    \includegraphics[width=0.49\textwidth]{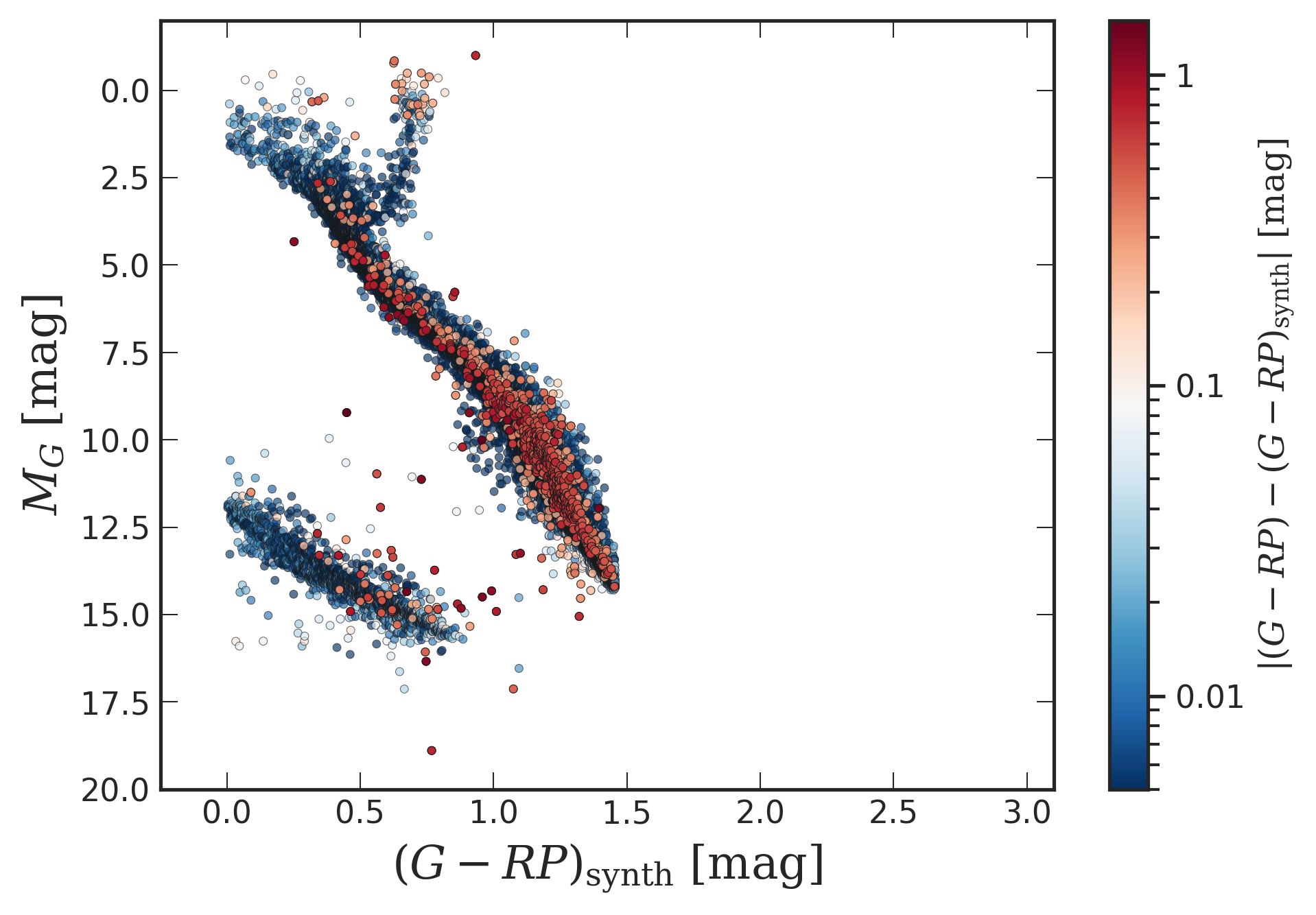}
    \caption{Difference between the measured and deblended $G-RP$ colours for objects in the 50~pc sample and their location on the CMD (using measured $G-RP$ colour). Only objects with $BP-RP$ colours within the applicability range $0.0\magrm < BP-RP < 4.25~{\rm mag}$ are plotted. The colour scale is logarithmic.}
    \label{fig:HRD_g_rp_synth_o_c}
\end{figure}

To illustrate the power of the suggested correction, CMDs before and after deblending are shown for the 50~pc sample (Fig.~\ref{fig:HRD_g_rp_synth_o_c}). 
The colour-coding indicates the difference between the catalogued and deblended values of $G-RP$.
When a CMD is plotted against the catalogued (i.e.\ uncorrected) $G-RP$ colour values, the objects with large residuals appear mostly at unexpected regions of the diagram (Fig.~\ref{fig:HRD_g_rp_synth_o_c}, \textit{top panel}) and their position in a CMD is not determined by intrinsic properties of an object, but is a consequence of inconsistency in photometry.
However, when the deblended colour is used (Fig.~\ref{fig:HRD_g_rp_synth_o_c}, \textit{bottom panel}), they are located in the expected regions of a CMD.
Here, we stress that the objects which have peculiar positions in the top panel were not removed, but are plotted with the deblended $G-RP$ colour and are now located on the main sequence, white dwarf sequence, and in the red clump region.

Deblended values of $G-RP$ are listed in the CNS5 for all objects with colours within the applicability range and with sufficient S/N of $BP$ and $RP$ fluxes.

We provide the derived spline coefficients and a python function to calculate deblended $G-RP$ values\footnote{\url{https://github.com/AlexGolovin/gaiaedr3_g_rp_synthetic}}.

Deblending of colours and magnitudes for objects with inconsistencies in photometry is crucial for correct object classification based on the position in the CMD, estimating effective temperatures and fitting isochrones.

\end{appendix}
\end{document}